\documentclass[aps,prd,twocolumn,tightenlines,preprintnumbers,showpacs,superscriptaddress,notitlepage,nofootinbib,eqsecnum,floatfix,longbibliography,10pt]{revtex4-1}

\usepackage{amsmath, amssymb}
\usepackage{graphicx}  
\usepackage{braket}
\usepackage{slashed}
\usepackage{hyperref}
\hypersetup{
    colorlinks=true,       
    linkcolor=blue,          
    citecolor=blue,        
    filecolor=blue,      
    urlcolor=blue           
}
\usepackage{placeins}
\usepackage[dvipsnames]{xcolor}
\usepackage{multirow}
\usepackage[caption=false]{subfig}
\usepackage[normalem]{ulem}

\newcommand{\BaMS}{0.5240}
\newcommand{\BaMSL}{17}
\newcommand{\BaMSP}{54}
\newcommand{\BbMS}{0.4794}
\newcommand{\BbMSL}{25}
\newcommand{\BbMSP}{35}
\newcommand{\BcMS}{0.746}
\newcommand{\BcMSL}{13}
\newcommand{\BcMSP}{17}
\newcommand{\BdMS}{0.897}
\newcommand{\BdMSL}{02}
\newcommand{\BdMSP}{10}
\newcommand{\BeMS}{0.6882}
\newcommand{\BeMSL}{78}
\newcommand{\BeMSP}{94}

\newcommand{\RbMS}{-18.90}
\newcommand{\RbMSL}{12}
\newcommand{\RbMSP}{17}
\newcommand{\RcMS}{5.92}
\newcommand{\RcMSL}{05}
\newcommand{\RcMSP}{13}
\newcommand{\RdMS}{41.94}
\newcommand{\RdMSL}{44}
\newcommand{\RdMSP}{46}
\newcommand{\ReMS}{10.64}
\newcommand{\ReMSL}{14}
\newcommand{\ReMSP}{15}

\graphicspath{./fig/}

\newcommand{\Fig}[1]{Figure~{#1}}
\newcommand{\Tab}[1]{Table~{#1}}
\newcommand{\be}{\begin{equation}}
\newcommand{\ee}{\end{equation}}

\newcommand\g[1]{\gamma_{#1}}
\newcommand{\matel}[2]{\braket{\bar{#2}|#1|#2}}
\newcommand{\matrixel}[3]{\ensuremath{ \left< #1 \vphantom{#2#3} \right| #2 \left| #3 \vphantom{#1#2} \right>} }

\newcommand{\mobius}{M\"{o}bius }
\newcommand{\MGamma}[1]{M_{\Gamma_{#1}}^{s_{#1}}}
\newcommand{\bMGamma}[1]{{\bar M}_{\Gamma_{#1}}^{s_{#1}}}
\newcommand{\opP}{\mathbb{P}}

\newcommand{\la}{\langle}
\newcommand{\ra}{\rangle}

\newcommand{\s}[1]{\slashed{#1}}
\def\pdv#1#2{\frac{\partial #1}{\partial #2}}

\makeatletter
\def\fslash#1{{\mathpalette\c@ncel{#1}}} 
\makeatother

\newcommand\cern{CERN, Theoretical Physics Department, Geneva, Switzerland}
\newcommand\bnl{Brookhaven National Laboratory, Upton, NY 11973, USA}

\newcommand\higgs{Higgs Centre for Theoretical Physics, School of Physics \& Astronomy, The University of
  Edinburgh, Edinburgh EH9 3FD, UK}
\newcommand\soton{School of Physics and Astronomy, University of
  Southampton,  Southampton SO17 1BJ, UK}
\newcommand\livpool{Theoretical Physics Division, Department of Mathematical Sciences, University of Liverpool, Liverpool L69 3BX, UK}
\newcommand\hope{School of Mathematics, Computer Science and Engineering, Liverpool Hope University, Hope Park, Liverpool L16 9JD, UK}

\begin{document}

\title{Kaon mixing beyond the standard model with physical masses}
\author{P.~A.~Boyle}\affiliation{\bnl}\affiliation{\higgs}
\author{F.~Erben}\affiliation{\cern}
\author{J.~M.~Flynn}\affiliation{\soton}
\author{N.~Garron}\affiliation{\hope}\affiliation{\livpool}
\author{J.~Kettle}\affiliation{\higgs}
\author{R.~Mukherjee}\affiliation{\soton}
\author{J.~T.~Tsang}\thanks{Corresponding author}\affiliation{\cern}

\collaboration{RBC and UKQCD Collaborations}\noaffiliation

\begin{abstract}
We present non-perturbative results for beyond the standard model kaon mixing
matrix elements in the isospin symmetric limit ($m_u=m_d$) of QCD, including a
complete estimate of all dominant sources of systematic error. Our results are
obtained from numerical simulations of lattice QCD with $N_f = 2+1$ flavours of
dynamical domain wall fermions. For the first time, these quantities are
simulated directly at the physical pion mass $m_\pi$~$\sim$~$139\,\mathrm{MeV}$
for two different lattice spacings. We include data at three lattice spacings in
the range $a = 0.11 $ -- $ 0.07\,\mathrm{fm}$ and with pion masses ranging from
the physical value up to 450$\,\mathrm{MeV}$. Compared to our earlier work, we
have added both direct calculations at physical quark masses and a third lattice
spacing making the removal of discretisation effects significantly more precise
and eliminating the need for any significant mass extrapolation beyond the range
of simulated data. We renormalise the lattice operators non-perturbatively using
RI-SMOM off-shell schemes. These schemes eliminate the need to model and
subtract non-perturbative pion poles that arises in the RI-MOM scheme and, since
the calculations are performed with domain wall fermions, the unphysical mixing
between chirality sectors is suppressed. Our results for the bag parameters in
the $\overline{\mathrm{MS}}$ scheme at $3\,\mathrm{GeV}$ are
$B_K~\equiv~\mathcal{B}_1 = \BaMS(\BaMSL)(\BaMSP)$, $\mathcal{B}_2 =
\BbMS(\BbMSL)(\BbMSP)$, $\mathcal{B}_3 = \BcMS(\BcMSL)(\BcMSP)$, $\mathcal{B}_4
= \BdMS(\BdMSL)(\BdMSP)$ and $\mathcal{B}_5 = \BeMS(\BeMSL)(\BeMSP)$, where the
first error is from lattice uncertainties and the second is the uncertainty due
to the perturbative matching to $\overline{\mathrm{MS}}$.
\end{abstract}

\maketitle
\preprint{CERN-TH-2024-040}
\preprint{LTH 1366}
\pagenumbering{gobble}

\pagenumbering{arabic}

\makeatletter
\def\l@subsection#1#2{}
\def\l@subsubsection#1#2{}
\makeatother

\setcounter{tocdepth}{0}
\tableofcontents
\refstepcounter{section}
\setcounter{section}{0}

\section{Introduction}
\label{sec:Introduction}
\subsection{Standard model kaon mixing}
\begin{figure}
  \centering
  \includegraphics[width=.3\textwidth]{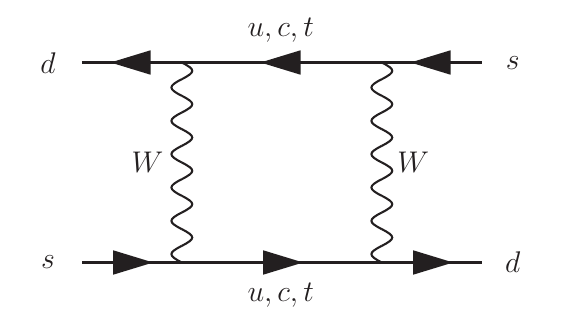}
  \includegraphics[width=.3\textwidth]{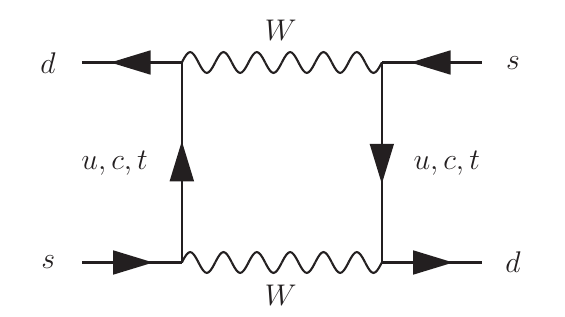}
  \caption{W exchange box diagrams mediating neutral kaon mixing in the standard model.}
  \label{fig:box}
\end{figure}

Neutral kaon mixing has long been an important area of study in standard model
(SM) particle physics. Most famously, CP violation was first observed in the
Nobel-prize-winning Christenson-Cronin-Fitch-Turlay
experiment~\cite{PhysRevLett.13.138}. Interested readers are referred to
Refs.~\cite{Buras:2020xsm,Bigi:2000yz,Branco:1999fs} and references therein.
Kaon mixing is mediated by a flavour changing neutral current interaction, which
is absent at tree-level, whereby the neutral kaon oscillates with its
antiparticle.  The leading-order SM processes are the well-known W exchange box
diagrams shown in \Fig{\ref{fig:box}}.

Typically, one separates the short and long distance contribution to this
process using the operator product expansion (OPE). This isolates the
non-perturbative matrix element which can be computed using lattice QCD,
$\braket{\bar{K}^0|O_1|K^0}$, of the (vector$-$axial)$\times$(vector$-$axial)
four-quark left handed operator $O_1$ from the perturbatively computed Inami-Lim
functions~\cite{Inami:1980fz}. Conventionally, the hadronic contribution to the
matrix element is parametrised by the kaon bag parameter and has been the
subject of many lattice calculations~\cite{Aoki:2010pe,RBC:2014ntl, Bae:2013tca,
  SWME:2015oos, Bae:2014sja, Carrasco:2015pra, Bertone:2012cu}. It is now known
at the percent-level and is reported in the
FLAG~\cite{FlavourLatticeAveragingGroupFLAG:2021npn} review with consistent
results from multiple collaborations.

The SM operator matrix element is a factor in the expression for the dominant
short-distance contribution to the indirect CP-violation parameter
$\varepsilon_K$. Lattice computations of the matrix element have reached the
point where isospin and electromagnetic effects are as large as the total error
quoted in the isospin-symmetric pure QCD theory. However, including these
effects would need to be accompanied by reduced uncertainty in the
Cabibbo-Kobayashi-Maskawa (CKM) quark-mixing matrix element $|V_{cb}|$, also
appearing in the SM short-distance contribution to
$\varepsilon_K$~\cite{Buchalla:1995vs,FlavourLatticeAveragingGroupFLAG:2021npn}.

For greater precision in the short-distance contribution to be meaningful, we
would in addition need to include long-distance effects from bi-local V--A
currents where two weak Hamilton insertions are connected by quark
loops. Initial progress has been made in
references~\cite{Christ:2012np,Christ:2015phf,Bai:2016gzv}.  The precision on
the computation of $\Delta M_K$, including all long-distance effects, is already
at the 10\% level~\cite{Wang:2022lfq}. A first full computation of the
long-distance part of $\varepsilon_K$ at the level of 40\% precision has
recently been reported in Ref.~\cite{Bai:2023lkr}. For recent broader reviews of
the status and prospects of kaon physics we refer the interested reader to
Refs.~\cite{Anzivino:2023bhp,Buras:2023qaf,Goudzovski:2022scl}.

\subsection{Beyond the standard model kaon mixing}
Beyond the standard model (BSM), new mediating particles could contribute to
neutral kaon mixing. These mediators are not restricted to the V--A Dirac
structure of the W boson and new four-quark operators would be allowed in the
effective Hamiltonian with Wilson coefficients suppressed by the new mass scale.
Eight such four-quark operators are allowed, but for computing their $K^0\bar
K^0$ matrix elements only five parity-even operators are needed, shown in
Eq.~\eqref{eq:opbasis_susy}, which are model independent and whose hadronic
matrix elements can be calculated using lattice QCD. Any BSM contributions to
the V--A structure would be very hard to distinguish from the SM signal in
experiments.  However the new colour-Dirac structure operators have no SM
contribution at this order in the weak forces and thus new physics arising from
these operators would be easier to detect in experiment. In addition, the matrix
elements of these BSM operators are enhanced in the chiral limit compared to the
SM operator, as can be seen from their chiral expansions.  Therefore our results
can be combined with the experimental value of $\varepsilon_K$ to constrain the
parameter space of specific BSM theories and the scale of New Physics, see for
example Ref.~\cite{ETM:2012vvy}.

Lattice QCD appears as a natural candidate for computing the BSM operator matrix
elements. However the mixing pattern of these four-quark operators makes this
more challenging than the computation for $B_K$ alone.  We refer the interested
reader to the pedagogical review~\cite{Lellouch:2011qw}. Here we take advantage
of the good chiral-flavour properties of the domain wall fermion formulation to
constrain the mixing to be the same as in the continuum theory. In practice,
this is only true up to lattice artefacts that are exponentially suppressed in
the extent of the domain wall fifth dimension and that we must keep under
control.

Early studies of BSM kaon mixing~\cite{Allton:1998sm, Donini:1999nn,
  Babich:2006bh} were performed in the quenched approximation.  They were
followed by dynamical simulations with $N_f$ quark flavours by several
collaborations: RBC-UKQCD $(N_f = 2+1)$~\cite{Boyle:2012qb, Garron:2016mva},
SWME $(N_f = 2+1)$~\cite{Bae:2013tca, Jang:2014aea, SWME:2015oos}, and ETM $(N_f
= 2)$~\cite{Bertone:2012cu} and $(N_f=2+1+1)$~\cite{Carrasco:2015pra}.  In
contrast to results for the SM operator, there are tensions between the
different collaborations' results for some of the BSM operators, as shown in
table~\ref{tab:compBSM} --- in which we already anticipate the results of this
work --- and summarised in the FLAG
report~\cite{FlavourLatticeAveragingGroupFLAG:2021npn}. We note that a similar
discrepancy is observed in neutral $B_{(s)}$-mixing~\cite{Tsang:2023nay,
  FlavourLatticeAveragingGroupFLAG:2021npn}.

\begin{table*}
\caption{Results from calculations of BSM bag parameters in
  $\overline{\text{MS}}(\mu=3\,\textrm{GeV})$ from RBC-UKQCD, SWME and ETM show
  tensions for $\mathcal{B}_4$ and $\mathcal{B}_5$. The results obtained by ETM,
  which were renormalised via RI-MOM, agree with RBC-UKQCD's results obtained
  via RI-MOM.  The SWME results, obtained via a 1 loop intermediate scheme agree
  with RBC-UKQCD's results obtained via RI-SMOM, for both $\gamma_{\mu}$ and
  $\slashed{q}$~\cite{Boyle:2017jwu}. This suggests tensions arise from the
  implementation of intermediate schemes, in particular caused by RI-MOM
  exhibiting exceptional infrared behaviour which is absent in RI-SMOM. All
  results are shown in the SUSY basis.}
	\label{tab:compBSM}
	\begin{tabular}{ c | c c | c | c | c c || c}
		\hline \hline
		& ETM12~\cite{Bertone:2012cu} & ETM15~\cite{Carrasco:2015pra} & RBC-UKQCD12~\cite{Boyle:2012qb} &  SWME15~\cite{SWME:2015oos}& \multicolumn{2}{c||}{RBC-UKQCD16~\cite{Garron:2016mva}} & THIS WORK \\
		\hline \hline
		$N_f$ & 2 & 2+1+1 & 2+1 & 2+1  & 2+1 & 2+1 & 2+1\\
		scheme & RI-MOM & RI-MOM & RI-MOM & 1 loop & RI-SMOM & RI-MOM & RI-SMOM\\
		\hline
		$\mathcal{B}_2$ & 0.47(2) & 0.46(3)(1) & 0.43(5) & 0.525(1)(23) & 0.488(7)(17) & 0.417(6)(2) & \BbMS(\BbMSL)(\BbMSP)\\
		$\mathcal{B}_3$ & 0.78(4) & 0.79(5)(1) & 0.75(9) & 0.773(6)(35) & 0.743(14)(65) & 0.655(12)(44) & \BcMS(\BcMSL)(\BcMSP) \\
		$\mathcal{B}_4$ & 0.76(3) & 0.78(4)(3) & 0.69(7) & 0.981(3)(62) & 0.920(12)(16) & 0.745(9)(28) & \BdMS(\BdMSL)(\BdMSP) \\
		$\mathcal{B}_5$ & 0.58(3) & 0.49(4)(1) & 0.47(6) & 0.751(7)(68) & 0.707(8)(44) & 0.555(6)(53) & \BeMS(\BeMSL)(\BeMSP) \\
		\hline \hline
	\end{tabular} 
\end{table*}

In references~\cite{Garron:2016mva, Boyle:2017skn}, it was proposed that the
source of these tensions was the choice of the intermediate renormalisation
scheme.  Specifically, it was proposed that the symmetric momentum subtraction
scheme RI-SMOM (which has non-exceptional kinematics) advocated by RBC-UKQCD has
several beneficial features compared to the previously used RI-MOM (which has
exceptional kinematics). This is likely due to the exceptional (divergent in the
massless limit), infrared non-perturbative ``pion pole'' behaviour in the RI-MOM
vertex functions, which must be correctly modelled and subtracted, while the
mass is simultaneously taken to zero to establish a mass independent
scheme. This behaviour is absent in the RI-SMOM scheme, giving greater
theoretical control as it avoids the possibility of imperfect modelling of the
non-perturbative pole systematically affecting the result. The results obtained
from two RI-SMOM schemes are in agreement with each other and with the
perturbatively renormalised results from the SWME
collaboration~\cite{Bae:2013tca, Jang:2014aea, SWME:2015oos}, while the
calculation with RI-MOM agreed with previous RBC-UKQCD~\cite{Boyle:2012qb} and
ETM~\cite{Bertone:2012cu,Carrasco:2015pra} results which also used RI-MOM.

This paper improves upon our most recent RBC-UKQCD BSM kaon mixing
calculation~\cite{Garron:2016mva,Boyle:2017skn} by adding a third lattice
spacing and including two data points at the physical light quark mass.  We
present results in the isospin symmetric limit of pure $N_f=2+1$ QCD with
sufficient precision that further work on this topic must address the strong and
electromagnetic isospin breaking effects. The status of this work has been
previously reported in Refs.~\cite{Boyle:2017ssm,Boyle:2018eor}.  Finally, it is
worth noting that a similar analysis performed in the pion sector allows to
extract the matrix elements which could dominate the short-distance contribution
to neutrino-less double beta decays, see for example
Refs.~\cite{Nicholson:2018mwc, Detmold:2022jwu}. In particular, the
renormalisation factors computed here could be employed for such a study.

\section{Background}
\label{sec:Background}
\subsection{Effective weak Hamiltonian and BSM basis}
By integrating out heavy particles such as the $W$ boson we can
separate the long- and short-distance effects into matrix elements,
$\braket{\bar{K}^0|O_i|K^0}$, and Wilson coefficients respectively.
Beyond the standard model a generic effective weak $\Delta S =
2$ Hamiltonian can be constructed, in which the standard model
operator, $O_1$ below, and seven additional four-quark operators
appear
\begin{equation}
  \mathcal{H}^{\Delta S = 2} = \sum_{i=1}^5 C_i(\mu) O_i(\mu) +
    \sum_{i=1}^3 \tilde{C_i}(\mu) \tilde{O_i}(\mu)\,,
    \label{eq:ds2Hamilton}
\end{equation}
where $\mu$ is a renormalisation scale. The Wilson coefficients
$C_i(\mu)$ depend on the BSM physics, but the QCD matrix elements
$\langle\bar K^0|O_i|K^0\rangle$ do not. The operators $O_i$ in the
so-called ``SUSY basis'' introduced in Ref.~\cite{Gabbiani:1996hi} are
\begin{equation}
\begin{aligned}
O_1 &= \bar{s}_a \gamma_{\mu}(1-\gamma_{5})d_a\;
       \bar{s}_b \gamma_{\mu}(1-\gamma_{5})d_b, \\
O_2 &= \bar{s}_a (1-\gamma_{5})d_a\;\bar{s}_b (1-\gamma_{5})d_b,\\
O_3 &= \bar{s}_a (1-\gamma_{5})d_b\;\bar{s}_b (1-\gamma_{5})d_a,\\
O_4 &= \bar{s}_a (1-\gamma_{5})d_a\;\bar{s}_b (1+\gamma_{5})d_b,\\
O_5 &= \bar{s}_a (1-\gamma_{5})d_b\;\bar{s}_b (1+\gamma_{5})d_a,
\label{eq:opbasis_orig_susy}
\end{aligned}
\end{equation}
where $a,b$ are colour indices. The $\tilde O_{1,2,3}$ are parity
partners of $O_{1,2,3}$ obtained by swapping $1{-}\gamma_5 \to
1{+}\gamma_5$, while $O_{4,5}$ are parity-even. Owing to parity
invariance of QCD, only the parity even parts, denoted with a $+$
superscript, contribute in the $\braket{\bar{K}^0|O_i|K^0}$ matrix
elements:
\begin{equation}
\begin{aligned}
O^+_1 &= \bar s_a\gamma_\mu d_a\; \bar s_b\gamma_\mu d_b +
           \bar s_a\gamma_\mu\g5 d_a\; \bar s_b\gamma_\mu\g5 d_b,\\
O^+_2 &= \bar s_a d_a\;\bar s_b d_b + \bar s_a\g5 d_a\;\bar s_b\g5 d_b,\\
O^+_3 &= \bar s_a d_b\;\bar s_b d_a + \bar s_a\g5 d_b\;\bar s_b\g5 d_a,\\
O^+_4 &= \bar s_a d_a\;\bar s_b d_b - \bar s_a\g5 d_a\;\bar s_b\g5 d_b,\\
O^+_5 &= \bar s_a d_b\;\bar s_b d_a - \bar s_a\g5 d_b\;\bar s_b\g5 d_a.
\label{eq:opbasis_susy}
\end{aligned}
\end{equation}
In practice we find it convenient to work in a different basis, referred to as the ``lattice'' or ``NPR'' basis~\cite{Boyle:2017skn}. This comprises colour-unmixed operators obtained by Fierz transforming the equivalent colour-mixed operators, as detailed in section~\ref{sec:Fierz} of the appendix, with $Q_1=O_1$ and
\begin{equation}
\begin{aligned}
Q_2 &= \bar s_a \gamma_\mu(1-\g5)d_a\;
       \bar s_b \gamma_\mu(1+\g5)d_b, \\
Q_3 &= \bar s_a (1-\g5)d_a\;\bar s_b (1+\g5)d_b,\\
Q_4 &= \bar s_a (1-\g5)d_a\;\bar s_b (1-\g5)d_b,\\
Q_5 &= \frac14 \bar s_a \sigma_{\mu\nu} (1-\g5)d_a\;\bar s_b \sigma_{\mu\nu}(1+\g5)d_b.
\label{eq:opbasis_orig_npr}
\end{aligned}
\end{equation}
Again we need to consider only the parity-conserving parts which read
\begin{equation}
  \begin{aligned}
Q_1^{+} &= \bar s_a \gamma_\mu d_a\;\bar s_b \gamma_\mu d_b
          + \bar s_a \gamma_\mu\g5d_a\;\bar s_b \gamma_\mu\g5d_b,\\
Q_2^{+} &= \bar s_a \gamma_\mu d_a\;\bar s_b \gamma_\mu d_b
          - \bar s_a \gamma_\mu\g5d_a\;\bar s_b \gamma_\mu\g5d_b,\\
Q_3^{+} &= \bar s_a  d_a\;\bar s_b d_b
          - \bar s_a \g5d_a\;\bar s_b \g5d_b,\\
Q_4^{+} &= \bar s_a  d_a\;\bar s_b d_b
          + \bar s_a \g5d_a\;\bar s_b \g5d_b,\\
Q_5^{+} &= \sum_{\nu > \mu} \bar s_a \gamma_\mu\gamma_\nu d_a\;
     \bar s_b \gamma_\mu\gamma_\nu d_b.
\label{eq:opbasis_unmixed_even}
  \end{aligned}
\end{equation}
We perform the lattice calculations and renormalisation in this basis and
transform to the SUSY basis prior to performing the required chiral and
continuum limit extrapolations.\footnote{Unless stated otherwise all results in
  this paper are quoted in the SUSY basis.} Observe that under
$\mathrm{SU}(3)_L\times \mathrm{SU}(3)_R$ quark flavour symmetry, $O_1^{(+)}$
transforms as $(27,1)$, $O_{2,3}^{(+)}$ as $(6,\bar 6)$ and $O_{4,5}^{(+)}$ as
$(8,8)$, while $Q_1^{(+)}$ is $(27,1)$, $Q_{2,3}^{(+)}$ are $(8,8)$ and
$Q_{4,5}^{(+)}$ are $(6,\bar 6)$.

\subsection{Bag parameters}

The conventional way to parameterise the hadronic matrix elements of
the four-quark operators is through the so-called bag parameters, defined as the
ratio of the matrix elements over their vacuum saturation approximation (VSA)
value:
\begin{equation}
  \label{eq:BagGen}
  \mathcal{B}_i(\mu) = \frac{ \matel{O_i(\mu)}{K^0} }{\matel{O_i(\mu)}{K^0}}_{\textrm{VSA}}.
\end{equation}
For the standard model operator, $B_K \equiv \mathcal{B}_1$ is given by,
\begin{equation}
  \label{eq:BagSM}
  \mathcal{B}_1(\mu)= \frac{ \matel{O_1(\mu)}{K^0} }{\frac{8}{3}m_K^2 f_K^2},
\end{equation}
where $m_K$ is the mass of the kaon and $f_K$ is the kaon decay constant defined by the
coupling of the kaon to the renormalised axial-vector current $\mathbb{A}^{\mathrm{R}}_\mu$,
\begin{equation}
\braket{0|{\mathbb A}^{\rm R}_\mu(x)|K(p)} =
i f_K \,p_\mu \,e^{-ip \cdot x} \;,
\label{eq:decay_const_def}
\end{equation}
where $p_\mu$ is the 4-momentum of the kaon.
The BSM bag parameters are,
\begin{equation} 
  \label{eq:BagBSM}
  \mathcal{B}_i(\mu) = \frac{ (m_s(\mu) + m_d(\mu))^2 }{N_i m_K^4 f_K^2}   \matel{O_i(\mu)}{K^0}, \quad i>1 \;,
\end{equation}
and the factors $N_i$ in the SUSY basis are $N^{\rm{SUSY}}_{i>1} =
{-\frac{5}{3},\frac{1}{3},2,\frac{2}{3}}$. The corresponding factors in the
parity-even NPR basis are $N^{\rm{NPR}}_{i>1} =
{-\frac{4}{3},2,-\frac{5}{3},-1}$.  The VSA replaces the four-quark matrix
elements with products of two-quark matrix elements. When ``chirally enhanced''
matrix elements of the pseudoscalar density $\bar s \g5 d$ appear, matrix
elements of axial vector and tensor currents are dropped. This leads to the
appearance of the square of the ratio $(m_s(\mu)+m_d(\mu))/m_K^2f_K$ in the
$\mathcal{B}_i$ for $i>1$.

\subsection{Ratios $R_i$ of BSM to SM matrix elements}
The bag parameters are not the only way to parameterise these four-quark
operators; other quantities have been defined, for example in
Refs.~\cite{Donini:1999nn,Babich:2006bh,SWME:2013laf}.  Here we choose to
consider the simple ratios of the BSM to SM matrix elements
\begin{equation}
  \label{eq:RatiosBSMtoSM-naive}
  R_i\left(\mu\right) = \frac{\matel{O_i(\mu)}{P}}{
    \matel{O_1(\mu)}{P} },\qquad i=2,\ldots,5\,.
\end{equation}

There are some clear advantages: there is no explicit quark-mass dependence in
the expression, so that the BSM matrix elements can be recovered from knowledge
of $R_i$, the SM bag parameter $\mathcal{B}_1$ and the experimentally measured
kaon mass and decay constant (cf. Eq.~\eqref{eq:BagSM}). Additionally, the
similarity of the numerator and denominator leads to partial cancellation of
systematic and statistical errors. Another approach originally proposed in
Ref.~\cite{Becirevic:2004qd} is to consider quantities in which the chiral logs
cancel out (either at all orders or at NLO). This strategy has been employed for
example in Refs.~\cite{SWME:2013laf, Carrasco:2015pra, Garron:2016mva}.

\section{Simulation details and ensemble properties}
\label{sec:SimulationDetails}
\begin{table*}
  \caption{Summary of the main parameters of the ensembles used in this work.
    In the ensemble name the first letter (C, M or F) stand for coarse, medium
    and fine, respectively. The last letter (M or S) stands for M\"obius and
    Shamir kernels, respectively. The column $N_\mathrm{conf}$ denotes the
    number of de-correlated gauge field configurations, $N_\mathrm{src}$ the
    number of equivalent measurements per configuration. Ensembles labelled with
    `$\dagger$' only enter the analysis in order to constrain the chiral
    extrapolation of the renormalisation constants described in
    Section~\ref{sec:Renormalisation}.}
  \label{tab:enspar}
  \begin{tabular}{lcrcccllcl}
	 \hline
	 \hline
	 name & $L/a$ & $T/a$ & $a^{-1}[\textrm{GeV}]$ & $m_\pi[\textrm{MeV}]$ & $N_\mathrm{conf}\times N_\mathrm{src}$ & $am_l^{uni} $  & $am_s^{sea}$ & $am_s^{val}$ & $am_s^{phys}$  \\
	 \hline
	 C0M & $48$ & $96$  & 1.7295(38) & 139 & 90$\times$48  & 0.00078 & 0.0362 & 0.0358  & 0.03580(16) \\
	 C1S & $24$ & $64$ & 1.7848(50) & 340 & 100$\times$32 & 0.005   & 0.04   & 0.03224 & 0.03224(18) \\
	 C2S & $24$ & $64$ & 1.7848(50) & 431 &  99$\times$32 & 0.01    & 0.04   & 0.03224 & 0.03224(18) \\
	 \hline
	 M0M & $64$ & $128$& 2.3586(70) & 139 & 82$\times$64 & 0.000678& 0.02661 & 0.0254  & 0.02539(17) \\
	 M1S & $32$ & $64$ & 2.3833(86) & 304 & 83$\times$32 & 0.004   & 0.03    & 0.02477 & 0.02477(18) \\
	 M2S & $32$ & $64$ & 2.3833(86) & 361 & 76$\times$32 & 0.006   & 0.03    & 0.02477 & 0.02477(18) \\
	 M3S & $32$ & $64$ & 2.3833(86) & 411 & 80$\times$32 & 0.008   & 0.03    & 0.02477 & 0.02477(18) \\
	 \hline
	 F1M & $48$& $96$ & 2.708(10)  & 232 & 72$\times$48 & 0.002144 & 0.02144 & 0.02144 & 0.02217(16)  \\
	 \hline\hline
	 C1M$^\dagger$ & $24$ & $64$ & 1.7295(38) & 276   & -            & 0.005   & 0.0362 & -       & 0.03580(16) \\
	 M1M$^\dagger$ & $32$ & $64$ & 2.3586(70) & 286   & -            & 0.004   & 0.02661 & -       & 0.02539(17) \\
	 \hline\hline
  \end{tabular}
\end{table*}

\subsection{Simulation parameters}

We use RBC-UKQCD's $N_f = 2+1$ gauge ensembles~\cite{Allton:2008pn, Aoki:2010pe,
Aoki:2010dy, RBC:2014ntl, Boyle:2017jwu, Boyle:2018knm} generated with the
Iwasaki gauge action~\cite{Iwasaki:1984cj,Iwasaki:1985we} and a domain wall
fermion (DWF) action with either \mobius (M)~\cite{Brower:2004xi, Brower:2005qw,
Brower:2012vk} or Shamir (S)~\cite{Shamir:1993zy} kernel. In our work
the \mobius and Shamir kernels differ only in their approximation of the sign
function, and agree in the limit $L_s \rightarrow \infty$, where $L_s$ is the
size of the fifth dimension~\cite{RBC:2014ntl}. We hence assume that
the \mobius and Shamir kernels lie on the same scaling trajectory.  The set of
ensembles contains three lattice spacings in the range $a = 0.11 -
0.07\,\mathrm{fm}$, labelled C(oarse), M(edium) and F(ine), and includes two
physical pion mass ensembles. The remaining ensembles have heavier pion masses
ranging up to $m_\pi \approx 450\,\mathrm{MeV}$, which are used to guide the
small chiral extrapolation on the finest ensemble.

On each ensemble, the light valence-quark mass ($am_l^\mathrm{uni}$) was chosen
identical to the light-quark mass in the sea.  The strange valence quark mass
($am_s^\mathrm{val}$) was simulated near its physical value
($am_s^\mathrm{phys}$) which typically differs from the sea quark mass
($am_s^\mathrm{sea}$). The main ensemble properties and the simulated masses are
listed in \Tab~\ref{tab:enspar}. Large parts of the data were generated using
the Grid and Hadrons framework~\cite{Boyle:2016lbp, Yamaguchi:2022feu,
antonin_portelli_2023_8023716}.

The lattice scale and the physical light and strange quark masses were set using
the physical values of $m_\pi$, $m_K$ and $m_\Omega$~\cite{RBC:2014ntl} before
the ensemble F1M was generated. This fit was repeated including the ensemble F1M
in reference~\cite{Boyle:2018knm} where more details about this ensemble are
described. We also introduce two new ensembles ``C1M'' and ``M1M'', which are
the M\"obius equivalents of the C1S and M1S. Since they share the same gauge
coupling and M\"obius scale as the C0M and M0M, respectively, they have the same
lattice spacing and physical strange quark mass. More details about these
ensembles are summarised in Appendix~\ref{sec:newens}.

\subsection{Correlation functions}
The quark propagators $S_F(y,x)$, where we write $x=({\bf x},t_x)$, are obtained
by inverting the domain wall Dirac operators on ${\mathbb Z}_2$ noise wall
sources $\eta(x)$~\cite{Foster:1998vw,McNeile:2006bz,Boyle:2008rh}. In order to
improve the overlap with the ground state, these sources are Gaussian-smeared
following a Jacobi procedure, i.e.
\begin{equation}
    \eta_\omega({\bf x},t_\mathrm{src})
    = \sum_{\bf y} \omega^S_{\mathrm{src}}({\bf x},{\bf y})\, \eta({\bf y},t_y) \,\delta_{t_y,t_\mathrm{src}} \;,
\end{equation}
where we omit spin-colour indices for simplicity. Further details about the
smearing parameters defining $\omega$ can be found in
reference~\cite{Boyle:2018knm}.

At the sink, we consider both the local (L) and smeared (S) case,
\begin{equation}
   S_F^{L,S}(x,y) = \sum_{\bf z} \omega_\mathrm{snk}^{L,S}({\bf x},{\bf z}) S(z,y) \delta_{t_x,t_z} \;,
\end{equation}
where $\omega^{L}({\bf x},{\bf y})=\delta_{{\bf x},{\bf y}}$.  From these
propagators, we construct two-point functions which are defined by

\begin{equation}
  \begin{aligned}
    C_{\Gamma_1,\Gamma_2}^{s_1,s_2}(t) &\equiv 
    \sum_{{\bf x}} \Big{\langle} \Big( O_{\Gamma_2}^{s_2}({\bf x},t) \Big) \Big(  O_{\Gamma_1}^{s_1}({\bf 0},0) \Big)^\dagger \Big{\rangle} \\
    &= \sum_n \frac{\big( M_{\Gamma_2}^{s_2}\big)_n  \big( M_{\Gamma_1}^{s_1}\big)^*_n}{2 E_n} \bigg(e^{-E_n t}\pm e^{-E_n(T-t)}\bigg)\;,
    \label{eq:2ptdef}
  \end{aligned}
\end{equation}
where $O_{\Gamma}^{s}$ is a bilinear with the flavour content of a kaon, defined by
\begin{equation}
    O_{\Gamma}^{s}({\bf x},t) = \bigg( \bar{q}_2({\bf x},t) \sum_{\bf y} \omega_s({\bf x}, {\bf y}) \Gamma q_1({\bf y},t) \bigg) \; .
\end{equation}
The Dirac structure is represented by $\Gamma_i$.  The hadronic matrix elements
are denoted by $(\MGamma{})_n=\langle X_n | (O_{\Gamma}^{s})^\dagger | 0 \rangle $ (so that
$(\bMGamma{})_n=({\MGamma{}})_n^*=\langle 0 | O_{\Gamma}^{s} |
X_n \rangle$) with the $n^\mathrm{th}$ excited meson states $| X_n \rangle$ with
corresponding energy $E_n$. For the bilinear, we only consider pseudoscalars
($\Gamma = \gamma_5 \equiv P$) and the temporal component of the axial current ($\Gamma
= \gamma_0 \gamma_5 \equiv A$). The smearing operator $\omega_s$ and the superscripts
$s_{1},s_{2}$ label the type of smearing. In our setup, we use local ($L$) and
smeared ($S$) propagators at source and sink. At the source, all our quark
fields are smeared, $s_1=SS$. We also require the smearing at the sink to be
the same for both the strange and the down quark, $s_2 \in \{ SS, LL \}$. An
exception to this is the ensemble F1M, where we keep both the source and sink
local, $s_1=s_2=LL$.

For three-point correlation functions, in contrast to the two-point functions, we
consider only pseudoscalar operators $\bar{\opP}$ ($\opP$) inducing the quantum
numbers of a $\bar{K}$ ($K$) at the source (sink). These operators are smeared
($s=SS$) on all ensembles apart from the F1M, where they are local ($s=LL$).
For notational convenience we drop the smearing indices for the
operators. Dropping around-the-world effects, we obtain
\begin{widetext}
\begin{equation}
  \begin{aligned}
    \label{eq:3ptdef}
    C^i_3(t, \Delta T) &\equiv \Big{\langle} \opP(\Delta T) Q^+_{i}(t) \bar{\opP}^\dagger(0) \Big{\rangle}
    = \sum_{n,n'} \frac{(M^s_P)_n}{4E_nE_{n'}} \langle X_n | Q^+_{i}(t) | X_{n'} \rangle  (M^s_P)^*_{n'} e^{-(\Delta T-t) E_n} e^{-t E_{n'}}  \,.
  \end{aligned}
\end{equation}
\end{widetext}
$ C^i_3(t, \Delta T) $ describes a three-point correlation function with a
source at $t=0$, sink at $t=\Delta T$ and a four-quark operator insertion
$Q^+_i$ at $t$.

By placing sources on every second time plane, we compute the above correlation
functions for $(T/a)/2$ time translations, where $T/a$ is the integer number of
time slices for a given ensemble. We time-translate and average equivalent
measurements on a given configuration into a single effective measurement prior
to any further analysis. This helps to reduce the variance of the measured
correlation functions. The only exception is that we use all available
measurements to estimate the correlation matrix, as outlined in
Appendix~\ref{app:fit-strat}.

\subsection{Combined fits to two-point and three-point functions}
\begin{figure*}
  \includegraphics[width=\textwidth]{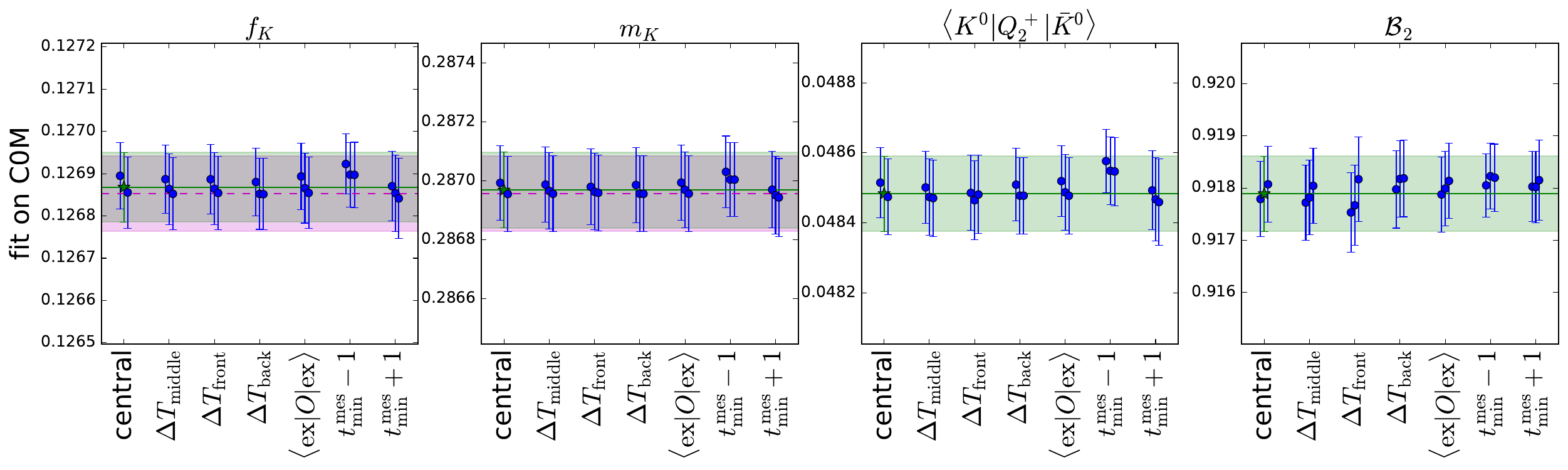} \caption{Stability
  of correlation function fits, illustrated on the example of the C0M ensemble
  for $Q^+_2$. All numbers are quoted in lattice units and the NPR
  basis.}  \label{fig:corfitstab}
\end{figure*}

For each of the operators $Q_i^+$, we extract the desired masses and matrix
elements from a combined fit to several two-point and three-point functions.  In
particular, we jointly fit $C_{PP}^{SL}$ $C_{PP}^{SS}$ and $C_{PA}^{SL}$
($C_{PP}^{LL}$ and $C_{PA}^{LL}$ on F1M) and $C_3^i(t;\Delta T)$ for multiple
choices of $\Delta T$, typically parameterising the ground state and the first
excited state. From these fits we extract the main quantities of interest: the
bare bag parameters $\mathcal{B}_i^\mathrm{bare}$; and the ratios of operators
$R_i^\mathrm{bare}$. They are determined and quoted in the NPR basis (see
Tables~\ref{tab:bareresults3pt_bag} and~\ref{tab:bareresults3pt_rat}) but can
subsequently be translated into the SUSY basis. For completeness we also quote
the meson masses and bare decay constants at our simulation points for the pion
and the kaon in Table~\ref{tab:bareresults2pt} in Appendix~\ref{sec:corfits}.
\begin{table*}
  \caption{Bare bag parameters on all ensembles quoted in the NPR basis.}
  \begin{tabular}{l|lllll}
\hline\hline
$ N_i \mathcal{B}^\mathrm{bare}_i|_\mathrm{NPR}$  & $i=1$ & $i=2$ & $i=3$ & $i=4$ & $i=5$\\ \hline
C0M&  1.5565(17) &  -1.22385(96) &  1.8631(15) &  -0.96015(86) &  -0.49446(43) \\
C1S&  1.5692(28) &  -1.2294(18) &  1.8574(27) &  -0.9907(17) &  -0.51051(86) \\
C2S&  1.5949(24) &  -1.2366(16) &  1.8586(23) &  -1.0118(15) &  -0.52241(74) \\
\hline
M0M&  1.4890(21) &  -1.2052(11) &  1.8461(17) &  -0.87841(86) &  -0.44361(42) \\
M1S&  1.5038(35) &  -1.2066(26) &  1.8374(41) &  -0.9051(23) &  -0.4577(11) \\
M2S&  1.5101(24) &  -1.2119(20) &  1.8409(31) &  -0.9133(17) &  -0.46219(80) \\
M3S&  1.5223(45) &  -1.2129(30) &  1.8380(47) &  -0.9252(26) &  -0.4684(13) \\
\hline
F1M&  1.4776(34) &  -1.1901(20) &  1.8218(30) &  -0.8691(14) &  -0.43601(70) \\

\end{tabular}

  \label{tab:bareresults3pt_bag}
\end{table*}

\begin{table}
  \caption{Bare ratio parameters on all ensembles quoted in the NPR basis.}
  \begin{tabular}{l|llll}
\hline\hline
$R^\mathrm{bare}_i|_\mathrm{NPR}$ & $i=2$ & $i=3$ & $i=4$ & $i=5$\\ \hline
C0M&  -23.511(30) &  35.788(48) &  -18.439(28) &  -9.494(15) \\
C1S&  -20.587(55) &  31.098(85) &  -16.592(59) &  -8.549(31) \\
C2S&  -18.368(35) &  27.604(56) &  -15.001(43) &  -7.747(23) \\
\hline
M0M&  -28.144(39) &  43.113(60) &  -20.516(29) &  -10.361(15) \\
M1S&  -25.125(72) &  38.25(11) &  -18.848(72) &  -9.531(36) \\
M2S&  -23.505(51) &  35.704(81) &  -17.742(67) &  -8.979(33) \\
M3S&  -22.107(56) &  33.501(88) &  -16.866(65) &  -8.538(33) \\
\hline
F1M&  -28.621(80) &  43.82(13) &  -20.929(70) &  -10.500(36) \\

\end{tabular}

  \label{tab:bareresults3pt_rat}
\end{table}

We pursue two independent fit strategies and systematically vary the fit ranges
of the two-point and three-point functions (including the choice of which
source-sink separations enter the fit) until we see stability in all fit
parameters. Figure~\ref{fig:corfitstab} demonstrates this stability for the
example of $Q_2^+$ on the C0M ensemble for the first strategy. The superimposed
dashed lines and magenta bands in the first two panels correspond to the chosen
fit if only the two-point functions are fitted. The green bands illustrate our
preferred choice of fit. Each set of three data points corresponds to variations
in the fit range for the three-point functions of $-1, 0, +1$ compared to the
chosen fit. Finally, the different blocks correspond to our chosen fit; the same
fit but only to the middle (first, last) half of the source-sink separations;
additionally including an excited-to-excited matrix element; the same fit but
for a varied choice of $t_\mathrm{min}$ for the two point functions which enter
the fit. We find that the ground state fit results are insensitive to any of
these choices. Further details are provided in Appendix~\ref{app:fit-strat}.

\subsection{Valence strange quark correction}
\begin{figure}
  \includegraphics[width=.48\textwidth]{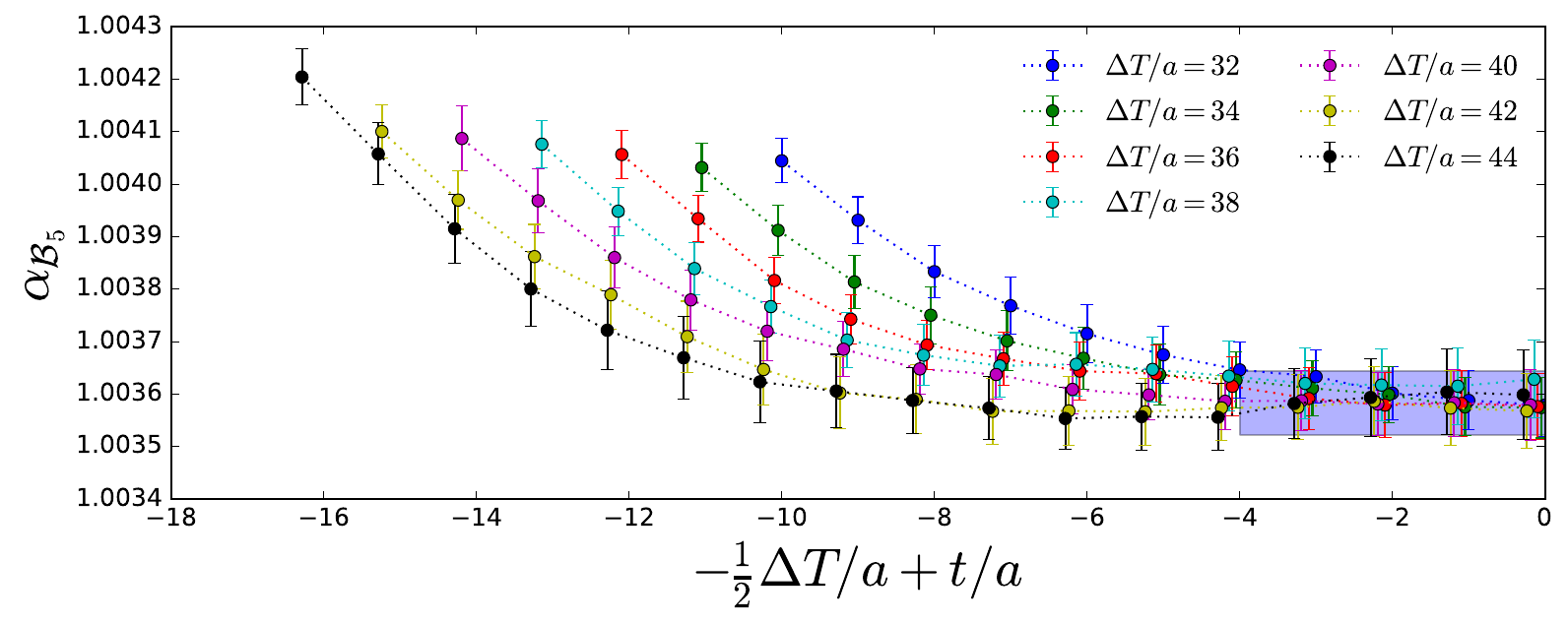}
  \includegraphics[width=.48\textwidth]{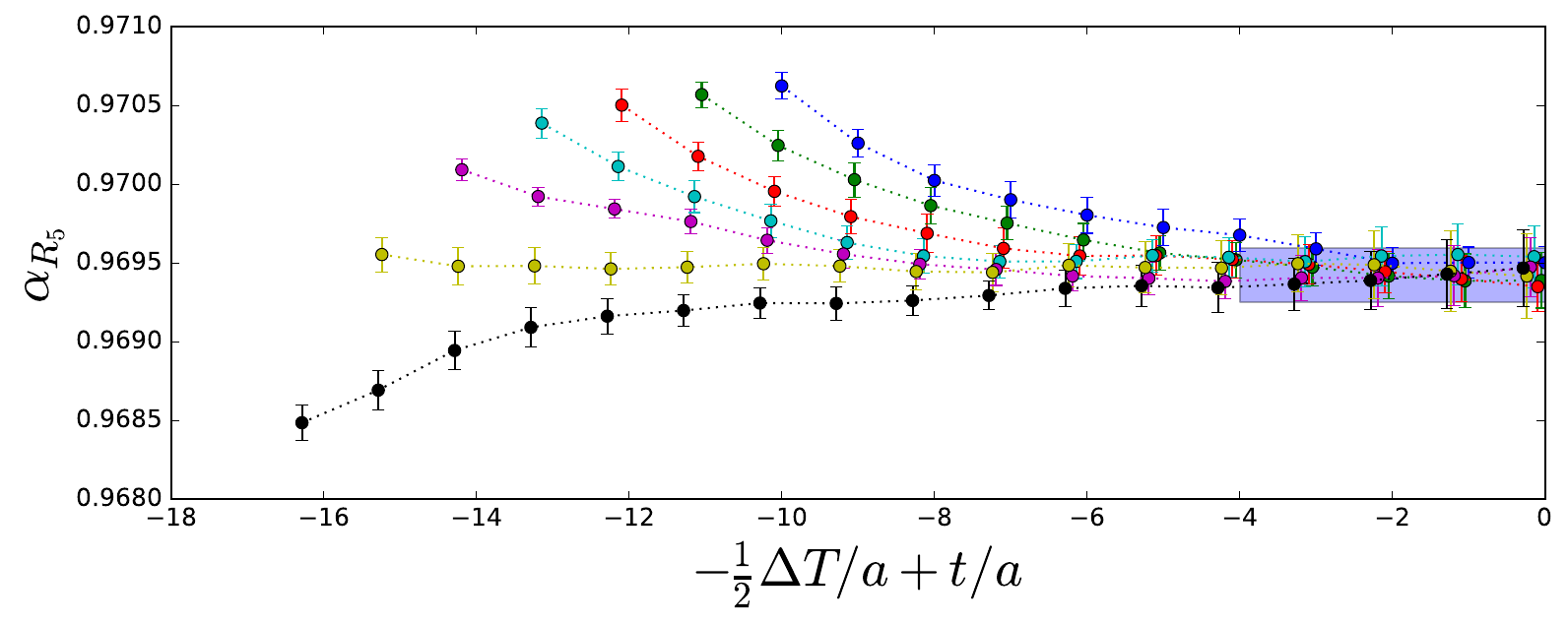}
  \vspace{-0.5cm}
  \caption{Strange valence mistuning quark correction on the F1M ensemble for the example of $\mathcal{B}_5$ (top) and $R_5$ (bottom).}
  \label{fig:ms_mistuningR5}
\end{figure}
\begin{table*}
  \caption{Correction factors to be applied to the bare values of $R_i$ and
    $\mathcal{B}_i$ on the F1M ensemble in the NPR basis in order to correct the
    observables to the physical strange quark
    mass. \label{tab:F1M_ms_correction}}
  \begin{tabular}{c|lllll}
    \hline\hline
    $\alpha_X$ & $i=1$ & $i=2$ &  $i=3$ &  $i=4$ & $i=5$ \\\hline
    $X=\mathcal{B}_i$ &  1.004983(99) &  1.004231(66) & 1.003036(65)&  1.003193(71)&  1.003583(61)\\
    $X=R_i$    & &   0.97005(18) & 0.96890(18) &  0.96904(17) & 0.96942(17)\\
    \hline\hline
  \end{tabular}
\end{table*}

As is evident from Table~\ref{tab:enspar}, the valence strange quark mass on the
F1M ensembles is slightly mistuned from the physical strange quark mass
value. We account for this effect by repeating the simulation at the physical
strange quark mass on an eighth of the full statistics. We then compute the
appropriate correction factors as
\begin{equation}
  \begin{aligned}
    \alpha_{R_i} &\equiv \frac{R_i^\mathrm{phys}}{R_i^\mathrm{uni}} = \lim_{0 \ll t \ll \Delta T} \frac{R_i^\mathrm{eff} (t,\Delta T)\big|_{m_s^{\mathrm{phys}}}}{R_i^{\mathrm{eff}}(t,\Delta T)\big|_{m_s^{\mathrm{uni}}}}\,, \\
    \alpha_{\mathcal{B}_i} &\equiv \frac{\mathcal{B}_i^\mathrm{phys}}{\mathcal{B}_i^\mathrm{uni}} = \lim_{0 \ll t \ll \Delta T} \frac{\mathcal{B}_i^\mathrm{eff} (t,\Delta T)\big|_{m_s^{\mathrm{phys}}}}{\mathcal{B}_i^{\mathrm{eff}}(t,\Delta T)\big|_{m_s^{\mathrm{uni}}}}\,, \\
  \end{aligned}
\end{equation}
where
\begin{equation}
  \begin{aligned}
    R_i^\mathrm{eff}(t,\Delta T) &= \frac{C^i_3(t,\Delta T)}{C^1_3(t,\Delta T)}\,,\\
    \mathcal{B}_i^\mathrm{eff}(t,\Delta T) &= \frac{C^i_3(t,\Delta T)}{N_iC_2(t) C_2(\Delta T - t)}\,.
  \end{aligned}
\end{equation}

We find that the effect of the $\sim 3 \%$ mistuning of the strange quark
valence mass leads to a $\sim 0.3-0.5\%$ correction for the bag parameters and a
$\sim 3\%$ correction for the ratio of operators --- see Table
\ref{tab:F1M_ms_correction}. Given that the relative uncertainty of the
correction factor is more than an order of magnitude smaller than that of the
values it is applied to, we treat this correction factor as
uncorrelated. Figure~\ref{fig:ms_mistuningR5} illustrates this correction
factor for the case of $\mathcal{B}_5$ and $R_5$.

\section{Non-perturbative renormalisation}
\label{sec:Renormalisation}
In order to obtain a well-defined value in the continuum limit, it is necessary
to renormalise the matrix elements $\matrixel{\bar{K}^0}{Q^+_i}{K^0}$. We
determine the matrix of renormalisation factors $Z_{ij}$ using the
Rome-Southampton method~\cite{Martinelli:1994ty} with non exceptional kinematics
(RI-SMOM)~\cite{Sturm:2009kb}.  At some renormalisation scale $\mu$ the
renormalised matrix element is then given by
\begin{equation}
\matrixel{\bar{K}^0}{Q^+_i}{K^0}^{\textrm{RI}}(\mu,a) = Z_{ij}^{\textrm{RI}}(\mu,a) \matrixel{\bar{K}^0}{Q^+_j}{K^0}^{\textrm{bare}}(a).
\label{eq:renorMatEl}
\end{equation}
Provided chiral symmetry breaking effects are negligible the matrix
$Z_{ij}^{\textrm{RI}}(\mu,a)$ has a block diagonal structure -- which is the
case for the set-up at hand~\cite{Boyle:2017skn}. The scale $\mu$ should fall
within the ``Rome-Southampton window''
\begin{equation}
    \Lambda_{\textrm{QCD}}^2 \ll \mu^2 \ll \bigg( \frac{\pi}{a} \bigg)^2\,,
    \label{eq:renormWindow}
\end{equation}
in which the upper limit is relevant to control discretisation effects and the
lower limit ensures accurate perturbative matching to $\overline{\textrm{MS}}$.

For the technical definitions, numerical values of the renormalisation constants
and details of the extrapolation of the renormalisation constants to the
massless limit, we refer the interested reader to
Sections~\ref{subsec:nprdefs},~\ref{subsec:Zijnumerical}
and~\ref{subsec:Zijtomassless} in Appendix~\ref{app:npr}. There we also discuss
two choices, called $\gamma_\mu$ and $\slashed q$, of the renormalisation
conditions imposed to precisely define the renormalisation scheme.

\subsection{Analysis of non-perturbative renormalisation data}\label{sec:npranalysis}
\begin{figure}
   \includegraphics[width=.32\columnwidth]{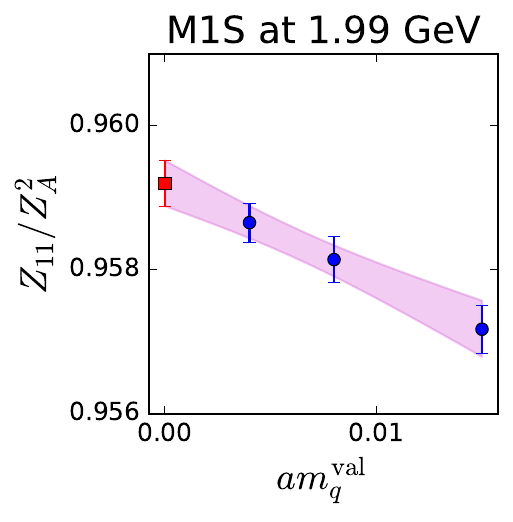}
   \includegraphics[width=.32\columnwidth]{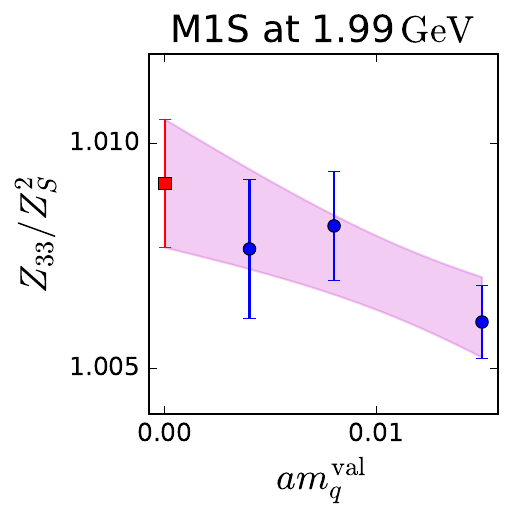}
   \includegraphics[width=.32\columnwidth]{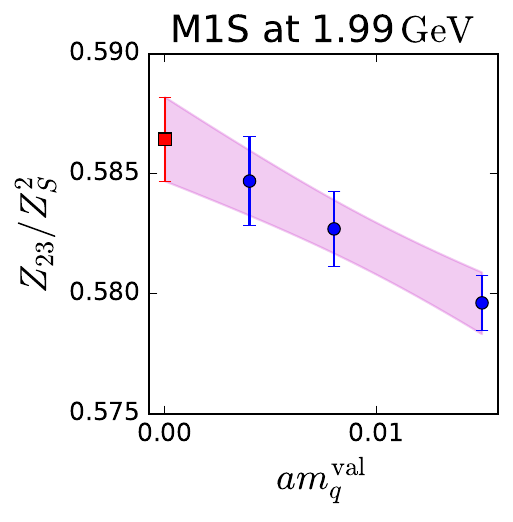}
   \caption{Extrapolation of the renormalisation constants to massless valence-quark
   limit for the example of the (11), (33) and (23) elements of the M1S ensemble
   close to 2$\,\mathrm{GeV}$. Results are presented in the RI-SMOM$^{(\gamma_\mu,\gamma_\mu)}$ scheme in the NPR basis.}
   \label{fig:Zijamqto0val}
\end{figure}

In practice, our data covers many values in the range
$2\,\mathrm{GeV} \lesssim \mu \lesssim 3\,\mathrm{GeV}$ (see
Fig~\ref{fig:NPRmuvals} in Appendix~\ref{subsec:Zijnumerical} for details).  On
the computationally most expensive ensembles C0M and M0M we simulate at
$am_s^\mathrm{sea}/2$ in the valence sector. On all other ensembles we have
additional simulation points at $am_l^\mathrm{sea}$ and $2am_l^\mathrm{sea}$. We
use this data ensemble-by-ensemble to extrapolate the renormalisation constants
to the massless limit in which the renormalisation constants are formally
defined (see Fig~\ref{fig:Zijamqto0val}). The extrapolation of the data on the
physical pion mass ensembles is performed by applying the slope of the C1M (M1M)
ensemble to the C0M (M0M) data.

\begin{figure}
   \includegraphics[width=.32\columnwidth]{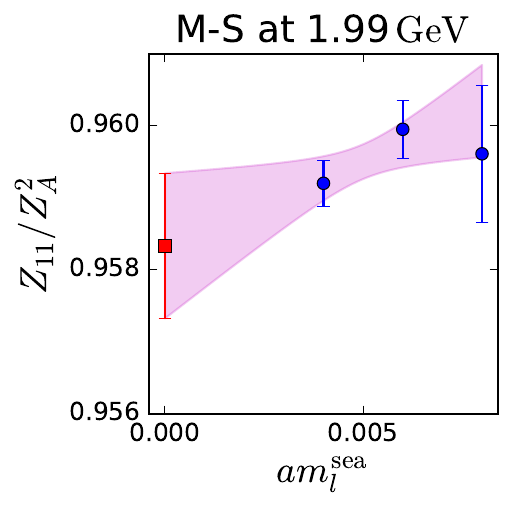}
   \includegraphics[width=.32\columnwidth]{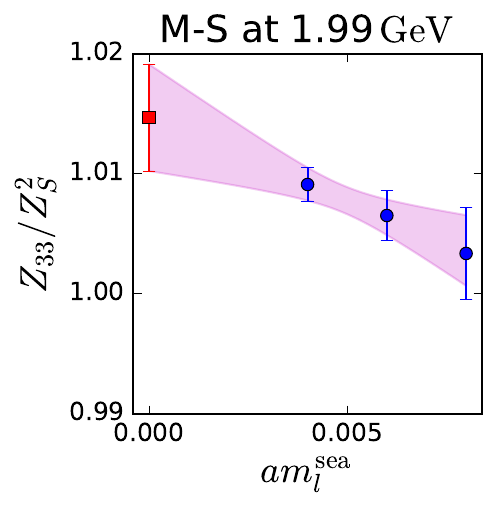}
   \includegraphics[width=.32\columnwidth]{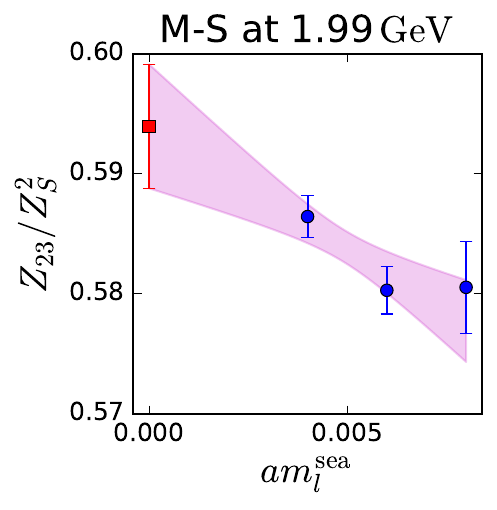}
   \caption{Extrapolation of the renormalisation constants to the zero light-sea
   quark mass limit for the example of the (11), (33) and (23) elements of the
   M-S ensembles close to 2$\,\mathrm{GeV}$.  Results are presented in the
   RI-SMOM$^{(\gamma_\mu,\gamma_\mu)}$ scheme in the NPR
   basis.}  \label{fig:Zijamqto0sea}
\end{figure}
In a subsequent step we extrapolate the results for each action to zero sea
light quark mass as is illustrated in Figure~\ref{fig:Zijamqto0sea} for the M-S
ensembles at a momentum point close to $2\,\mathrm{GeV}$.  Since for the case of
the F1M ensemble we only have a single sea quark mass data point, in practice we
first interpolate the results on all ensembles to a common renormalisation scale
and then perform the sea-light quark mass to zero limit for each choice of
distinct lattice spacing. For the F-M ensemble this is done by applying each of
the four slopes (obtained from C-S, C-M, M-S and M-M) in turn and assigning a
systematic uncertainty of half the spread of these results. 

We list the chirally extrapolated renormalisation constants for each lattice
spacing at $\mu=2\,\mathrm{GeV}$ in Table~\ref{tab:ch-extrap-2.0} (results for
other values of $\mu$ can be found in Tables~\ref{tab:ch-extrap-2.5}
- \ref{tab:ch-extrap-3.0} in Appendix~\ref{subsec:Zijtomassless}). Since these
numbers contain information from multiple ensembles and the NPR calculations are
based on a subset of the configurations, we propagate these small uncertainties
in an uncorrelated fashion. To this end we add statistical and systematic
uncertainties in quadrature and generate bootstrap samples for each of the
$Z_{ij}$ by drawing from a Gaussian distribution with the appropriate mean and
width.

\begin{table*}
\caption{Elements of $Z_{ij}/Z_{A/S}^2$ extrapolated to the
massless limit. All results are provided in RI-SMOM$^{(\gamma_\mu,\gamma_\mu)}$
at $\mu={2.0}\,\mathrm{GeV}$ in the SUSY basis. The first parenthesis is the
statistical error and the second is the systematic error. More detail is
provided in App.~\ref{subsec:Zijtomassless}. \label{tab:ch-extrap-2.0}}
\begin{tabular}{c|ccccc}
\hline
\hline
$a^{-1}$ [GeV] & 1.7295(38) & 1.7848(50) & 2.3586(70) & 2.3833(86) & 2.708(10) \\
\hline
$Z_{11}/Z_A^2$ & 0.93258(26)(0) & 0.93444(77)(1) & 0.96021(51)(2) & 0.9579(12)(0) & 0.97120(69)(47) \\
\hline
$Z_{22}/Z_S^2$ & 1.0703(26)(0) & 1.0788(37)(2) & 1.1185(18)(0) & 1.1237(64)(1) & 1.1405(27)(18) \\
$Z_{23}/Z_S^2$ & -0.06092(49)(8) & -0.0603(10)(0) & -0.04045(55)(5) & -0.0387(10)(0) & -0.03054(62)(27) \\
$Z_{32}/Z_S^2$ & 0.0343(31)(3) & 0.0403(74)(1) & 0.1135(47)(1) & 0.120(11)(0) & 0.1522(57)(6) \\
$Z_{33}/Z_S^2$ & 1.6094(53)(13) & 1.643(13)(0) & 1.9107(85)(9) & 1.948(21)(0) & 2.064(10)(6) \\
\hline
$Z_{44}/Z_S^2$ & 1.0113(21)(2) & 1.0161(35)(2) & 1.0134(11)(0) & 1.0160(51)(0) & 1.0122(21)(16) \\
$Z_{45}/Z_S^2$ & -0.07270(48)(4) & -0.07384(98)(6) & -0.06217(46)(7) & -0.06268(98)(0) & -0.05774(49)(44) \\
$Z_{54}/Z_S^2$ & -0.23507(99)(38) & -0.2417(31)(1) & -0.2851(20)(2) & -0.2940(46)(2) & -0.3119(22)(15) \\
$Z_{55}/Z_S^2$ & 1.4762(38)(9) & 1.4994(96)(6) & 1.6718(64)(5) & 1.699(15)(0) & 1.7670(74)(55) \\
\hline
\hline
\end{tabular}
\end{table*}

Finally, we use these values to renormalise the quantities of interest. In
particular we find
\begin{equation}
\begin{aligned}
R_i(\mu)^{\textrm{ren}} &= \frac{Z_{ij}}{Z_{11}} R_j(\mu,a)|_\mathrm{bare}\,,\\
\mathcal{B}_1(\mu)^{\textrm{ren}} &= \frac{Z_{11}}{Z_A^2} \mathcal{B}_1(\mu,a)|_\mathrm{bare}\,,\\
N_i \mathcal{B}_i(\mu)^{\textrm{ren}} &= \frac{Z_{ij}}{Z_P^2}N_j\mathcal{B}_j(\mu,a)|_\mathrm{bare}\,\qquad i=2,\cdots,5.
\label{eq:renormfactors}
\end{aligned}
\end{equation}
Here the $N_i$ are the appropriate normalisation factors defined above and in
practice we make use of the relation $Z_S \approx Z_P$ due to chiral symmetry.

\subsection{Step-scaling}
When performing the renormalisation we have the freedom to choose the
renormalisation scale $\mu$ within the Rome-Southampton window of our ensembles,
which includes $2\,\mathrm{GeV} \lesssim \mu \lesssim
3\,\mathrm{GeV}$. We note that higher scales are more susceptible to
discretisation effects, whilst lower scales face larger errors when matching
perturbatively to e.g. $\overline{\textrm{MS}}$.

We can scale the value of an operator renormalised at one scale to another with
the use of a scale evolution matrix, $\sigma(\mu_2,\mu_1)$, in a procedure
called step-scaling~\cite{Arthur:2010ht,Arthur:2011cn,Boyle:2011cc}. We define the continuum
scale evolution matrix for the renormalisation of the four-quark operators as
\begin{equation}
   \sigma(\mu_2,\mu_1) = \lim_{a^2 \rightarrow 0}  Z({\mu}_2,a)Z^{-1}({\mu}_1,a),
   \label{eq:stepscaling}
\end{equation}
where $Z(\mu,a)$ is the $5\times 5$ block-diagonal matrix described above.
Therefore it is possible to scale our operators, once renormalised and
extrapolated to the continuum limit, from $\mu_1$ to $\mu_2$. By renormalising
at $\mu = 2\,\textrm{GeV}$, where lattice artefacts are less significant, but
step-scaling our results in RI-SMOM to $\mu = 3\,\textrm{GeV}$ before
perturbatively matching to $\overline{\textrm{MS}}$ we also avoid the higher
errors associated with the truncation of the perturbative series at lower
scales. Since we have mapped out the region
$2\,\mathrm{GeV} \lesssim \mu \lesssim 3\,\mathrm{GeV}$, we can further
split \eqref{eq:stepscaling} into multiple smaller steps $\Delta =
(\mu_2-\mu_1)/N$, i.e. we can compute the product
\begin{equation}
\prod_{k=0}^{N-1}{\sigma(\mu_1+k\Delta +\Delta,\mu_1+k\Delta)}\,.
   \label{eq:stepscaling_multi}
\end{equation}
Alongside directly renormalising at $3\,\mathrm{GeV}$, we can also renormalise
the result at $2\,\mathrm{GeV}$ and step scale to $3\,\mathrm{GeV}$ in one step,
or in multiple steps as described above. This allows us to probe the effect the
scale of the renormalisation has. Details of the computation and numerical
values for the step-scaling matrices are provided in
Section~\ref{subsec:Stepscaling}.  The numerical values for the step-scaling
matrices in the $\text{RI-SMOM}^{(\gamma_\mu,\gamma_\mu)}$-scheme and in the
SUSY basis are given by
\begin{widetext}
   \begin{align}
       \sigma(3\,\mathrm{GeV},2\,\mathrm{GeV}) &= \begin{bmatrix}
0.98021(53) & 0.0 & 0.0 & 0.0 & 0.0 \\
0.0 & 0.9194(22) & -0.0630(16) & 0.0 & 0.0 \\
0.0 & -0.00284(35) & 0.6846(19) & 0.0 & 0.0 \\
0.0 & 0.0 & 0.0 & 0.9988(24) & 0.0784(25) \\
0.0 & 0.0 & 0.0 & 0.00838(59) & 0.7542(24) \\
\end{bmatrix} \label{eq:step2_3_bag}\,, \\
      \sigma(3\,\mathrm{GeV}\xleftarrow{\Delta=0.5\,\mathrm{GeV}}2\,\mathrm{GeV}) &= \begin{bmatrix}
0.98030(35) & 0.0 & 0.0 & 0.0 & 0.0 \\
0.0 & 0.9199(22) & -0.0634(15) & 0.0 & 0.0 \\
0.0 & -0.00260(31) & 0.6863(17) & 0.0 & 0.0 \\
0.0 & 0.0 & 0.0 & 0.9990(25) & 0.0778(24) \\
0.0 & 0.0 & 0.0 & 0.00860(44) & 0.7552(23) \\
\end{bmatrix}\,. \label{eq:step2_3_multi_bag}
   \end{align}
\end{widetext}

\section{Chiral continuum fits and final results}
\label{sec:FitResults}
\subsection{Fit ansatz}
\label{sec:CombinedChiralFits}
\label{subsec:fitansatz}
To recover continuum results at physical quark masses we perform a simultaneous
chiral-continuum limit fit. Our fit ansatz is based on NLO SU(2) chiral
perturbation theory ($\chi_{\textrm{PT}}$), covered in more detail in
Ref.~\cite{Garron:2016mva}, and includes a chiral logarithm term.  Furthermore
our fit function is linear in $a^2$ and $m_\pi^2$ and the mistuning of the
strange quark mass $\delta^\mathrm{sea}_{m_s}$. It is given by
\begin{widetext}
  \begin{equation}
  Y_i(a^2,m_{\pi}^2,m_s^\mathrm{sea}) = Y_i^\mathrm{phys}\left(1 + \alpha_i (a\Lambda)^2 + \beta_i \frac{m_\pi^2-(m_\pi^\mathrm{phys})^2}{(m_\pi^\mathrm{phys})^2} +\gamma_i \delta^\mathrm{sea}_{m_s} + L^Y_i(m_\pi) - L^Y_i(m^\mathrm{phys}_\pi)\right).
  \label{eq:fitform}
  \end{equation}
  \end{widetext}
where $Y \in \{\mathcal{B}, R\}$ is the quantity of interest. $\Lambda$ is a
typical QCD-scale and we take the isospin-averaged pion mass to be
$m_\pi^\mathrm{phys}=(2m_\pi^\pm+m_\pi^0)/3 \approx
138\,\mathrm{MeV}$~\cite{ParticleDataGroup:2022pth}.  $Y_i^\mathrm{phys}$,
$\alpha$, $\beta$ and $\gamma$ are fit parameters, $\delta^\mathrm{sea}_{m_s} =
(m_s^\mathrm{sea} - m_s^\mathrm{phys})/m_s^\mathrm{phys}$ parameterises the
mistuning of the sea strange quark mass and the chiral logarithms are given by
$L^Y_i(m_\pi) = C^Y_i\, m_\pi^2\, \log(m^2_\pi/\Lambda^2)/(16\pi^2f_\pi^2)$.
The coefficients $C_i$ are known constants with numerical values
$C^\mathcal{B}_i = -0.5$ for $i=1,2,3$ and $C^\mathcal{B}_i = 0.5$ for $i=4,5$
and we take $f_\pi =
130.41(23)\,\mathrm{MeV}$~\cite{ParticleDataGroup:2022pth}. For the ratios $R_2$
and $R_3$ the chiral logarithms vanish ($C^R_2=C^R_3=0$) and finally
$C^R_4=C^R_5=1$.

As stated in section~\ref{sec:SimulationDetails} the lattice spacings were
determined in Ref.~\cite{RBC:2014ntl} (and updates thereof in
Refs~\cite{Boyle:2017jwu,Boyle:2018knm}) from some of the same ensembles
included in this work, hence a correlation between the data exists. However we
perform an uncorrelated fit to decouple this work from the previous work. We
propagate the error on the lattice spacing by generating bootstrap samples
following a Gaussian distribution with width equal to the error on the lattice
spacing.  Given that the errors on the lattice spacings are of order 0.5\% and
all extrapolated quantities are dimensionless, we believe that neglecting these
correlations has a negligible effect.

Our central fit results in the two intermediate schemes are obtained from a
chiral-continuum limit fit at $\mu=2\,\mathrm{GeV}$ performed in the SUSY
basis. These results are then step-scaled to $3\,\mathrm{GeV}$ (e.g. using the
matrix provided in Eq.~\eqref{eq:step2_3_bag}) and perturbatively matched to
$\overline{\mathrm{MS}}$. In the following sections we will present the results
of the chiral-continuum limit fits and assemble the full uncertainty budget
relating to the lattice computation in the intermediate RI-SMOM schemes at
$3\,\mathrm{GeV}$. Only subsequently do we match these results perturbatively to
$\overline{\mathrm{MS}}$. This allows us to cleanly separate the uncertainties
due to the perturbative matching to $\overline{\mathrm{MS}}$ from those arising
in the lattice calculation.
\subsection{Results of the chiral-continuum limit fits}
\begin{figure*}
    \centering
    \includegraphics[width=.9\textwidth]{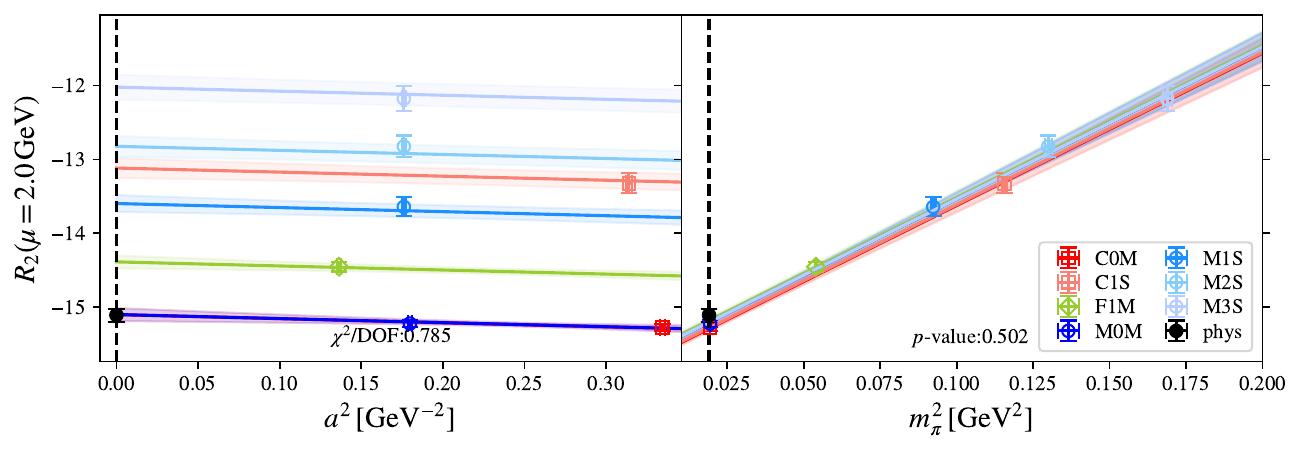}
    \includegraphics[width=.9\textwidth]{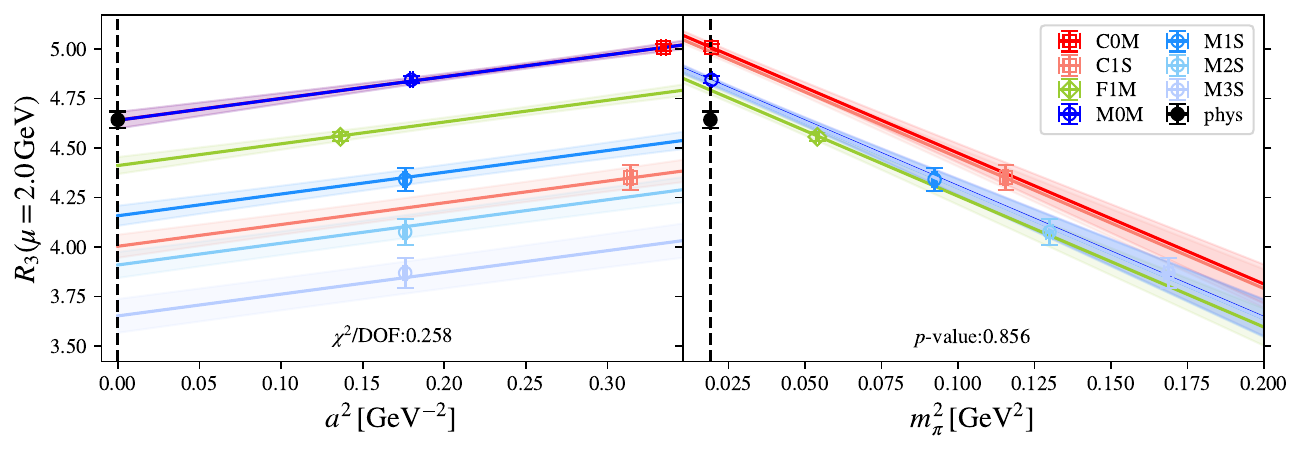}
    \caption{Chiral-continuum limit fit to BSM ratio parameters $R_2$ (top) and $R_3$ (bottom) in the SUSY basis, renormalised in the $\text{RI-SMOM}^{(\gamma_\mu, \gamma_\mu)}$ scheme.}
    \label{fig:fit_R2}
    \label{fig:fit_R3}
\end{figure*}
\begin{figure*}
    \centering
    \includegraphics[width=.9\textwidth]{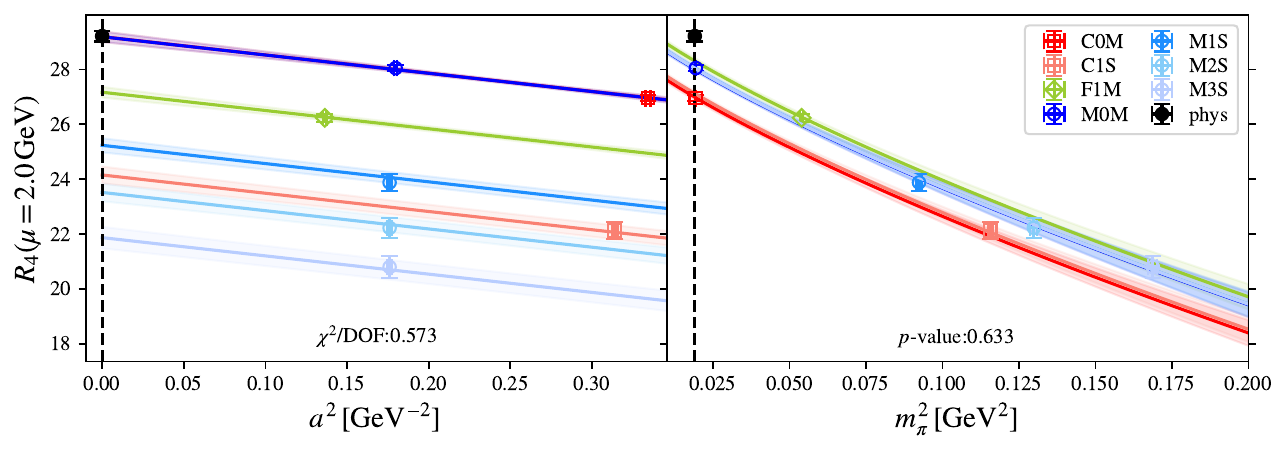}
    \includegraphics[width=.9\textwidth]{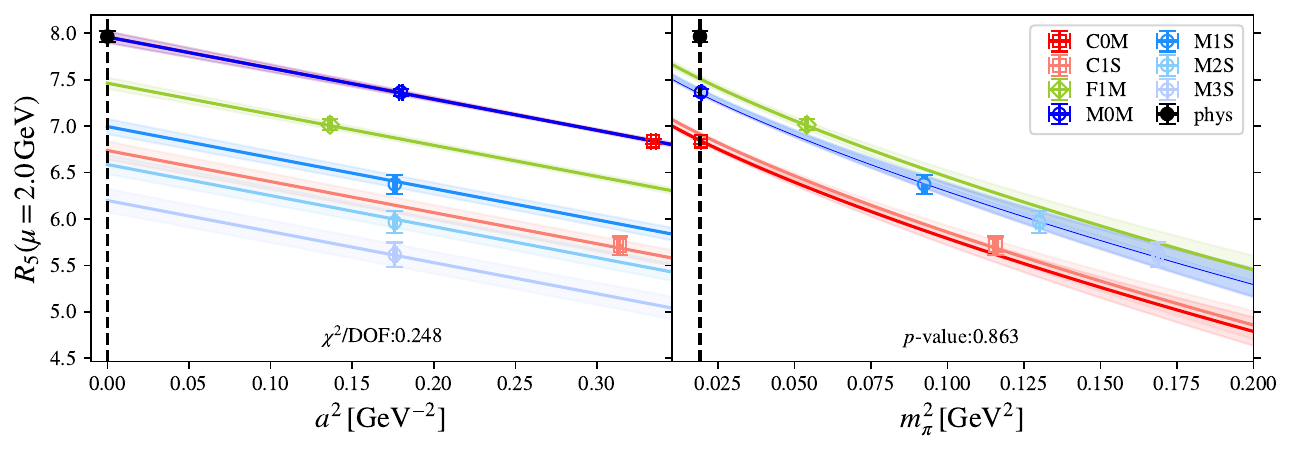}
    \caption{Chiral-continuum limit fit to BSM ratio parameters $R_4$ (top) and $R_5$ (bottom) in the SUSY basis, renormalised in the $\text{RI-SMOM}^{(\gamma_\mu, \gamma_\mu)}$ scheme.}
    \label{fig:fit_R4}
    \label{fig:fit_R5}
\end{figure*}

\begin{figure*}
    \centering
    \includegraphics[width=.9\textwidth]{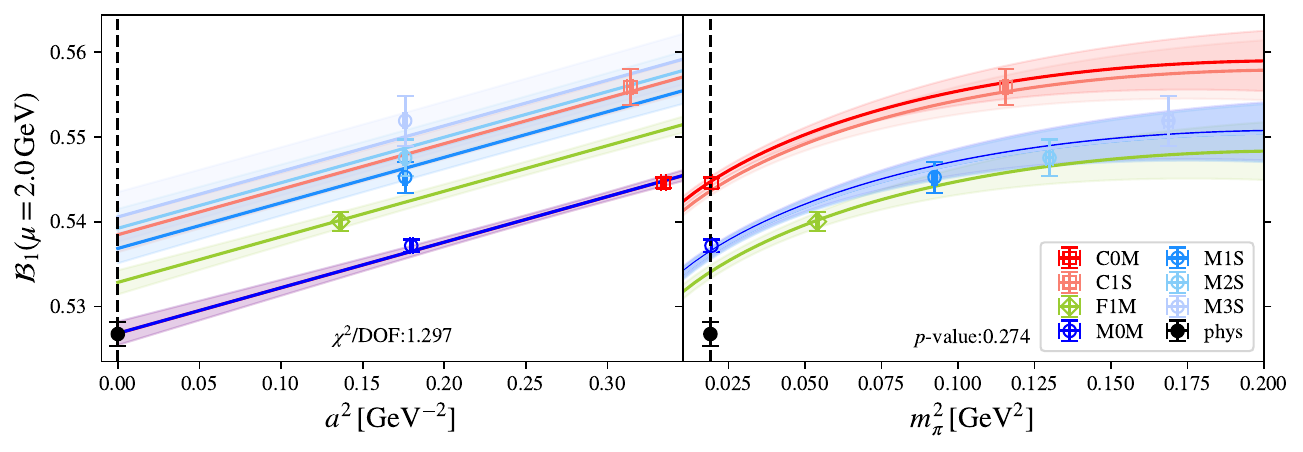}
    \includegraphics[width=.9\textwidth]{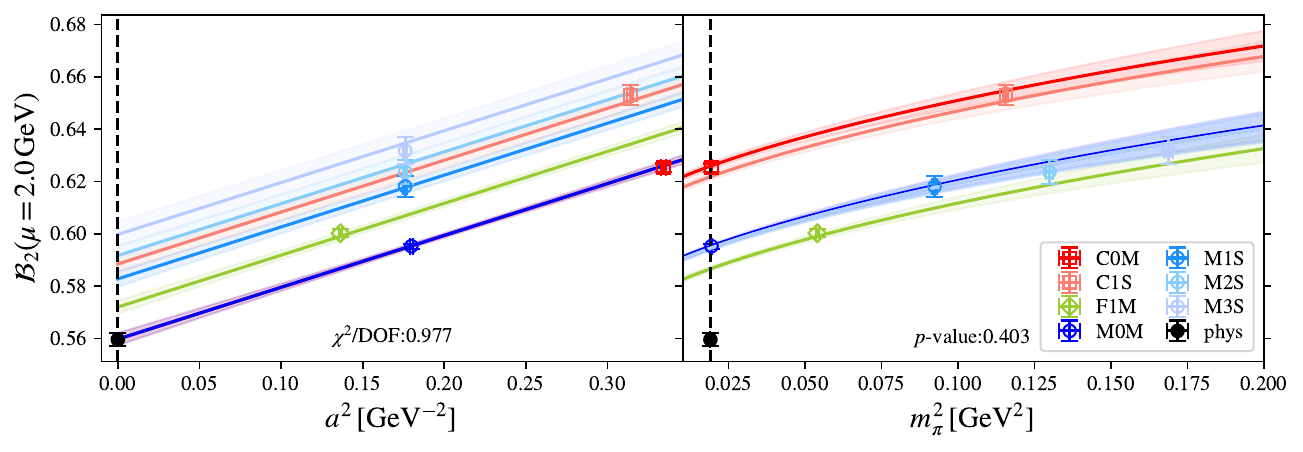}
    \includegraphics[width=.9\textwidth]{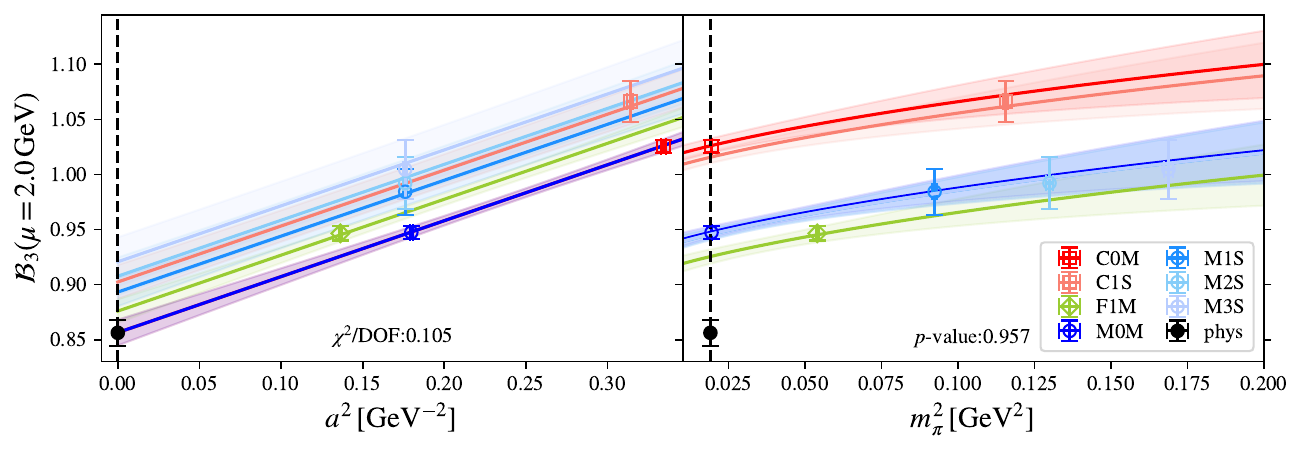}
    \caption{Chiral-continuum limit fit to the standard model bag parameter
      $\mathcal{B}_1$ (top) and BSM bag parameters $\mathcal{B}_2$ (middle) and
      $\mathcal{B}_3$ (bottom) in the SUSY basis, renormalised in the $\text{RI-SMOM}^{(\gamma_\mu, \gamma_\mu)}$ scheme. }
    \label{fig:fit_B1}
    \label{fig:fit_B2}
    \label{fig:fit_B3}
\end{figure*}

\begin{figure*}
    \centering
    \includegraphics[width=.9\textwidth]{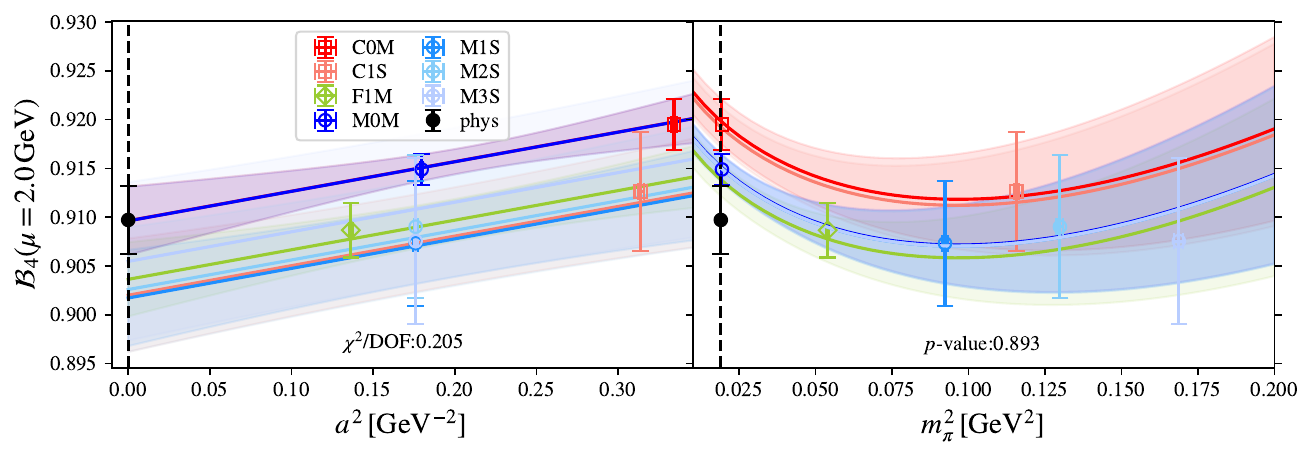}
    \includegraphics[width=.9\textwidth]{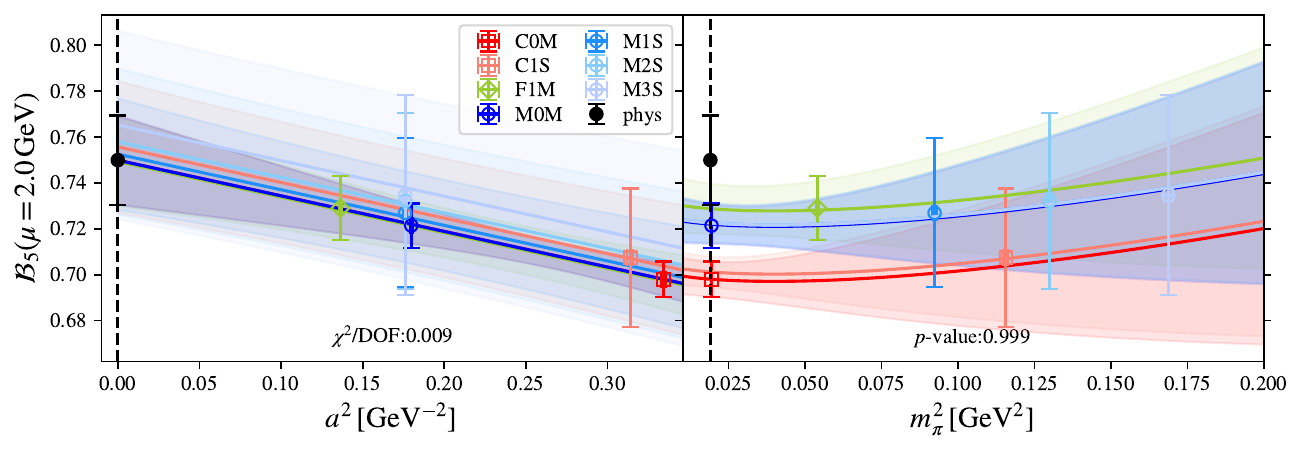}
    \caption{Chiral-continuum limit fit to BSM bag parameters $\mathcal{B}_4$ and $\mathcal{B}_5$ in the SUSY basis, renormalised in the $\text{RI-SMOM}^{(\gamma_\mu, \gamma_\mu)}$ scheme.}
    \label{fig:fit_B4}
    \label{fig:fit_B5}
\end{figure*}

In this section we present the chiral-continuum limit fits in the
RI-SMOM$^{(\gamma_\mu,\gamma_\mu)}$ scheme at $\mu=2\,\mathrm{GeV}$ and in the
SUSY basis.  We show these fits for the ratios $R_i$ in Figures~\ref{fig:fit_R2}
and \ref{fig:fit_R5} and for the bag parameters $\mathcal{B}_i$ in
Figures~\ref{fig:fit_B1} and \ref{fig:fit_B5}. Since we find that the C2S
ensemble --- which is at the heaviest pion mass and the coarsest lattice spacing
--- is not always well described by the fit ansatz, we remove it from our
central fits. The data is well described by the fit function \eqref{eq:fitform}
in all cases with acceptable $p$-values ($> 5\%$) for all fits presented.


\subsection{Error budget\label{subsec:budget}}
In the following we quantify all relevant sources of uncertainties. We consider
variations to the data and the fit ansatz; variations of the renormalisation
procedure; and uncertainties stemming from the perturbative matching.  We
quantify the uncertainties for the variations by considering
\begin{equation}
\delta^\mathrm{var}_i(\mu) = \frac{ | Y_i^{\textrm{central}} - Y_i^{\mathrm{variation}}|} {\frac{1}{2} (Y_i^{\textrm{central}} + Y_i^{\mathrm{variation}})}\,,
\label{eq:delta}
\end{equation}
where $Y \in \{\mathcal{B}, R\}$.

\subsubsection{Chiral extrapolation}
The two precise data points at physical pion masses make the mass extrapolation
element of the fit very benign. We quantify the associated uncertainty by
varying the pion mass cut to the data, removing terms from our fit form and by
repeating the fits using the alternative correlator fit results (see
App.~\ref{section:fit_strategy_alt}). For each of these variations we compute
the associated $\delta$ (see Eq.~\eqref{eq:delta}) which measures the shift in
central value and list the corresponding values in
Table~\ref{tab:fit_systematics}. For all ratios and bag parameters this error is
well below 1\% and typically sub-statistical. For each observable we assign the
maximum of those values as the systematic uncertainty associated to the chiral
extrapolation listed as ``chiral'' in Table~\ref{tab:error}.

\begin{table*}
    \centering
    \caption{Chiral-continuum limit fit systematics depending on choice of
      ansatz at $\mu=2.0$ GeV in RI-SMOM$^{(\gamma_\mu,\gamma_\mu)}$ in the SUSY
      basis. The first column shows the central value with statistical
      uncertainty, whilst the remaining columns quantify variations arising from
      different choices in the data that enters the fit as well as the model to
      which the chiral dependences is fitted. The last column illustrates the
      effect of using the alternative choice of correlation function fits
      underlying the analysis.\label{tab:fit_systematics}}
    \begin{tabular}{c|c|c|c|c|c|c|c}
\hline
\hline
 & Central fit  & no $\delta_{m_s}$ & no chiral logs  & $m_\pi<440$ MeV & $m_\pi<370$ MeV & $m_\pi<350$ MeV & alternate fit\\

\hline
$R_2$ & \textcolor{black}{$-15.106(87)$} & \textcolor{black}{$0.22$\%} & - & \textcolor{black}{$0.46$\%} & \textcolor{black}{$0.05$\%} & \textcolor{black}{$0.08$\%} & \textcolor{black}{$0.45$\%}\\
$R_3$ & \textcolor{black}{$4.643(41)$} & \textcolor{black}{$0.42$\%} & - & \textcolor{black}{$0.28$\%} & \textcolor{black}{$0.10$\%} & \textcolor{black}{$0.11$\%} & \textcolor{black}{$0.21$\%}\\
$R_4$ & \textcolor{black}{$29.22(19)$} & \textcolor{black}{$0.51$\%} & \textcolor{black}{$0.59$\%} & \textcolor{black}{$0.58$\%} & \textcolor{black}{$0.04$\%} & \textcolor{black}{$0.06$\%} & \textcolor{black}{$0.49$\%}\\
$R_5$ & \textcolor{black}{$7.965(62)$} & \textcolor{black}{$0.10$\%} & \textcolor{black}{$0.47$\%} & \textcolor{black}{$0.50$\%} & \textcolor{black}{$0.00$\%} & \textcolor{black}{$0.05$\%} & \textcolor{black}{$0.13$\%}\\
\hline
$\mathcal{B}_1$ & \textcolor{black}{$0.5268(13)$} & \textcolor{black}{$0.10$\%} & \textcolor{black}{$0.21$\%} & \textcolor{black}{$0.49$\%} & \textcolor{black}{$0.02$\%} & \textcolor{black}{$0.06$\%} & \textcolor{black}{$0.40$\%}\\
$\mathcal{B}_2$ & \textcolor{black}{$0.5596(23)$} & \textcolor{black}{$0.05$\%} & \textcolor{black}{$0.17$\%} & \textcolor{black}{$0.06$\%} & \textcolor{black}{$0.01$\%} & \textcolor{black}{$0.08$\%} & \textcolor{black}{$0.02$\%}\\
$\mathcal{B}_3$ & \textcolor{black}{$0.856(11)$} & \textcolor{black}{$0.06$\%} & \textcolor{black}{$0.28$\%} & \textcolor{black}{$0.02$\%} & \textcolor{black}{$0.07$\%} & \textcolor{black}{$0.06$\%} & \textcolor{black}{$0.00$\%}\\
$\mathcal{B}_4$ & \textcolor{black}{$0.9097(35)$} & \textcolor{black}{$0.03$\%} & \textcolor{black}{$0.17$\%} & \textcolor{black}{$0.08$\%} & \textcolor{black}{$0.01$\%} & \textcolor{black}{$0.02$\%} & \textcolor{black}{$0.06$\%}\\
$\mathcal{B}_5$ & \textcolor{black}{$0.750(19)$} & \textcolor{black}{$0.02$\%} & \textcolor{black}{$0.24$\%} & \textcolor{black}{$0.18$\%} & \textcolor{black}{$0.02$\%} & \textcolor{black}{$0.03$\%} & \textcolor{black}{$0.13$\%}\\
\hline
\hline
\end{tabular}

\end{table*}

\subsubsection{Discretisation effects}
The good chiral symmetry of domain wall fermions constrains $O(a)$ and $O(a^3)$
terms to be small. The $O(a^2)$ effects are controlled and removed by our three
lattice spacings present in the fit. Power counting suggests that $O(a^4)$
effects for hadronic physics scales with a $1.73\,\mathrm{GeV}$ coarsest inverse
lattice spacing will remain small on all data points.  However, the same is not
necessarily true for hard, off-shell vertex functions where the momenta are
chosen as the best compromise for a Rome-Southampton window.  The leading
unremoved discretisation effects are thus likely to come from the
non-perturbative renormalisation, and may be probed by comparing different ways
of renormalising our data. Our central chiral-continuum limit fit is based on
data renormalised at $\mu = 2\,\mathrm{GeV}$ which is then step-scaled to
$3\,\mathrm{GeV}$ by the step-scaling function
$\sigma(3\,\mathrm{GeV},2\,\mathrm{GeV})$ presented in
Eq.~\eqref{eq:step2_3_bag} for the bag parameters.  We compare the results
obtained this way to using the alternative prescription to obtain the scaling
function (cf. Eq.~\eqref{eq:stepscaling_multi}) with $N=2,3$ and to performing
the continuum limit to data renormalised directly at $\mu = 3\,\mathrm{GeV}$.
We compute and report the associated values for $\delta_i$ in
Table~\ref{tab:scaling_systematics}.

For our main analysis we extract the bare matrix elements and renormalisation
factors in the NPR basis, transform them to the SUSY basis and then perform the
various analysis steps. Performing the entire analysis in the NPR basis and
converting the final values to the SUSY basis causes a reshuffling of
discretisation effects. The corresponding $\delta_i$ are presented in the column
labelled SUSY $\leftarrow$ NPR in Table~\ref{tab:scaling_systematics}.

We take the maximum of these variations as estimate for the systematic
uncertainties due to higher order discretisation effects, labelled ``discr'' in
Table~\ref{tab:error}.

\begin{table*}
    \centering
    \caption{Bag and ratio parameters at $\mu=3$ GeV in
      RI-SMOM$^{(\gamma_\mu,\gamma_\mu)}$ in the SUSY basis. Central value comes
      from performing the chiral-continuum limit fit at $\mu=2\,\mathrm{GeV}$
      and non-perturbative scaling the result to $\mu=3\,\mathrm{GeV}$ using
      $\sigma(3\,\mathrm{GeV}, 2\,\mathrm{GeV})$. We also list variations where
      the continuum step-scaling is obtained in steps, or the data is
      renormalised directly at $3\,\mathrm{GeV}$. The central value uses
      $Z$-factors with chirally vanishing elements removed (masked) from
      $(P\Lambda)^T$ before the inversion $Z = F((P\Lambda)^T)^{-1}$. We list
      the percent shift in the result in foregoing this step (no mask). We also
      compare with performing the entire analysis in the NPR basis and then
      rotating to the SUSY basis. \label{tab:scaling_systematics}}
    \begin{tabular}{c|c|c|c|c|c|c}
\hline
\hline
 & $\sigma(3\,\mathrm{GeV},2\,\mathrm{GeV})$ & $\sigma(3\,\mathrm{GeV}\xleftarrow{\Delta=0.5}2\,\mathrm{GeV})$ & $\sigma(3\,\mathrm{GeV}\xleftarrow{\Delta=0.33}2\,\mathrm{GeV})$ & NPR at 3 GeV & rcsb & SUSY$\leftarrow$NPR \\
\hline
$R_2$ & \textcolor{black}{$-18.37(10)$} & \textcolor{black}{$0.12$\%} & \textcolor{black}{$0.13$\%} & \textcolor{black}{$0.17$\%} & \textcolor{black}{$0.11$\%} & \textcolor{black}{$0.02$\%}\\
$R_3$ & \textcolor{black}{$5.485(36)$} & \textcolor{black}{$0.18$\%} & \textcolor{black}{$0.50$\%} & \textcolor{black}{$0.37$\%} & \textcolor{black}{$0.14$\%} & \textcolor{black}{$0.15$\%}\\
$R_4$ & \textcolor{black}{$38.60(27)$} & \textcolor{black}{$0.02$\%} & \textcolor{black}{$0.02$\%} & \textcolor{black}{$0.48$\%} & \textcolor{black}{$0.09$\%} & \textcolor{black}{$0.01$\%}\\
$R_5$ & \textcolor{black}{$10.932(47)$} & \textcolor{black}{$0.11$\%} & \textcolor{black}{$0.01$\%} & \textcolor{black}{$0.97$\%} & \textcolor{black}{$0.03$\%} & \textcolor{black}{$1.22$\%}\\
\hline
$\mathcal{B}_1$ & \textcolor{black}{$0.5164(14)$} & \textcolor{black}{$0.00$\%} & \textcolor{black}{$0.01$\%} & \textcolor{black}{$0.01$\%} & \textcolor{black}{$0.04$\%} & \textcolor{black}{$0.01$\%}\\
$\mathcal{B}_2$ & \textcolor{black}{$0.5150(12)$} & \textcolor{black}{$0.04$\%} & \textcolor{black}{$0.20$\%} & \textcolor{black}{$0.45$\%} & \textcolor{black}{$0.03$\%} & \textcolor{black}{$0.05$\%}\\
$\mathcal{B}_3$ & \textcolor{black}{$0.7624(52)$} & \textcolor{black}{$0.32$\%} & \textcolor{black}{$0.24$\%} & \textcolor{black}{$1.51$\%} & \textcolor{black}{$0.06$\%} & \textcolor{black}{$0.15$\%}\\
$\mathcal{B}_4$ & \textcolor{black}{$0.9107(19)$} & \textcolor{black}{$0.02$\%} & \textcolor{black}{$0.16$\%} & \textcolor{black}{$0.02$\%} & \textcolor{black}{$0.01$\%} & \textcolor{black}{$0.02$\%}\\
$\mathcal{B}_5$ & \textcolor{black}{$0.7792(79)$} & \textcolor{black}{$0.11$\%} & \textcolor{black}{$0.24$\%} & \textcolor{black}{$0.38$\%} & \textcolor{black}{$0.00$\%} & \textcolor{black}{$0.26$\%}\\
\hline
\hline
\end{tabular}
\end{table*}

\subsubsection{Residual chiral symmetry breaking}
Domain wall fermions provide a good approximation to chiral symmetry, however a
small degree of residual chiral symmetry breaking is present in the data.
Chiral symmetry restricts the allowed mixing pattern to be block-diagonal. For
our central analysis we impose this, by setting the chirally forbidden elements
of $Z_{ij}$ to zero which we refer to as ``masking''. To test the effect
residual chiral symmetry breaking has on our results, we repeat the entire
analysis without masking. We find that the deviations are well below the percent
level, indicating that our approximation to chiral symmetry is very well
controlled. We report the associated systematic uncertainties in
Table~\ref{tab:error} as ``rcsb''.

\begin{table*}
    \centering
    \caption{Central values and combined systematic errors for ratio and bag
      parameters in the SUSY basis at $\mu=3\,\mathrm{GeV}$ in the two RI-SMOM
      schemes --- $(\gamma_\mu, \gamma_\mu)$ and $(\slashed{q}, \slashed{q})$
      --- as well as in $\overline{\text{MS}}$. For the RI-SMOM schemes we list
      the errors arising from statistics, chiral extrapolation, residual chiral
      symmetry breaking and discretisation effects and combine them into total
      uncertainties. For the $\overline{\mathrm{MS}}$ values we list the
      separate conversions from $(\gamma_\mu,\gamma_\mu)$ and $(\slashed
      q,\slashed q)$. The central values are defined as the average of those two
      numbers and the perturbative truncation error as half their
      difference. The lattice error is taken from the $(\gamma_\mu,\gamma_\mu)$
      scheme (see Table~\ref{tab:MSconv} for scheme-wise error
      budget) \label{tab:error}.} \begin{tabular}{c|c|cccc|ccccc}
\hline
\hline
scheme & & $R_2$ & $R_3$ & $R_4$ & $R_5$ & $\mathcal{B}_1$ & $\mathcal{B}_2$ & $\mathcal{B}_3$ & $\mathcal{B}_4$ & $\mathcal{B}_5$\\
\hline
\multirow{6}{*}{$\textrm{RI-SMOM}^{(\gamma_\mu,\gamma_\mu)}$} & central & $-18.37$ & $5.485$ & $38.60$ & $10.93$ & $0.5164$ & $0.5150$ & $0.762$ & $0.9107$ & $0.7792$ \\
\cline{2-11}
 & stat & 0.59\% & 0.66\% & 0.72\% & 0.44\% & 0.28\% & 0.24\% & 0.69\% & 0.22\% & 1.02\% \\
 & chiral & 0.22\% & 0.42\% & 0.59\% & 0.47\% & 0.21\% & 0.17\% & 0.28\% & 0.17\% & 0.24\% \\
 & rcsb & 0.11\% & 0.14\% & 0.09\% & 0.03\% & 0.04\% & 0.03\% & 0.06\% & 0.01\% & 0.00\% \\
 & discr & 0.17\% & 0.50\% & 0.48\% & 1.22\% & 0.01\% & 0.45\% & 1.51\% & 0.16\% & 0.38\% \\
\cline{2-11}
 & total & 0.66\% & 0.94\% & 1.05\% & 1.38\% & 0.35\% & 0.54\% & 1.68\% & 0.31\% & 1.11\% \\
\hline
\multirow{6}{*}{$\textrm{RI-SMOM}^{(\slashed{q}, \slashed{q})}$} & central & $-19.53$ & $5.818$ & $40.99$ & $10.49$ & $0.5342$ & $0.5155$ & $0.765$ & $0.9137$ & $0.7078$ \\
\cline{2-11}
 & stat & 0.68\% & 0.90\% & 0.81\% & 0.83\% & 0.29\% & 0.42\% & 1.20\% & 0.36\% & 2.19\% \\
 & chiral & 0.47\% & 0.77\% & 1.21\% & 1.23\% & 0.24\% & 0.27\% & 0.43\% & 0.28\% & 0.53\% \\
 & rcsb & 0.28\% & 0.20\% & 0.23\% & 0.13\% & 0.08\% & 0.19\% & 0.28\% & 0.03\% & 0.01\% \\
 & discr & 0.35\% & 0.66\% & 0.20\% & 2.25\% & 0.11\% & 0.63\% & 1.88\% & 0.19\% & 0.10\% \\
\cline{2-11}
 & total & 0.94\% & 1.37\% & 1.48\% & 2.69\% & 0.40\% & 0.83\% & 2.29\% & 0.49\% & 2.25\% \\
\hline
\multirow{6}{*}{$\overline{\text{MS}}$} & $(\gamma_\mu,\gamma_\mu)$ & $-18.73$ & $5.781$ & $41.45$ & $10.80$ & $0.5185$ & $0.4759$ & $0.728$ & $0.8862$ & $0.6977$ \\
 & $(\slashed{q},\slashed{q})$ & $-19.07$ & $6.059$ & $42.43$ & $10.49$ & $0.5295$ & $0.4829$ & $0.764$ & $0.9070$ & $0.6788$ \\
 & central & $-18.90$ & $5.920$ & $41.94$ & $10.64$ & $0.5240$ & $0.4794$ & $0.746$ & $0.8966$ & $0.6882$ \\
\cline{2-11}
 & lattice & 0.66\% & 0.96\% & 1.06\% & 1.40\% & 0.34\% & 0.52\% & 1.75\% & 0.32\% & 1.14\% \\
 & PT & 0.91\% & 2.35\% & 1.17\% & 1.47\% & 1.05\% & 0.74\% & 2.40\% & 1.16\% & 1.38\% \\
\cline{2-11}
 & total & 1.12\% & 2.54\% & 1.57\% & 2.03\% & 1.10\% & 0.90\% & 2.97\% & 1.20\% & 1.79\% \\
\hline
\hline
\end{tabular}
\end{table*}

\subsubsection{Finite Volume Effects}
Finite volume effects (FVEs) could be neglected in our previous studies, but at
this level of precision we need to revisit this assumption. We estimate these
effects using chiral perturbation theory and note that the finite volume
corrections appear with the same pre-factors $C_i^Y$ as the chiral
logarithms~\cite{Becirevic:2004qd}. The FVEs are given
by~\cite{Becirevic:2003wk}
\begin{equation}
  C_i^Y  \frac{m_\pi^2}{(4\pi f_\pi)^2} \frac{12 \sqrt{2\pi} \exp{\left(-m_\pi L\right)}}{(m_\pi L)^{3/2}}\,.
  \label{eq:FVEs}
\end{equation}
The leading order FVEs cancel in the ratios $R_2$ and $R_3$. Numerically
evaluating Eq.~\eqref{eq:FVEs} for our ensembles, we find that the largest
effect is observed on the M1S ensemble, where the estimate of finite size
effects is 1.1 per-mille for the bag parameters and 2.1 per-mille for $R_4$ and
$R_5$. Noting that this is a sub-leading effect (c.f.~Tab.~\ref{tab:error}) and
that the FVEs on the ensembles which are most constraining for the fit (C0M,
M0M, F1M) are more than a factor three smaller than this, we conclude that FVEs
remain negligible at our current level of precision.

\subsubsection{Perturbative matching}
The dominant source of uncertainty arises in the conversion of our results to
$\overline{\textrm{MS}}$ where the matching is done in perturbation theory to
one-loop. The truncation of the perturbative series leads to an uncertainty.  We
have defined two intermediate RI-SMOM schemes, differentiated by their
projectors and use these to estimate the size of this error. We expect results
in $\overline{\textrm{MS}}$ to be independent of the intermediate
renormalisation scheme. We take our central value as the average between the
results obtained from the two intermediate schemes and associate a truncation
uncertainty of half their difference. For definiteness we assign the relative
error from the $(\gamma_\mu,\gamma_\mu)$ scheme to quantify the combined lattice
uncertainty in our final results. The estimate of the perturbative truncation
uncertainty is quoted as ``PT'' in the last column of Table~\ref{tab:error}.

\subsection{Self consistency check}
\begin{figure}
  \centering
  \includegraphics[width=.48\textwidth]{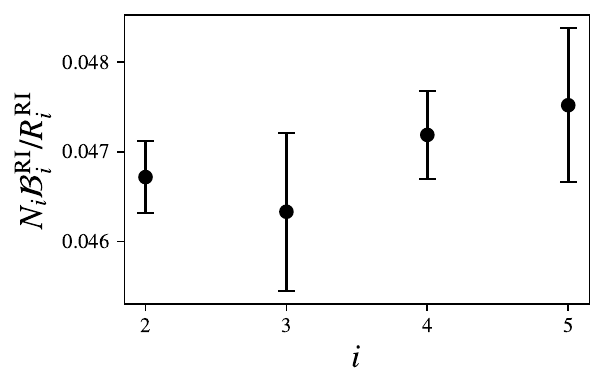}
  \includegraphics[width=.48\textwidth]{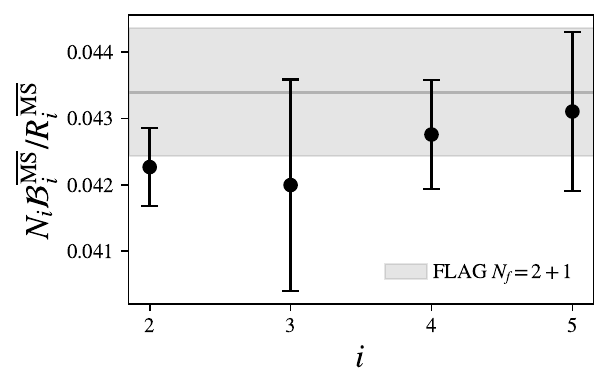}
  \caption{\label{fig:consistency}Self-consistency check by forming the ratio
    Eq.~\eqref{eq:consistency} at $\mu=3\,\mathrm{GeV}$. The data points are
    from our calculations in the $\text{RI-SMOM}^{(\gamma_\mu,\gamma_\mu)}$ scheme (top) and in the $\overline{\text{MS}}$ scheme (bottom). For the $\overline{\text{MS}}$ plot we show the expected value using FLAG inputs as the grey horizontal band.}
\end{figure}

Having determined the $R_i$ and the $\mathcal{B}_i$ parameters we can perform a
self-consistency check. Recalling the definitions in Eqs.~\eqref{eq:BagSM},
\eqref{eq:BagBSM} and \eqref{eq:RatiosBSMtoSM-naive} we consider
\begin{equation}
  \frac{N_i \mathcal{B}_i}{R_i} = \frac{8}{3}\frac{(m_s(\mu)+m_d(\mu))^2}{m_K^2} \mathcal{B}_1 \,, \quad i=2,\cdots,5\,.
  \label{eq:consistency}
\end{equation}
The right hand side is independent of $i$ and hence the ratios
from each operator should give compatible results. The black data points in
Figure~\ref{fig:consistency} display this comparison for the results at
$\mu = 3\,\mathrm{GeV}$ in RI-SMOM$^{(\gamma_\mu,\gamma_\mu)}$ (top) and
$\overline{\mathrm{MS}}$ (bottom). The $R_i$ and $\mathcal B_i$ have notably different --- and sometimes steep --- approaches to the continuum limit. The good
agreement between the different results gives us confidence that
uncertainties in general and discretisation effects in particular have
been well estimated.

We compare our $\overline{\text{MS}}$ results to the value obtained by
evaluating the right hand side using external inputs. We use the
isospin-symmetrised kaon mass $ m_K = (m_{K^0}+m_{K^\pm})/2 =
496.144(9)\,\mathrm{MeV}$~\cite{ParticleDataGroup:2022pth}. We take
FLAG~\cite{FlavourLatticeAveragingGroupFLAG:2021npn} values for the
$N_f=2+1+1$~\cite{ExtendedTwistedMass:2021gbo, FermilabLattice:2018est,
  EuropeanTwistedMass:2014osg, Giusti:2017dmp, Lytle:2018evc,
  Chakraborty:2014aca, Carrasco:2015pra} and $N_f=2+1$ ~\cite{Bruno:2019vup,
  RBC:2014ntl, Durr:2010aw, Durr:2010vn, Bazavov:2010yq, Bruno:2019xed,
  MILC:2009ltw, McNeile:2010ji, RBC:2014ntl, BMW:2011zrh, Laiho:2011np,
  SWME:2015oos} isospin-symmetrised light quark mass and strange quark mass in
$\overline{\mathrm{MS}}$ at $\mu = 2\,\mathrm{GeV}$, together with $\hat{B}_K$,
the renormalisation group invariant (RGI) value for $\mathcal{B}_1$,
\begin{widetext}
\begin{equation}
  \begin{aligned}
    N_f=2{+}1{+}1: &\quad& m_{ud}=3.410(43)\,\mathrm{MeV} && m_s = 93.44(68)\,\mathrm{MeV} && \hat{B}_K=0.717(24)\hphantom{0},\\
    N_f=2{+}1:   && m_{ud}=3.364(41)\,\mathrm{MeV} && m_s = 92.03(88)\,\mathrm{MeV} && \hat{B}_K=0.7625(97)\,.\\
  \end{aligned}
\end{equation}
\end{widetext}
and run them to $\mu = 3\,\mathrm{GeV}$, allowing us to construct the right hand side of Eq.~\eqref{eq:consistency} (the conversion of the four-quark operators at a given scale to RGI operators is shown in Eq.~\eqref{eq:rgi-ops}). This is shown as the grey band in the lower plot in Figure~\ref{fig:consistency}.


Furthermore, we can use the constant value in both the RI-SMOM and
$\overline{\mathrm{MS}}$ schemes, combining it with our value for
$\mathcal{B}_1$, to predict the sum of the quark masses (see also the discussion
in Ref.~\cite{Babich:2006bh}). From our result for $i=2$ we find:
\begin{equation}
  \begin{aligned}
    (m_s + m_{ud})^\mathrm{RI}(3\,\mathrm{GeV}) &= 91.38(41)\,\mathrm{MeV}, \\
    (m_s + m_{ud})^{\overline{\mathrm{MS}}}(3\,\mathrm{GeV}) &= 86.29(79)\,\mathrm{MeV}.
  \end{aligned}
\end{equation}
We compare this to the corresponding FLAG values
\begin{equation}
    \begin{aligned}
         N_f=2{+}1{+}1: & (m_s + m_{ud})^{\overline{\mathrm{MS}}}(3\,\mathrm{GeV})
         = 88.18(63)\,\text{MeV},\\
         N_f=2{+}1: & (m_s + m_{ud})^{\overline{\mathrm{MS}}}(3\,\mathrm{GeV}) = 86.34(79)\, \text{MeV}.
    \end{aligned}
\end{equation}

\subsection{Comparison to our previous work}
\begin{figure*}
    \centering
    \includegraphics[width=\textwidth]{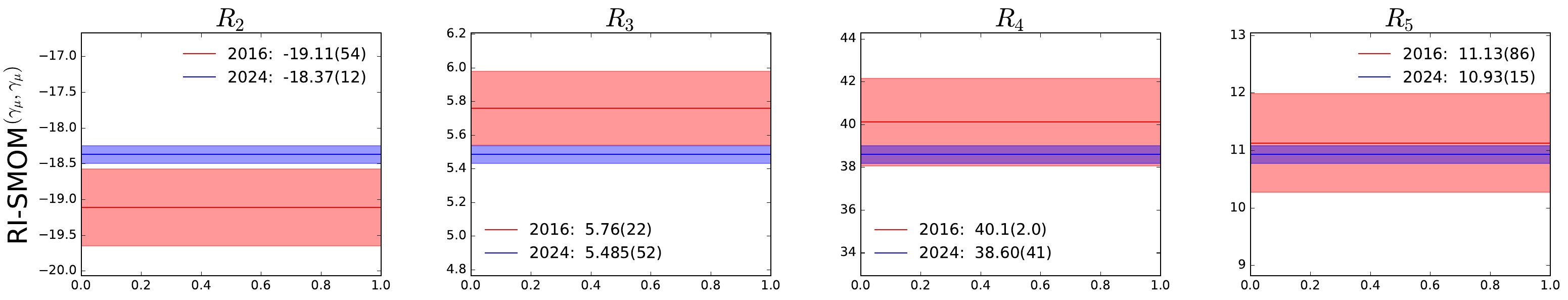}
    \caption{Comparison of the $R_i$ in RI-SMOM$^{(\gamma_\mu,\gamma_\mu)}$
      at $\mu=3\,\mathrm{GeV}$ to our previous
      work~\cite{Garron:2016mva}.\label{fig:compare2016-RI-SMOM-Ri}}
\end{figure*}

In Figure~\ref{fig:compare2016-RI-SMOM-Ri} we compare our results in the RI-SMOM
scheme at $\mu=3\,\mathrm{GeV}$ to our previous
determination~\cite{Garron:2016mva}. The addition of two physical pion mass
ensembles and a third lattice spacing helps to constrain the chiral and
continuum limit extrapolations respectively, yielding a significantly reduced
uncertainty. Given the significantly different data set, we find good agreement
between our previous result and this work.

\begin{figure*}
    \centering
    \includegraphics[width=\textwidth]{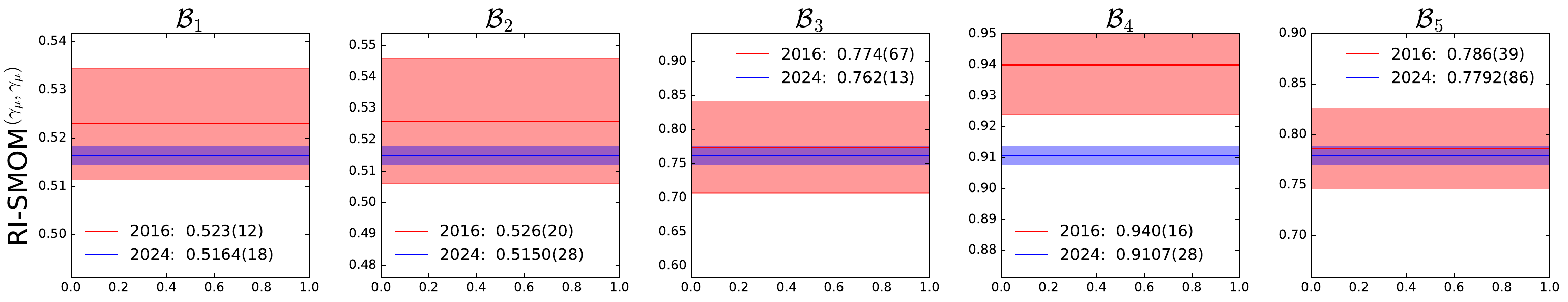}
    \caption{Comparison of the $\mathcal{B}_i$ in
      RI-SMOM$^{(\gamma_\mu,\gamma_\mu)}$ at $\mu=3\,\mathrm{GeV}$ to our
      previous work~\cite{Garron:2016mva}.\label{fig:compare2016-RI-SMOM-Bi}}
\end{figure*}

\subsection{Correlation between the different fit parameters}
We provide the statistical correlation matrix between the $\mathcal{B}_i$, $R_i$
and $\matrixel{K}{O_i^+}{\bar{K}}$ as an ancillary hdf5 file. For completeness
we include results for
RI-SMOM$^{(\gamma_\mu,\gamma_\mu)}(2\,\mathrm{GeV})$,
RI-SMOM$^{(\gamma_\mu,\gamma_\mu)}(3\,\mathrm{GeV})$,
RI-SMOM$^{(\slashed q,\slashed q)}(2\,\mathrm{GeV})$ and
RI-SMOM$^{(\slashed q,\slashed q)}(3\,\mathrm{GeV})$ as well as
$\overline{\mathrm{MS}} \leftarrow \mathrm{RI-SMOM}^{(\gamma_\mu,\gamma_\mu)}(3\,\mathrm{GeV})$ and 
$\overline{\mathrm{MS}} \leftarrow \mathrm{RI-SMOM}^{(\slashed q,\slashed q)}(3\,\mathrm{GeV})$.
\begin{figure}
  \centering
  \includegraphics[width=.8\columnwidth]{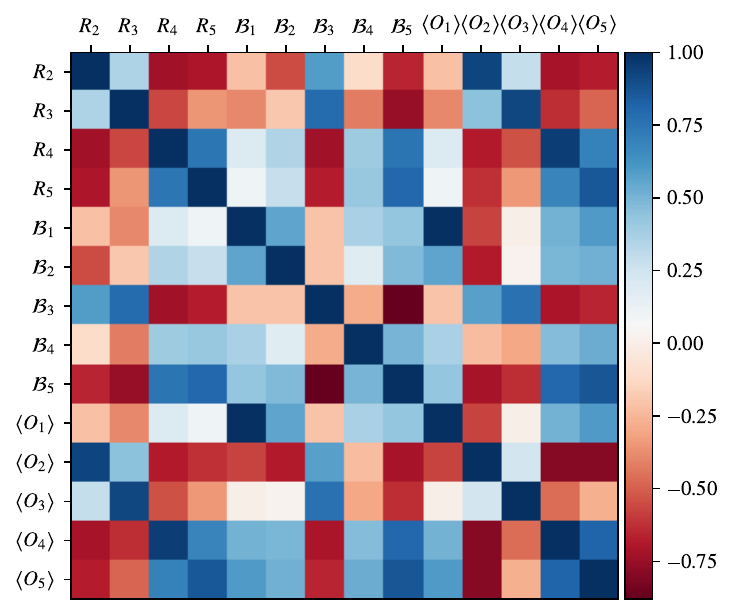}
  \caption{Heat-map of the statistical correlation matrix between the
    $\mathcal{B}_i$, $R_i$ and the $\matrixel{K}{O_i^+}{\bar{K}}$ in
    $\overline{\mathrm{MS}}$ at $3\,\mathrm{GeV}$.}
\end{figure}

\section{Conclusions}
\label{sec:Conclusions}
In this paper we have performed the first calculation of the non-standard model
neutral kaon mixing matrix elements with data directly simulated with physical
quark masses. Using an increased level of volume averaging, with many
$\mathbb{Z}_2$ wall sources on each configuration, we have been able obtain much
reduced statistical errors compared to our previous publications, even with
physical quark masses.

All sources of systematic uncertainties have been estimated. For each of the bag
parameters and ratios of matrix elements a simultaneous fit has been performed
to the mass and lattice spacing dependence. Direct simulation at physical quark
masses leaves the mass dependence of this extrapolation a negligible
systematic. With the inclusion of a third lattice spacing we can test the
validity of $a^2$ scaling and find that in the range covered by our data it
works well. We assess discretisation uncertainties by considering different
renormalisation points and/or different ways of obtaining the
non-perturbative scaling matrix. The self-consistency check of
comparing ratios $N_i \mathcal B_i/R_i$ increases our confidence that
the discretisation effects have been well estimated, since those
ratios approach the continuum limit in notably different ways.

The dominant systematic error comes from perturbative matching from the RI-SMOM
scheme to $\overline{\textrm{MS}}$ at the $3\,\mathrm{GeV}$ renormalisation
point. This key error was assessed by comparing two different intermediate RI-SMOM
schemes after continuum extrapolation. If the matching were non-perturbative the
intermediate scheme would be irrelevant, but with truncated, perturbative
matching the results differ due to the truncation error. The differences are of
the order 1-3\%.

Our final results in the $\overline{\mathrm{MS}}$ scheme at $3\,\mathrm{GeV}$
where the first error is the RI-SMOM error and the second the uncertainty from
the matching to $\overline{\mathrm{MS}}$ are:

\begin{equation}
   \begin{aligned}
    \mathcal{B}_1^{\overline{\textrm{MS}}} &= \hphantom{+}\BaMS(\BaMSL)(\BaMSP)\\
    \mathcal{B}_2^{\overline{\textrm{MS}}} &= \hphantom{+}\BbMS(\BbMSL)(\BbMSP)\\
    \mathcal{B}_3^{\overline{\textrm{MS}}} &= \hphantom{+}\BcMS(\BcMSL)(\BcMSP)\\
    \mathcal{B}_4^{\overline{\textrm{MS}}} &= \hphantom{+}\BdMS(\BdMSL)(\BdMSP)\\
    \mathcal{B}_5^{\overline{\textrm{MS}}} &= \hphantom{+}\BeMS(\BeMSL)(\BeMSP)\\
    R_2^{\overline{\textrm{MS}}} &= \RbMS(\RbMSL)(\RbMSP)\\
    R_3^{\overline{\textrm{MS}}} &= \hphantom{+}\RcMS(\RcMSL)(\RcMSP)\\
    R_4^{\overline{\textrm{MS}}} &= \hphantom{+}\RdMS(\RdMSL)(\RdMSP)\\
    R_5^{\overline{\textrm{MS}}} &= \hphantom{+}\ReMS(\ReMSL)(\ReMSP)\\
\end{aligned}
\end{equation}

\begin{figure*}
    \centering
    \includegraphics[width=\textwidth]{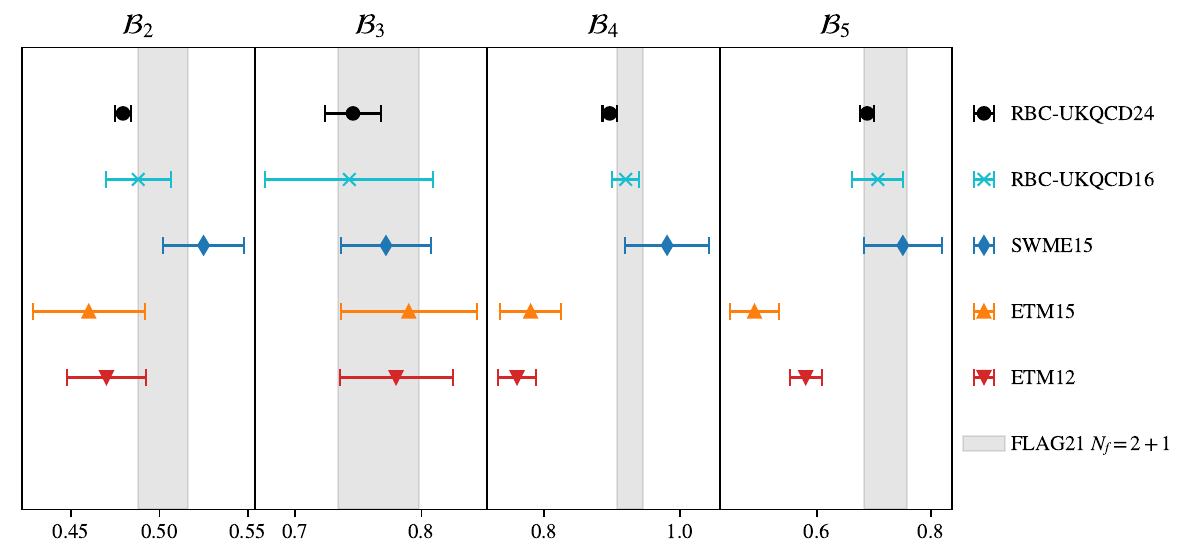}
    \caption{Comparison of our results for the BSM bag parameters in
      $\overline{\text{MS}}$ at 3 GeV with previous results (RBC-UKQCD16~\cite{Garron:2016mva}, SWME15~\cite{SWME:2015oos}, ETM12~\cite{Bertone:2012cu}, ETM15~\cite{Carrasco:2015pra}).}
\end{figure*}

\begin{figure}
    \centering
    \includegraphics[width=\columnwidth]{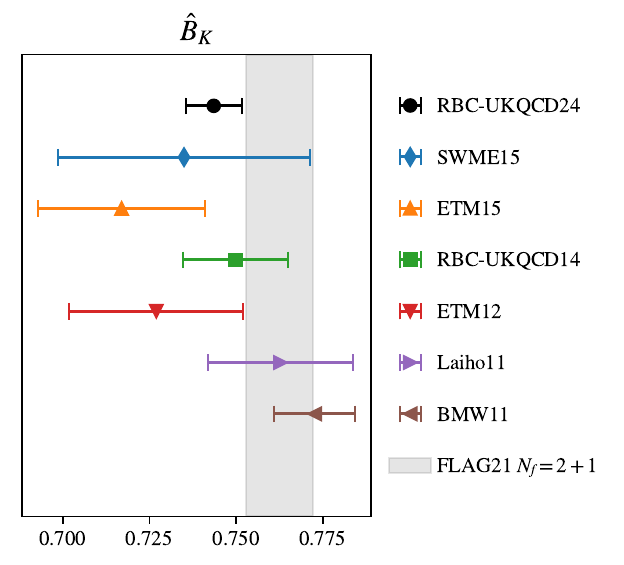}
    \caption{\label{fig:B_K_hat}Comparison of our results for the RGI SM bag
    parameter $\hat{B}_K$ with previous results.}
\end{figure}

We have also presented the value we obtained for the SM bag parameter $B_K$. We
achieve good consistency with our collaborations most up-to-date result,
$B_K^{\overline{\textrm{MS}}\leftarrow \textrm{SMOM}^{(\slashed{q},\slashed{q})}}(3\,\mathrm{GeV})
= 0.530(11)$~\cite{RBC:2014ntl}. A different fitting procedure in which the
physical point data is over-weighted is employed in Ref.~\cite{RBC:2014ntl}, and
while it also includes the coarse and medium ensembles included in this work, it
included a different third lattice spacing with a heavier pion.  Further, it
combined additional coarser ensembles with a different gauge action in a global
fit and reweighting factors to adjust the sea strange mass to the physical
values. In this work, we instead leave the sea strange dependence as a fit
parameter.  Given the differences in the underlying correlator data and the
various fitting procedures, the consistency of the results is reassuring. A
comparison of our results for the RGI value $\hat{B}_K$ with the literature is
shown in Figure~\ref{fig:B_K_hat} where good agreement with the literature is
seen.

The prospects for further improvements of this calculation are as follows: We
believe that the RI-SMOM scheme results are sufficiently precise that there is
no purpose in further reduction in the error within the isospin symmetric pure
QCD approximation. Instead, strong isospin breaking and QED must be addressed if
greater accuracy is required. For our final results a significant source of
error stems from the perturbative matching to $\overline{\textrm{MS}}$. This
could be addressed by raising the matching scale at which we convert
operators. The convergence is logarithmic in the energy scale and this will not
lead to a rapid improvement in the calculation. It would be better to accompany
this with a two-loop calculation of the scheme change factors presented in
Ref.~\cite{Boyle:2017skn}.  The quadratic suppression in $\alpha_s$ would be
more beneficial than an increase in the renormalisation scale towards the
$b$-threshold.



Consequently we believe our results are a robust determination that in the short
term may only be further improved with an additional loop in the perturbative
matching, or by the inclusion of isospin breaking effects.

This work is an important step towards the determination of the same observables
in the $B^0_{(s)}$-$\bar{B}^0_{(s)}$ systems which is also pursued by our
collaboration~\cite{Tsang:2022smt,Boyle:2021kqn}.

\section*{Acknowledgments}
We thank our colleagues in RBC and UKQCD for their contributions and discussion,
in particular Nikolai Husung for helpful discussions on perturbative
renormalisation, Fabian Joswig for important contributions to transitioning
Hadrons' NPR code to GPU architectures, Antonin Portelli for developing and
maintaining Hadrons, and Andreas J\"uttner for contributions at an early stage
of this analysis.

This work used the DiRAC Blue Gene Q Shared Petaflop system and the DiRAC
Extreme Scaling service at the University of Edinburgh, operated by the
Edinburgh Parallel Computing Centre on behalf of the STFC DiRAC HPC Facility
(www.dirac.ac.uk). This equipment was funded by BIS National E-infrastructure
capital grant ST/K000411/1, STFC capital grant ST/H008845/1, and STFC DiRAC
Operations grants ST/K005804/1 and ST/K005790/1.  DiRAC is part of the National
e-Infrastructure.

This work has received funding from an STFC funded studentship. F.E. has
received funding from the European Union’s Horizon Europe research and
innovation programme under the Marie Sk\l{}odowska-Curie grant agreement
No.\ 101106913. R.M. is funded by a University of Southampton Presidential
Scholarship. P.B. has been supported by US DOE Contract
DESC0012704(BNL). N.G. acknowledges support from STFC grant ST/X000699/1.

\FloatBarrier
\appendix
\section{C1M and M1M ensembles \label{sec:newens}}
RBC-UKQCD 2+1f configurations initially used the standard domain wall fermion
action with $L_s=16$~\cite{RBC:2010qam}, but when substantially lighter quark
masses at physical values were introduced~\cite{RBC:2014ntl}, it was desirable
to reduce the level of residual chiral symmetry breaking and the M\"obius domain
wall fermion framework~\cite{Brower:2012vk} was adopted with fixed independent
M\"obius parameter $b=1.5$ and $c=0.5$.  We take $H_W = \gamma_5 D_W$ as the
hermitian Wilson operator at negative domain wall mass, and with this choice of
$b$ and $c$ this yields a kernel $H_M$ entering the overlap sign function that
is identical to the kernel $H_T$ of standard domain wall fermions in the large
$L_s$ limit, with:
\begin{equation}
H_M = \frac{2 H_W}{2+D_W} = 2 H_T,
\end{equation}
entering the four dimensional effective overlap action via an approximate sign
function which (if exact) removes the rescaling factor:
\begin{equation}
D_{ov}  = \frac{1+m}{2} + \frac{1-m}{2}\gamma_5 \epsilon(H_M).
\end{equation}
Here the sign function approximation $\epsilon(H_M)$ is the tanh approximation
to the exact sign function,
\begin{equation}
\begin{split}
\epsilon(H_M)&=
\frac{ (1+ H_M)^{L_s}-(1 - H_M)^{L_s}}
     { (1+ H_M)^{L_s}+(1 - H_M)^{L_s}}\\
     &= \tanh \left( L_s \tanh^{-1} H_M \right).
\end{split}
\end{equation}
The four dimensional effective action of two actions coincide exactly when the
extent of the fifth dimension is infinite, and at finite $L_s$ differ only by
terms that are exponentially suppressed in the fifth dimension extent, a scale
which can by measured by the residual mass measured as the defect of the
M\"obius chiral Ward identity~\cite{RBC:2014ntl}.  Indeed on these two new
ensembles, we obtain residual masses which are very close to those of the
corresponding physical pion mass ensembles~\cite{RBC:2014ntl} as can be seen
from Table~\ref{tab:plaqs}.

\begin{table}
    \centering
    \begin{tabular}{c|c|c}
    \hline\hline
    Ensemble & C1M & M1M \\
    \hline
    Volume & $24^3\times64$ & $32^3\times 64$\\
    $\beta$ & 2.13& 2.25\\
    $b$ & 1.5 & 1.5\\
    $c$ & 0.5 & 0.5\\
    $L_s$ & 24 & 12\\
    $m_l$ & 0.005 & 0.004\\
    $m_s$& 0.0362 & 0.02661\\
    $m_{h}$&  $\{ 0.02,0.2,0.6\}$& $\{ 0.02,0.2,0.6\}$\\
    $N_{\rm Traj}$ & 1990 & 1950\\
    Fermion steps& 12 & 10\\
    Sexton-Weingarten ratio & 8& 8\\
    Integrator& Force Gradient& Force Gradient\\
    \hline\hline
    \end{tabular}
    \caption{Simulation and HMC parameters for the C1M and M1M ensembles.}
    \label{tab:c1mm1m}
\end{table}

Previously RBC-UKQCD have directly combined calculations on ensembles with the
Shamir formulation of domain wall fermions at heavier quark masses with physical
quark mass ensembles using the about M\"obius action. This cost saving measure
is reasonable at the percent scale level accuracy sought at the time, with
residual chiral symmetry breaking effects being of lower magnitude and around
$3\times 10^{-3}$ in the worst-case coarsest ensemble.  In these initial
ensembles, since the input quark mass that corresponds to the physical strange
quark mass is determined by simulation, the strange quark mass could only be
approximately tuned in advance of their generation and so differed from the
physical value.
However, longer term there is great simplicity in generating new ensembles so
that comparisons can be made and extrapolations performed with fewer variables
changing (volume, residual chiral symmetry breaking, strange quark masses),
while increased computing power made this now relatively easily affordable.

To an extent chiral effective approaches and other methods can both estimate
finite volume effects to be small and allow to correct for them.  We therefore
continue to use smaller volumes $24^3\times 64$ for C1M and $32^3\times 64$ for
M1M, but took the opportunity to eliminate two of these three confounding
effects that arise when using the older ensembles.  The first is to keep the
same residual chiral symmetry breaking approximation to the overlap operator,
and the second is to retune the dynamical strange quark mass to its physical
value on each ensemble, matching that used in the physical point ensembles C0M
and M0M.

The new ensembles were generated using the Grid library~\cite{Boyle:2016lbp},
with simulation parameters and intermediate two flavor determinant Hasenbusch
masses~\cite{Urbach:2005ji} that are given in Table~\ref{tab:c1mm1m}.  The exact
one flavor algorithm~\cite{Chen:2014hyy} was used for the strange quark using
the implementation in Grid~\cite{Jung:2017xef}.

For M1M, a two level nested integrator was used in the HMC with the Force
Gradient integrator~\cite{Clark:2011ir,Yin:2011np}, all pseudofermion action
fragments taking ten steps with timestep $\delta t=\frac{1}{10}$, and eight
gauge steps for each fermion step.  For C1M, a two level nested integrator was
used in the HMC with the Force Gradient integrator, all pseudofermion action
fragments taking twelve steps with timestep $\delta t=\frac{1}{12}$, and eight
gauge steps for each fermion step.

\begin{figure}
    \centering
    \includegraphics[width=0.49\textwidth]{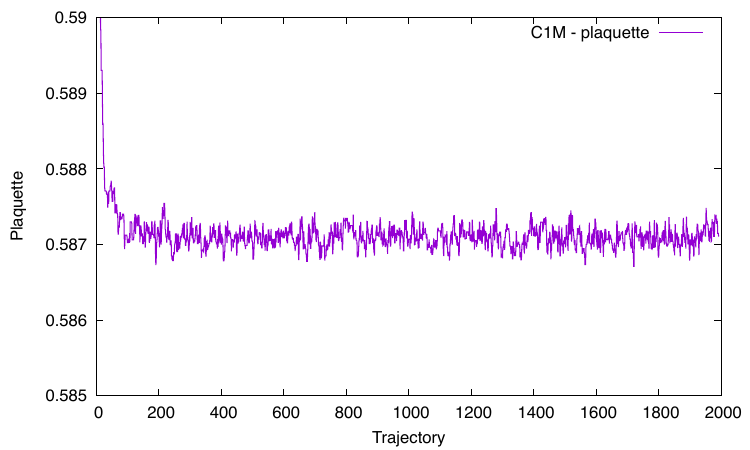}
    \includegraphics[width=0.49\textwidth]{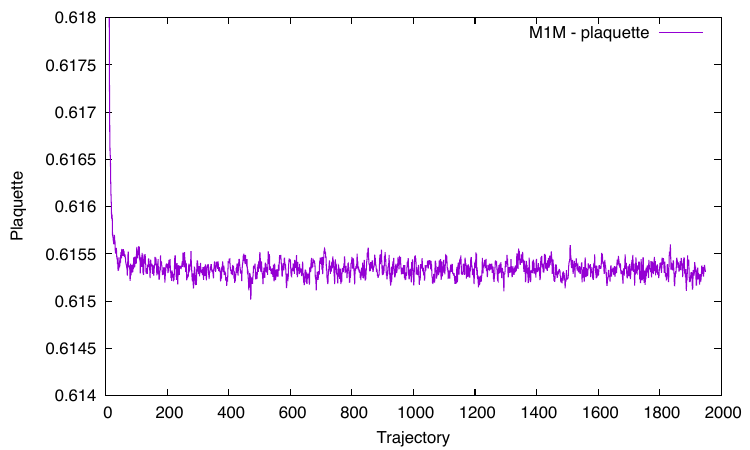}
    \caption{Plaquette molecular dynamics time history for the C1M (top) and M1M (bottom) ensembles.}
    \label{fig:plaqs}
\end{figure}

The first 200 trajectories were discarded from a hot start before
measurements. The plaquette histories are shown in Figure~\ref{fig:plaqs}, and
the average values are more than adequately consistent with those obtained more
precisely in a larger volume at physical quark mass, Table~\ref{tab:plaqs},
given that the quark mass and volume differ.

\begin{table}
    \centering
    \begin{tabular}{c|ll}
    \hline\hline
        Ensemble &  Plaquette &Residual mass\\\hline
        C0M & 0.5871119(25) & 0.0006102(40)\\
        C1M & 0.587091(7) & 0.000604(11)\\
        M0M & 0.6153342(21) & 0.0003116(23)\\
        M1M & 0.615337(4) & 0.0003063(53)\\
    \hline\hline
    \end{tabular}
    \caption{Average plaquettes for the new M1M and C1M ensembles are
    more than adequately consistent with the (more precise) values obtained on the corresponding large volume and physical quarm mass ensembles, given the volume and mass differ.}
    \label{tab:plaqs}
\end{table}

\section{Correlation function fits}
\label{sec:corfits}
\label{app:fit-strat}
This section discusses the choices required in the correlation function fits. In
Section~\ref{sec:covariance} we discuss the construction of the covariance
matrix used in these fits. To mitigate any bias stemming from fit range choices,
two of the authors independently did the analysis of the correlation function
fits. This resulted in two different fit strategies which will be outlined in
Sections~\ref{section:fit_strategy} and~\ref{section:fit_strategy_alt}.

\subsection{Correlated fits to data}\label{sec:covariance}
In all of this paper, our fits are correlated frequentist minimisations of the $\chi^2$-function
\begin{equation}
  \chi^2 = \sum_{i,j}\left(f(a;x_i)-y_i\right) \mathrm{cov}(y_i,y_j)^{-1} \left(f(a;x_j)-y_j\right)\,,
\end{equation}
where $y_i$ are our data, $f(a;x_i)$ is the model with parameters $a$ that is
being fitted and $\mathrm{cov}(y_i,y_j)$ is the covariance matrix of the
data. Resampling $N$ statistical estimators into $N_\mathrm{boot}$
bootstrap samples $\tilde{y}_i$ and denoting the mean of $y_i$ by $\bar{y}_i$, we
define the covariance matrix $\mathrm{cov}(y_i,y_j)$ as
\begin{equation}
\mathrm{cov}(y_i,y_j)=\frac{1}{N_{\mathrm{boot}}} \sum_{k=1}^{N_{\mathrm{boot}}} ((\tilde{y}_i)_k-\bar{y}_i) ((\tilde{y}_j)_k-\bar{y}_j)
\,.
\end{equation}
We relate the covariance matrix to the standard deviation $\sigma$ via $\sigma_i
= \sqrt{\mathrm{cov}(y_i,y_i)}$. Furthermore we define the \emph{normalised
  covariance matrix} or \emph{correlation matrix} $\mathrm{cor}(y_i,y_j)$ as
\begin{equation}
  \mathrm{cor}(y_i,y_j) = \mathrm{diag}(1/\sigma_i) \mathrm{cov}(y_i,y_j) \mathrm{diag}(1/\sigma_j)\,.
\end{equation}
Since we jointly fit multiple two-point and three-point functions, it is
important to be able to accurately invert the covariance matrix that appears in
the $\chi^2$-function.

On a given ensemble, we have $N_{\mathrm{meas}} = N_{\mathrm{conf}} \times
N_{\mathrm{src}}$ estimators for the $y_i$ (compare \Tab~\ref{tab:enspar}). When
estimating the correlation matrix and the standard deviations, we therefore need
to choose whether we treat measurements on different time translations on the
same configuration as independent or whether we bin them into an effective
measurement. From these two choices we obtain
$(\sigma_i,\mathrm{cor}(y_i,y_j))^\mathrm{unbinned}$ and
$(\sigma_i,\mathrm{cor}(y_i,y_j))^\mathrm{binned}$ based on
$N_\mathrm{conf}\times N_\mathrm{src}$ measurements and $N_\mathrm{conf}$ effective
measurements, respectively. We now construct the covariance matrix we use in the
fit to the correlation functions as
\begin{equation}
  \mathrm{cov}(y_i,y_j) = \mathrm{diag}(\sigma_i^\mathrm{binned}) \mathrm{cor}(y_i,y_j)^\mathrm{unbinned} \mathrm{diag}(\sigma_j^\mathrm{binned})\,.
\end{equation}
This has the benefit that it resolves the correlations but estimates the
statistical uncertainties without any assumption of independence for measurements
from different source positions on the same configurations.

\subsection{Correlation function fits - strategy and stability}
\label{section:fit_strategy}
We jointly fit several two-point and three-point functions directly to their
functional forms given by Eqs.~\eqref{eq:2ptdef} and \eqref{eq:3ptdef}. In a
first step, we start by only fitting the two-point functions. We determine fit
ranges $t^\mathrm{2pt}_\mathrm{min}$ and $t^\mathrm{2pt}_\mathrm{max}$ for the
two-point functions which produce stable ground and first excited state results
for masses and overlap factors. For each four-quark operator $Q^+_i$ we then
perform a joint fit to the same two-point functions but also the corresponding
three-point functions $C_3^i(t,\Delta T)$ for several values of $\Delta T$.  As
a first step, we keep $t^\mathrm{2pt}_\mathrm{min}$,
$t^\mathrm{2pt}_\mathrm{max}$ from above for the two-point functions. For the
three-point functions we use the same $t_\mathrm{min}^\mathrm{3pt}$ irrespective
of $\Delta T$. This is determined by choosing an integer $\delta$ to set
$t^\mathrm{3pt}_\mathrm{min} = t^\mathrm{2pt}_\mathrm{min} + \delta$ and
$t^\mathrm{3pt}_\mathrm{max} = \Delta T -
t^\mathrm{3pt}_\mathrm{min}$. Typically we have $\delta \in \{0,1\}$. We then
vary $\delta$ by $\pm 1$, vary the choice of which values of $\Delta T$ enter
the fit and vary $t_\mathrm{min}^\mathrm{2pt}$ by $\pm 1$. We adjust these
choices until we see stability in all fit parameters.

For completeness we summarise the meson masses $m_P=E_0$ and the bare decay
constants $f^\mathrm{bare}_P$ for the pion and the kaon in
Table~\ref{tab:bareresults2pt}.

\begin{table}
  \caption{Masses and bare decay constants of the pion and kaons for all of the
    ensembles used in this work. The ensembles labelled with `$\dagger$' only
    enter the analysis in order to constrain the chiral extrapolation of the
    renormalisation constants described in Section~\ref{sec:Renormalisation}.}
  \begin{tabular}{lllll}
\hline\hline
ens & $am_\pi$ & $af^\mathrm{bare}_\pi$ & $am_K$ & $af^\mathrm{bare}_K$\\ \hline
C0M&  0.08048(10) &  0.10654(12) &  0.28696(13) &  0.126852(89) \\
C1S&  0.19052(40) &  0.11902(27) &  0.30630(39) &  0.13201(22) \\
C2S&  0.24159(38) &  0.12743(20) &  0.32518(35) &  0.13737(18) \\
\hline
M0M&  0.059078(74) &  0.074620(86) &  0.21065(10) &  0.089081(60) \\
M1S&  0.12750(35) &  0.08292(28) &  0.22491(36) &  0.09379(20) \\
M2S&  0.15123(36) &  0.08680(22) &  0.23208(35) &  0.09578(17) \\
M3S&  0.17238(42) &  0.09023(25) &  0.23994(40) &  0.09775(20) \\
\hline
F1M&  0.08581(16) &  0.06768(15) &  0.18810(19) &  0.07821(15) \\
\hline\hline
C1M$^\dagger$&  0.15987(50) &  0.11659(60) &  0.30560(51) &  0.13261(52) \\
M1M$^\dagger$&  0.12116(52) &  0.07943(39) &  0.22778(62) &  0.09193(32) \\
\hline\hline

\end{tabular}

  \label{tab:bareresults2pt}  
\end{table}

\subsection{Alternative strategy}\label{section:fit_strategy_alt}
We define the ratios of two-point functions \eqref{eq:2ptdef} and three-point functions
\eqref{eq:3ptdef}
\begin{equation}
  \begin{aligned}
r^1(t, \Delta T) &= \frac{C^1_3(t, \Delta T)}{C_{PA}(t)C_{AP}(\Delta T - t)}\,,\\
r^i(t, \Delta T) &= \frac{C^i_3(t, \Delta T)}{C_{PP}(t)C_{PP}(\Delta T - t)} \, , \phantom{a} i>1  \label{eq:C3rat}
  \end{aligned}
\end{equation}
which are constructed to asymptotically approach the bag parameters
\begin{align}
r^i(t, \Delta T) &\xrightarrow[0 \ll t \ll \Delta T \ll T]{} N_i \mathcal{B}_i\,.
\end{align}
For simplicity we omit the smearing labels $s_1,s_2$, which are chosen to ensure
that only local matrix elements remain. Expanding numerator and denominator of
the first line of Eq.~\eqref{eq:C3rat} using Eqs.\eqref{eq:2ptdef}
and~\eqref{eq:3ptdef} taking into account the ground state ($\ket{0}$) and first
excited state ($\ket{1}$) contributions ($n=0,1$) (but neglecting the
excited-to-excited matrix elements) yields
\begin{widetext}
\begin{equation}
  \begin{aligned}
    r_1(t, \Delta T)&=\frac{ \langle 0 | \mathcal{O}_1 | 0 \rangle }{M_{A,0}^2} \bigg[1 + X_1(t, \Delta T) e^{-\Delta E\Delta T / 2} + Y_1(t, \Delta T) e^{-\Delta E\Delta T}  \bigg]\, .
  \end{aligned}
\end{equation}
where we defined 
\begin{equation}
  \begin{aligned}
    X_1(t, \Delta T) &= 2\frac{M_{P,1} E_0}{M_{P,0} E_1} \cosh\left[\Delta E(t - \Delta T /2)\right] \left(  \frac{\langle 0 | \mathcal{O}_1 | 1 \rangle}{\langle 0 | \mathcal{O}_1 | 0 \rangle} - \frac{M_{A,1}}{M_{A,0}} \right)\,, \\
    Y_1(t, \Delta T) &=  - 4\frac{M_{P,1}^2 E_0^2}{M_{P,0}^2 E_1^2} \mathrm{cosh}^2 \left[\Delta E(t - \Delta T /2)\right] \frac{\langle 0 | \mathcal{O}_1 | 1 \rangle}{\langle 0 | \mathcal{O}_1 | 0 \rangle}  \frac{M_{A,1}}{M_{A,0}}\,,
  \end{aligned}
\end{equation}
\end{widetext}
and $\Delta E = E_1-E_0$. The expression for $r^i$ ($i>1$) is very similar. 
Defining a summed version of the ratio

\begin{equation}
  \begin{aligned}
  r^i(t_c, \Delta T) &\equiv \sum_{t=t_c}^{\Delta T - t_c} r^i(t, \Delta T)
  \end{aligned}
\end{equation}
and using the identity
\begin{equation}
  \begin{aligned}
  \sum_{t=t_c}^{\Delta T-t_c} \cosh [\Delta E(t - \Delta T /2)] = \frac{\sinh [\Delta E/2(\Delta T-2t_c+1)]}{\sinh [ \Delta E/2]}
  \end{aligned}
\end{equation}
the summed ratio can be expressed as
\begin{widetext}
\begin{equation}
  \begin{aligned}
    r^1(t_c, \Delta T) &=\frac{ \langle 0 | \mathcal{O}_1 | 0 \rangle }{M_{A,0}^2} \bigg[\hat{t}+ 2\frac{M_{P,1} E_0}{M_{P,0} E_1} e^{-\Delta E\Delta T / 2}\frac{\sinh [\Delta E/2\hat{t}]}{\sinh [ \Delta E/2]}\bigg(  \frac{\langle 0 | \mathcal{O}_1 | 1 \rangle}{\langle 0 | \mathcal{O}_1 | 0 \rangle} - \frac{M_{A,1}}{M_{A,0}} \bigg)  \bigg]\,,
  \end{aligned}
\end{equation}
\end{widetext}
where $\hat{t}\equiv\Delta T-2t_c+1$. For a given operator $\mathcal{O}_i$ and
value of $t_c$ we then jointly fit the correlation functions and
$C_{PP}^{SL}(t)$, $C_{PA}^{SL}(t)$,$C_{PP}^{SS}(t)$ and ratios $r^i(t_c, \Delta
T)$ (and the LL equivalent for F1M). The fit ranges of $t \in
[t^{\mathrm{2pt}}_{\mathrm{min}},t^{\mathrm{2pt}}_{\mathrm{max}}]$ and $\Delta T
\in [\Delta T^{\mathrm{3pt}}_{\mathrm{min}},\Delta
  T^{\mathrm{3pt}}_{\mathrm{max}}]$ are chosen such that all fit parameters
remain stable when these ranges are varied by small amounts.

\section{Renormalisation factors}
  \label{app:npr}
\subsection{Definitions}\label{subsec:nprdefs}
We closely follow Ref.~\cite{Boyle:2017jwu} in which the reader will find more
details.  The renormalisation factors are defined by imposing the
renormalisation condition that the projected renormalised amputated-vertex
Green's function, in the Landau gauge, for some chosen external momenta, is equal to its
tree level value (denoted by $F$).  Using the SMOM kinematics
\begin{equation}
 (p_1 - p_2)^2 = p_1^2 = p_2^2 = \mu^2\,,\label{eq:smom}
\end{equation}
the $Z$-factors are defined in the massless limit ($m_q \to 0$) and extracted by
imposing
\begin{equation}
\label{eq:firstZ}
    \lim_{m_q \to 0} P_k\bigg[\frac{Z_{ij}^{\textrm{RI-SMOM}}(\mu,a)}{(Z_q^\text{RI-SMOM}(\mu,a))^2}\Pi_j^{\textrm{bare}}(a,p_1,p_2)\bigg]_{\mathrm{SMOM}}
    = P_k[\Pi_i^{(0)}] \;,
\end{equation}
and the tree-level value $F_{ik} \equiv  P_k[\Pi_i^{(0)}]$ 
is obtained by replacing the propagators by the identity in colour-Dirac space.
Explicitly, for a given four-quark operator $Q_i$, the vertex functions is defined 
as (with $\tilde{x}_i = x_i - x$)
\begin{widetext}
\be
{\Pi_i^{\textrm{bare}}}(a,p_1,p_2)^{\bar \delta \bar \gamma;  \bar \beta \bar \alpha}
=
\la  \bar G(p_2)^{-1} \ra^{\bar \delta \delta } \;
\la G(p_1)^{-1} \ra ^{\gamma \bar \gamma} \;
\la  \bar G(p_2)^{-1} \ra^{\bar \beta \beta } \;
\la G(p_1)^{-1} \ra ^{\alpha \bar \alpha} \;
(M_i^{\textrm{bare}})^{\delta \gamma ;\beta \alpha} (a,p_1,p_2),
\ee
where
\begin{equation}
  \begin{aligned}
    (M_i^{\textrm{bare}})^{\delta \gamma ;\beta \alpha} (q^2)=&
    \sum_{x,x_1,\ldots,x_4}
    \la 0 | s^\delta (x_4) \bar d^\gamma (x_3) 
    \left[  Q_i(x)  \right]
    s^\beta (x_2) \bar d^\alpha (x_1)| 0 \ra
    {\rm e}^{ -i p_1.\tilde x_1 +i p_2.\tilde x_2 -ip_1.\tilde x_3 +ip_2.\tilde x_4},
    \\
    =&
    2\sum_x\left(\la
    \left[ \bar G_x(p_2) \Gamma^1 G_x(p_1) \right]^{\delta \gamma}
    \left[ \bar G_x(p_2) \Gamma^2 G_x(p_1) \right]^{\beta  \alpha}
    \ra \right. \\
    &\qquad - \left.
    \la 
    \left[ \bar G_x(p_2)  \Gamma^1 G_x(p_1) \right]^{\delta \alpha}
    \left[ \bar G_x(p_2) \Gamma^2 G_x(p_1) \right]^{\beta  \gamma}
    \ra
    \right).
    \label{eq:Wick}
  \end{aligned}
\end{equation}
\end{widetext}

Here $G_x(p)$ represents an incoming quark with momentum $p$ and $\bar G_x(-p)$
an outgoing quark with momentum $p$. In addition we have also introduced the 
inverse of the ``full momentum''  propagators
\be
G(p) = \sum_x G_x(p) \quad \text{and} \quad \bar{G}(p) = \sum_x \bar{G}_x(p) \;.
\ee
In Eq.~\eqref{eq:Wick}, the Dirac structure of the four-quark operator
$Q_i$ is encoded in $\Gamma^1 \times \Gamma^2$.

In this work we use two different sets of projectors, $P^{(\gamma_\mu)}$ and
$P^{(\slashed{q})}$, as defined in Ref.~\cite{Boyle:2017jwu}.  The
$P^{(\gamma_\mu)}$ follow the same structure as the four-quark operators. We
split them in three groups according to their chiral-flavour properties. For the
standard model, we have
\be
  \left[ P_1^{(\gamma^\mu)}\right]_{\beta\alpha ; \delta \gamma}^{ba;dc}
=
 \left[ (\gamma^\mu)_{\beta\alpha} (\gamma^\mu)_{\delta\gamma} 
        + (\gamma^\mu \gamma^5)_{\beta\alpha}(\gamma^\mu\gamma^5)_{\delta\gamma} \right] \delta^{ba}\delta^{dc}
\;.
\label{P_gamma1}
\ee
For the $(8,8)$ doublet we define 
\be
\begin{aligned}
\left[ P_2^{(\gamma^\mu)}\right]_{\beta\alpha ; \delta \gamma}^{ba;dc} 
&=
\left[ (\gamma^\mu)_{\beta\alpha} (\gamma^\mu)_{\delta\gamma} 
  - (\gamma^\mu \gamma^5)_{\beta\alpha}(\gamma^\mu\gamma^5)_{\delta\gamma} \right]
\delta^{ba}\delta^{dc}\,,
\\
\left[ P_3^{(\gamma^\mu)}\right]_{\beta\alpha ; \delta \gamma}^{ba;dc}
&=
\left[  
\delta_{\beta\alpha} \delta_{\delta\gamma} - (\gamma^5)_{\beta\alpha}(\gamma^5)_{\delta \gamma}  \right]
\delta^{ba}\delta^{dc}\,,
\end{aligned}
\ee
and similarly for the $(\bar 6,6)$:
\be
\begin{aligned}
\left[ P_4^{(\gamma^\mu)}\right]_{\beta\alpha ; \delta \gamma}^{ba;dc} 
&=
\left[ \delta_{\beta\alpha} \delta_{\delta\gamma} + 
  (\gamma^5)_{\beta\alpha}(\gamma^5)_{\delta \gamma} \right] \delta^{ba}\delta^{dc}\,,
\\
\left[ P_5^{(\gamma^\mu)}\right]_{\beta\alpha ; \delta \gamma}^{ba;dc} 
&=
\left[ \sum_{\nu>\mu}(\gamma^\mu \gamma^\nu)_{\beta\alpha} (\gamma^\mu \gamma^\nu)_{\delta \gamma} \right] \delta^{ba}\delta^{dc} \;.
\end{aligned}
\ee
To define the $P^{(\slashed{q})}$ projectors we replace $\gamma_\mu$ by
$\slashed{q}/q^2$ in the previous equations~\cite{Aoki:2010pe}.  When there is
no explicit $\gamma_\mu$, we take advantage of some Fierz identities, which
relate the colour-unmixed four-quark operators of certain Dirac structure to
colour-mixed four-quark operators of a different Dirac
structure~\cite{Fierz:1937wjm}. Following
Refs.~\cite{Boyle:2017skn,Garron:2018tst} we define
\begin{widetext}
\be
\begin{aligned}
\left[ P_1^{(\s{q})}\right]_{\beta\alpha ; \delta \gamma}^{ba;dc}
    &= \frac{1}{q^2}
      \left[ (\s{q})_{\beta\alpha} (\s{q})_{\delta\gamma} 
        + (\s{q} \gamma^5)_{\beta\alpha}(\s{q}\gamma^5)_{\delta\gamma} \right] \delta^{ba}\delta^{dc}
\;,\\
\left[ P_2^{(\s{q})}\right]_{\beta\alpha ; \delta \gamma}^{ba;dc} 
&=  \frac{1}{q^2}
\left[ (\s{q})_{\beta\alpha} (\s{q})_{\delta\gamma} - 
  (\s{q} \gamma^5)_{\beta\delta}(\s{q}\gamma^5)_{\delta \gamma} \right] \delta^{ba}\delta^{dc}\,,
\\
\left[ P_3^{(\s{q})}\right]_{\beta\alpha ; \delta \gamma}^{ba;dc} 
&= \frac{1}{q^2}
\left[ (\s{q})_{\beta\alpha} (\s{q})_{\delta\gamma} - 
  (\s{q} \gamma^5)_{\beta\delta}(\s{q}\gamma^5)_{\delta \gamma} \right]\delta^{bc}\delta^{da}\;.\\
\left[ P_4^{(\s{q})}\right]_{\beta\alpha ; \delta \gamma}^{ba;dc} 
&=  \frac{1}{p_1^2 p_2^2 - (p_1.p_2)^2}
\left[ 
  \left(p_1^\mu (\sigma^{\mu\nu} P_L )  p_2^{\nu} \right)_{\beta \alpha}
  \left(p_1^\rho (\sigma^{\rho \sigma} P_L) p_2^\sigma \right)_{\delta \gamma}
  \right] \delta^{bc}\delta^{da} \;,
\\
\left[ P_5^{(\s{q})}\right]_{\beta\alpha ; \delta \gamma}^{ba;dc} \,
&=  \frac{1}{p_1^2 p_2^2 - (p_1.p_2)^2}
\left[ 
  \left(p_1^\mu (\sigma^{\mu\nu} P_L )  p_2^{\nu} \right)_{\beta \alpha}
  \left(p_1^\rho (\sigma^{\rho \sigma} P_L) p_2^\sigma \right)_{\delta \gamma}
\right] \delta^{ba}\delta^{dc}\;,
\end{aligned}
\ee
\end{widetext}
where we used the standard definition $\sigma^{\mu\nu}=\frac{1}{2}[\gamma^\mu,\gamma^\nu]$.

In order to eliminate the explicit $Z_q$-dependence in Eq.\eqref{eq:firstZ}, it
is customary to divide the (amputated-projected) vertex function of the
four-quark operators by the one of a bilinear operator. Here we choose the
axial-vector current and we denote $\Pi_A, P_A, F_A$ the corresponding vertex
function, projector and tree level value, respectively. Finally, the choice of
projectors completes the definition of the non-perturbative scheme.  Using
$\mathcal{A}$ to indicate $\gamma_\mu$ or $\slashed{q}$ we define the
renormalisation factors $Z_{ij}^{(\mathcal{A},\mathcal{A})}/Z_A^2$
as\footnote{We can also define $Z_{ij}^{(\mathcal{A},\mathcal{B})}$ with
  $\mathcal{A}\ne\mathcal{B}$, but in this work we consider only
  $(\gamma_\mu,\gamma_\mu)$ and $(\slashed{q},\slashed{q})$}
\begin{widetext}
\begin{equation}
\label{eq:ZBK}
\frac{Z_{ij}^{(\mathcal{A},\mathcal{A})}(\mu,a)}{Z_A^2(a)} \times  \lim_{m_q \to 0} 
 \frac{ P_k^{(\mathcal{A})}  [ \Pi_j^{\textrm{bare}} (a,p_1,p_2) ] }
{ ( P_A^{ (\mathcal{A}) }   [\Pi_A^{ \textrm{bare}} (a,p_1,p_2) ] )^2  } 
\bigg|_{\mathrm{SMOM}}
=\frac{F_{ik}^{(\mathcal{A})}}{(F_A^{(\mathcal{A})})^2}\;.
\end{equation}
\end{widetext}
Our conventions are such that from Eq. \eqref{eq:ZBK} we can define
$Z_{\mathcal{B}_1} \equiv Z_{B_K} = Z_{11} / Z_A^2$.  Finally, to obtain the bag
parameters we also need to renormalise the quark mass. Making use of $Z_m = 1/Z_S$ we impose
\begin{equation}
  \label{eq:ZSoZA}
  \frac{Z_A^{(\mathcal{A})}(\mu,a)}{Z_S(\mu,a)} \times  \lim_{m_q \to 0} 
  \frac{ P_A^{(\mathcal{A})}  [ \Pi_A^{\textrm{bare}} (a,p_1,p_2) ] }
       { P_S [\Pi_S^{ \textrm{bare}} (a,p_1,p_2) ]} 
         \bigg|_{\mathrm{SMOM}}
         =\frac{F_{A}^{(\mathcal{A})}}{F_S}\;.
\end{equation}

The use of two schemes provides a way of estimating systematic errors in the
renormalisation by examining the spread of the results.  The vertex function's
external momenta $p_i^\mu$ are chosen to ensure non-exceptional SMOM kinematics
given in Eq.~\eqref{eq:smom}. Using a combination of Fourier momenta and
partially twisted boundary conditions, the behaviour of the renormalisation
factors as a function of the momentum scale $\mu$ is mapped out in the range
$\mu \in [2-3]\,\mathrm{GeV}$.  In the original exceptional kinematics (used in
RI-MOM), infrared effects fall only as $p^{-2}$ and pion pole subtraction is
required to tame these.  Using non-exceptional kinematics (used in RI-SMOM), the
infrared effects are far more suppressed falling with $p^{-6}$ as has been shown
in Ref.~\cite{Aoki:2007xm}.  Alternative renormalisation schemes such as the
massive RI-SMOM~\cite{Boyle:2016wis,DelDebbio:2023naa} and interpolating
MOM-schemes~\cite{Garron:2022fnn} are currently being explored by the
collaboration. Their application to Kaon mixing is left for future studies.

\subsection{Numerical results for the renormalisation factors ensemble by ensemble}
\label{subsec:Zijnumerical}

\begin{figure}
  \includegraphics[width=\columnwidth]{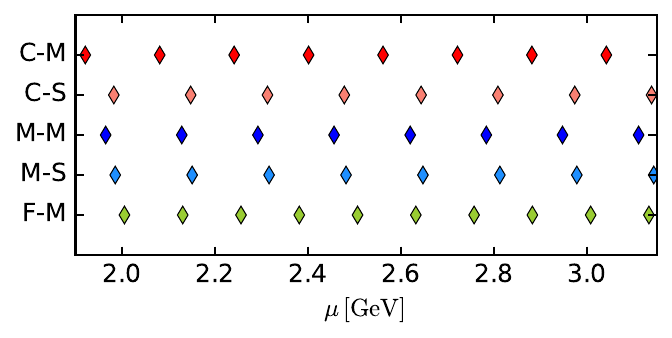}
  \caption{Simulation values of the scale $\mu$ on the various ensembles.}
  \label{fig:NPRmuvals}
\end{figure}
On each ensemble we simulate the NPR data points at a range of choices for the
renormalisation scale $\mu$. These are depicted in
Figure~\ref{fig:NPRmuvals}. We furthermore repeat the simulation at multiple
quark masses, in particular at $am_l^\mathrm{sea}$, $2 am_l^\mathrm{sea}$
and $am_s^\mathrm{sea}/2$ with the exception of the most expensive C0M and M0M
ensembles where we only simulate at $am_s^\mathrm{sea}/2$.

We follow the procedure outlined in Section~\ref{subsec:nprdefs} to calculate
the four-quark matrix elements' renormalisation constants $Z_{ij}/Z_A^2$. We
then normalise these to build the appropriate renormalisation constants listed
in Eq.~\eqref{eq:renormfactors}. In order to ensure reproducibility, we provide
numerical values of the renormalisation constants at some choice of lattice
momenta on the lightest pion mass ensemble for each distinct lattice spacing in
Tables~\ref{tab:C0renormvals}-\ref{tab:F1Mrenormvals}.

\begin{table*}
    \centering
    \caption{\label{tab:C0renormvals} Values of $Z_{ij}/Z_A^2$ for chirally
      non-vanishing matrix elements for a subset of the simulated momenta (in
      lattice units) on the C0M ensemble  for $am_q^\mathrm{val}=0.0181$. All values are given in the
      RI-SMOM$^{(\gamma_\mu,\gamma_\mu)}$ scheme and in the SUSY basis.}
    \begin{tabular}{c|c c c c c c}
\hline
\hline
$a\mu$ & 1.11072 & 1.3884 & 1.66608 & 1.94376 \\
\hline
$Z_{11}/Z_A^2$ & $0.93224(13)$ & $0.923561(68)$ & $0.915757(54)$ & $0.907465(34)$ \\
\hline
$Z_{22}/Z_A^2$ & $0.74937(13)$ & $0.82424(16)$ & $0.877220(37)$ & $0.917178(57)$ \\
$Z_{23}/Z_A^2$ & $-0.041062(66)$ & $-0.056518(65)$ & $-0.071844(45)$ & $-0.087583(32)$ \\
$Z_{32}/Z_A^2$ & $0.03145(13)$ & $-0.010829(52)$ & $-0.043507(40)$ & $-0.071655(36)$ \\
$Z_{33}/Z_A^2$ & $1.154966(63)$ & $1.106912(37)$ & $1.079801(38)$ & $1.063261(10)$ \\
\hline
$Z_{44}/Z_A^2$ & $0.70331(20)$ & $0.79997(14)$ & $0.867628(12)$ & $0.918149(54)$ \\
$Z_{45}/Z_A^2$ & $-0.050481(64)$ & $-0.060964(52)$ & $-0.073113(41)$ & $-0.086769(28)$ \\
$Z_{54}/Z_A^2$ & $-0.171645(50)$ & $-0.149960(37)$ & $-0.139921(39)$ & $-0.137573(37)$ \\
$Z_{55}/Z_A^2$ & $1.051566(66)$ & $1.049023(33)$ & $1.049000(28)$ & $1.050999(11)$ \\
\hline
$Z_A/Z_S$ & $1.197636(71)$ & $1.129805(84)$ & $1.088466(26)$ & $1.061227(33)$ \\
\hline
\hline
\end{tabular}
\end{table*}

\begin{table*}
    \centering
    \caption{\label{tab:C1renormvals} Same as Table \ref{tab:C0renormvals} but
      for the C1S ensemble for $am_q^\mathrm{val}=0.005$.}  \begin{tabular}{c|c c c c c c}
\hline
\hline
$a\mu$ & 1.11072 & 1.3884 & 1.66608 & 1.94376 \\
\hline
$Z_{11}/Z_A^2$ & $0.93303(13)$ & $0.924165(55)$ & $0.916271(66)$ & $0.907958(58)$ \\
\hline
$Z_{22}/Z_A^2$ & $0.75249(61)$ & $0.82630(28)$ & $0.87856(11)$ & $0.91832(12)$ \\
$Z_{23}/Z_A^2$ & $-0.04129(15)$ & $-0.056775(74)$ & $-0.071886(82)$ & $-0.087656(88)$ \\
$Z_{32}/Z_A^2$ & $0.03059(21)$ & $-0.01137(10)$ & $-0.043656(74)$ & $-0.071733(95)$ \\
$Z_{33}/Z_A^2$ & $1.15504(28)$ & $1.10673(11)$ & $1.079562(84)$ & $1.063174(48)$ \\
\hline
$Z_{44}/Z_A^2$ & $0.70746(75)$ & $0.80253(30)$ & $0.86921(11)$ & $0.91939(13)$ \\
$Z_{45}/Z_A^2$ & $-0.05062(20)$ & $-0.06120(10)$ & $-0.073071(83)$ & $-0.086787(86)$ \\
$Z_{54}/Z_A^2$ & $-0.17065(21)$ & $-0.149293(88)$ & $-0.139280(88)$ & $-0.137090(68)$ \\
$Z_{55}/Z_A^2$ & $1.05273(21)$ & $1.049571(82)$ & $1.049163(54)$ & $1.051141(43)$ \\
\hline
$Z_A/Z_S$ & $1.19361(86)$ & $1.12820(21)$ & $1.08753(10)$ & $1.060548(76)$ \\
\hline
\hline
\end{tabular}
\end{table*}

\begin{table*}
    \centering
    \caption{\label{tab:M0renormvals} Same as Table \ref{tab:C0renormvals} but
      for the M0M ensemble for $am_q^\mathrm{val}=0.0133$.}  \begin{tabular}{c|c c c c c c}
\hline
\hline
$a\mu$ & 0.83304 & 1.0413 & 1.24956 & 1.3884 \\
\hline
$Z_{11}/Z_A^2$ & $0.958988(97)$ & $0.948999(54)$ & $0.941580(40)$ & $0.937357(31)$ \\
\hline
$Z_{22}/Z_A^2$ & $0.701561(98)$ & $0.774046(79)$ & $0.826662(61)$ & $0.854299(47)$ \\
$Z_{23}/Z_A^2$ & $-0.023850(59)$ & $-0.034541(40)$ & $-0.044346(40)$ & $-0.050758(38)$ \\
$Z_{32}/Z_A^2$ & $0.075307(47)$ & $0.034965(33)$ & $0.005582(30)$ & $-0.010480(23)$ \\
$Z_{33}/Z_A^2$ & $1.212886(87)$ & $1.149630(45)$ & $1.111072(35)$ & $1.093473(27)$ \\
\hline
$Z_{44}/Z_A^2$ & $0.63286(11)$ & $0.727410(91)$ & $0.796468(58)$ & $0.832571(45)$ \\
$Z_{45}/Z_A^2$ & $-0.038240(46)$ & $-0.043612(32)$ & $-0.049903(29)$ & $-0.054545(30)$ \\
$Z_{54}/Z_A^2$ & $-0.182166(54)$ & $-0.154917(30)$ & $-0.137278(21)$ & $-0.129566(23)$ \\
$Z_{55}/Z_A^2$ & $1.057128(68)$ & $1.050256(31)$ & $1.045950(24)$ & $1.044326(22)$ \\
\hline
$Z_A/Z_S$ & $1.26232(13)$ & $1.181992(83)$ & $1.131563(41)$ & $1.107811(22)$ \\
\hline
\hline
\end{tabular}
\end{table*}

\begin{table*}
    \centering
    \caption{\label{tab:M1renormvals} Same as Table \ref{tab:C0renormvals} but
      for the M1S ensemble for $am_q^\mathrm{val}=0.004$.}  \begin{tabular}{c|c c c c c c}
\hline
\hline
$a\mu$ & 0.83304 & 1.0413 & 1.24956 & 1.45782 \\
\hline
$Z_{11}/Z_A^2$ & $0.95866(27)$ & $0.94835(17)$ & $0.94111(13)$ & $0.93495(10)$ \\
\hline
$Z_{22}/Z_A^2$ & $0.70006(63)$ & $0.77412(33)$ & $0.82662(14)$ & $0.86644(13)$ \\
$Z_{23}/Z_A^2$ & $-0.02394(22)$ & $-0.034736(81)$ & $-0.044604(78)$ & $-0.054232(65)$ \\
$Z_{32}/Z_A^2$ & $0.07552(21)$ & $0.03452(10)$ & $0.005247(63)$ & $-0.018170(55)$ \\
$Z_{33}/Z_A^2$ & $1.21448(44)$ & $1.14967(17)$ & $1.111076(76)$ & $1.086321(33)$ \\
\hline
$Z_{44}/Z_A^2$ & $0.63178(68)$ & $0.72810(36)$ & $0.79677(14)$ & $0.84859(14)$ \\
$Z_{45}/Z_A^2$ & $-0.03853(25)$ & $-0.043920(89)$ & $-0.050118(52)$ & $-0.057240(60)$ \\
$Z_{54}/Z_A^2$ & $-0.18333(20)$ & $-0.15511(13)$ & $-0.137472(66)$ & $-0.126927(60)$ \\
$Z_{55}/Z_A^2$ & $1.05863(24)$ & $1.05060(10)$ & $1.046175(54)$ & $1.044027(36)$ \\
\hline
$Z_A/Z_S$ & $1.2629(14)$ & $1.18148(46)$ & $1.13157(10)$ & $1.09803(10)$ \\
\hline
\hline
\end{tabular}
\end{table*}

\begin{table*}
    \centering
    \caption{\label{tab:F1Mrenormvals} Same as Table \ref{tab:C0renormvals} but
      for the F1M ensemble for $am_q^\mathrm{val}=0.0021$.}  \begin{tabular}{c|c c c c c}
\hline
\hline
$a\mu$ & 0.74048 & 0.87932 & 1.01816 & 1.43468 \\
\hline
$Z_{11}/Z_A^2$ & $0.97098(29)$ & $0.96182(11)$ & $0.955127(87)$ & $0.941078(33)$ \\
\hline
$Z_{22}/Z_A^2$ & $1.06173(45)$ & $1.05453(10)$ & $1.049270(35)$ & $1.041636(15)$ \\
$Z_{23}/Z_A^2$ & $0.37422(52)$ & $0.32971(35)$ & $0.29528(14)$ & $0.239419(67)$ \\
$Z_{32}/Z_A^2$ & $0.01748(19)$ & $0.018812(88)$ & $0.020455(43)$ & $0.026601(18)$ \\
$Z_{33}/Z_A^2$ & $0.61015(69)$ & $0.68079(75)$ & $0.73890(33)$ & $0.85407(11)$ \\
\hline
$Z_{44}/Z_A^2$ & $0.69653(77)$ & $0.75380(76)$ & $0.80124(37)$ & $0.89759(12)$ \\
$Z_{45}/Z_A^2$ & $-0.00933(14)$ & $-0.01280(13)$ & $-0.015992(62)$ & $-0.025037(17)$ \\
$Z_{54}/Z_A^2$ & $-0.36177(54)$ & $-0.31116(28)$ & $-0.27401(16)$ & $-0.215021(71)$ \\
$Z_{55}/Z_A^2$ & $1.23079(76)$ & $1.17174(16)$ & $1.130128(97)$ & $1.058518(33)$ \\
\hline
$Z_A/Z_S$ & $1.28592(66)$ & $1.22090(96)$ & $1.17291(37)$ & $1.093643(58)$ \\
\hline
\hline
\end{tabular}
\end{table*}

\subsection{Extrapolation of the renormalisation factors to the massless limit}
\label{subsec:Zijtomassless}
Formally the renormalisation constants are defined in the massless (zero quark
mass) limit. In order to perform this limit lattice-spacing-by-lattice-spacing
we proceed as follows. We first extrapolate the valence quark mass to zero
ensemble-by-ensemble as described in Sec.~\ref{sec:npranalysis}. We then
interpolate the renormalisation constants on all ensembles to a fixed scale
$\mu$. We either perform a linear fit to the two closest simulated values of
$\mu$ or a quadratic fit to the three closest points. We then perform a chiral
extrapolation in $(am_\pi)^2$ to all ensembles that share an identical lattice
spacing (C1S and C2S; C0M and C1M; M1S, M2S and M3S; M0M and M1M). Since we only
have data on a single ensemble (F1M) for the finest lattice spacing, we apply
each of the slopes with $m_\pi^2$ in turn. On F-M, we assign the central value
to be the mean of these four extrapolated results and take half the spread of
the results as a systematic uncertainty associated to the chiral
extrapolation. We provide numerical values of these chirally extrapolated
renormalisation constants for each of the lattice spacings at
$\mu=2\,\mathrm{GeV}$ (Table~\ref{tab:ch-extrap-2.0}), $\mu=2.5\,\mathrm{GeV}$
(Table~\ref{tab:ch-extrap-2.5}), and $\mu=3\,\mathrm{GeV}$
(Table~\ref{tab:ch-extrap-3.0}).

\begin{table*}
    \centering
    \caption{\label{tab:ch-extrap-2.5} Same as Table \ref{tab:ch-extrap-2.0} but
      at mass scale $\mu={2.5}\,\mathrm{GeV}$.}
    \begin{tabular}{c|ccccc}
\hline
\hline
$a^{-1}$ [GeV] & 1.7295(38) & 1.7848(50) & 2.3586(70) & 2.3833(86) & 2.708(10) \\
\hline
$Z_{11}/Z_A^2$ & 0.92273(29)(1) & 0.92491(52)(0) & 0.94913(20)(0) & 0.94752(86)(2) & 0.95950(43)(33) \\
\hline
$Z_{22}/Z_S^2$ & 1.05048(77)(4) & 1.0519(13)(0) & 1.08128(96)(1) & 1.0830(21)(0) & 1.0964(11)(6) \\
$Z_{23}/Z_S^2$ & -0.07511(48)(3) & -0.07280(82)(2) & -0.04932(29)(0) & -0.04900(66)(1) & -0.04007(40)(17) \\
$Z_{32}/Z_S^2$ & -0.0228(16)(1) & -0.0162(45)(0) & 0.0447(28)(1) & 0.0465(63)(1) & 0.0746(37)(0) \\
$Z_{33}/Z_S^2$ & 1.3816(30)(7) & 1.4014(83)(2) & 1.5913(50)(6) & 1.603(11)(0) & 1.6981(67)(17) \\
\hline
$Z_{44}/Z_S^2$ & 1.02458(44)(3) & 1.02303(97)(4) & 1.01873(49)(7) & 1.0201(14)(0) & 1.01595(76)(71) \\
$Z_{45}/Z_S^2$ & -0.07967(51)(2) & -0.07835(69)(1) & -0.06134(25)(1) & -0.06193(46)(0) & -0.05599(23)(24) \\
$Z_{54}/Z_S^2$ & -0.18458(58)(19) & -0.1880(18)(0) & -0.2130(11)(1) & -0.2162(26)(1) & -0.2305(15)(5) \\
$Z_{55}/Z_S^2$ & 1.3170(21)(5) & 1.3310(60)(1) & 1.4581(36)(3) & 1.4673(80)(5) & 1.5267(46)(17) \\
\hline
\hline
\end{tabular}
\end{table*}
\begin{table*}
    \centering
    \caption{\label{tab:ch-extrap-3.0} Same as Table \ref{tab:ch-extrap-2.0} but
      at mass scale $\mu={3.0}\,\mathrm{GeV}$.}
    \begin{tabular}{c|ccccc}
\hline
\hline
$a^{-1}$ [GeV] & 1.7295(38) & 1.7848(50) & 2.3586(70) & 2.3833(86) & 2.708(10) \\
\hline
$Z_{11}/Z_A^2$ & 0.91427(17)(0) & 0.91641(55)(0) & 0.94123(17)(1) & 0.94044(67)(0) & 0.95157(31)(27) \\
\hline
$Z_{22}/Z_S^2$ & 1.03795(18)(4) & 1.03874(73)(2) & 1.05744(45)(5) & 1.0596(10)(0) & 1.06947(54)(36) \\
$Z_{23}/Z_S^2$ & -0.08818(26)(2) & -0.08564(98)(0) & -0.05786(22)(0) & -0.05804(54)(0) & -0.04741(25)(25) \\
$Z_{32}/Z_S^2$ & -0.0589(10)(0) & -0.0532(31)(0) & 0.0036(18)(0) & 0.0052(44)(0) & 0.0300(24)(1) \\
$Z_{33}/Z_S^2$ & 1.2568(19)(2) & 1.2711(59)(2) & 1.4093(34)(4) & 1.4206(79)(1) & 1.4919(44)(11) \\
\hline
$Z_{44}/Z_S^2$ & 1.03009(18)(0) & 1.02856(45)(0) & 1.02098(20)(0) & 1.02223(60)(0) & 1.01811(29)(38) \\
$Z_{45}/Z_S^2$ & -0.08895(23)(2) & -0.08683(80)(0) & -0.06453(14)(1) & -0.06486(36)(0) & -0.05716(16)(22) \\
$Z_{54}/Z_S^2$ & -0.16199(30)(9) & -0.1636(11)(0) & -0.17286(72)(9) & -0.1757(17)(0) & -0.1842(10)(3) \\
$Z_{55}/Z_S^2$ & 1.2268(12)(1) & 1.2368(41)(1) & 1.3299(24)(2) & 1.3389(57)(1) & 1.3852(31)(10) \\
\hline
\hline
\end{tabular}
\end{table*}

\subsection{Step-scaling matrices}\label{subsec:Stepscaling}
\begin{figure*}
    \vspace{-1cm}
    \subfloat{\includegraphics[width=0.49\textwidth]{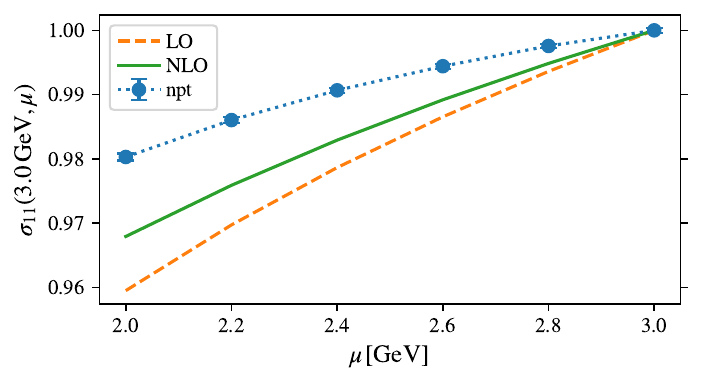}}\\
    \vspace{-0.3cm}
    \subfloat{\includegraphics[width=0.49\textwidth]{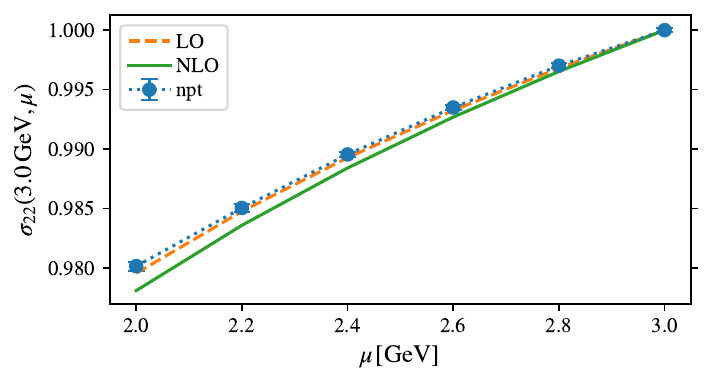}}
    \hfill
    \subfloat{\includegraphics[width=0.49\textwidth]{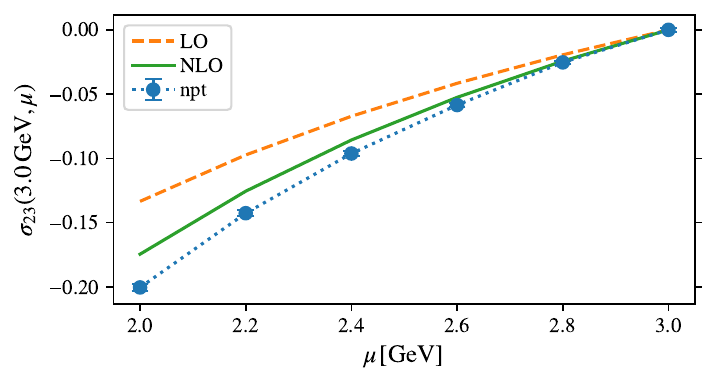}}\\
    \vspace{-0.3cm}
    \subfloat{\includegraphics[width=0.49\textwidth]{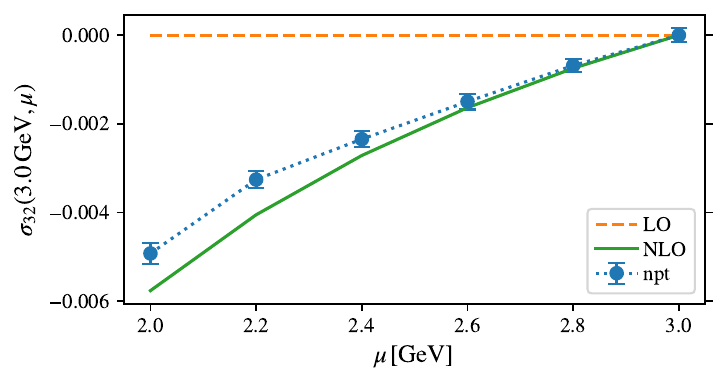}}
    \hfill
    \subfloat{\includegraphics[width=0.49\textwidth]{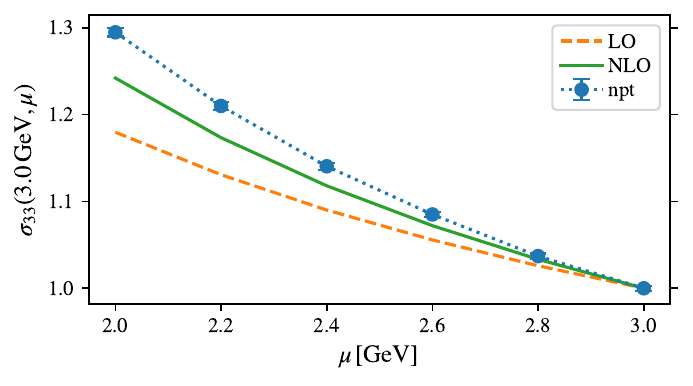}}\\
    \vspace{-0.3cm}
    \subfloat{\includegraphics[width=0.49\textwidth]{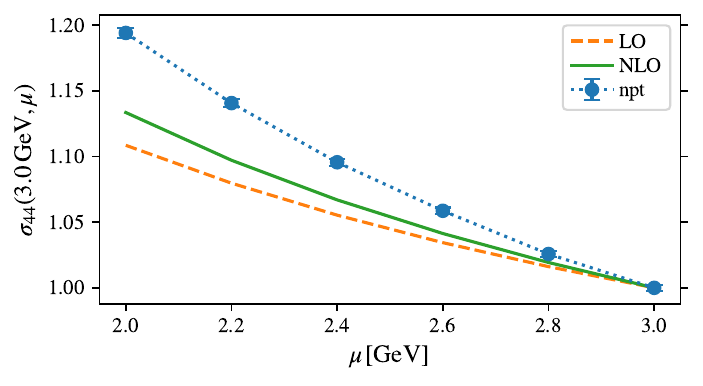}}
    \hfill
    \subfloat{\includegraphics[width=0.49\textwidth]{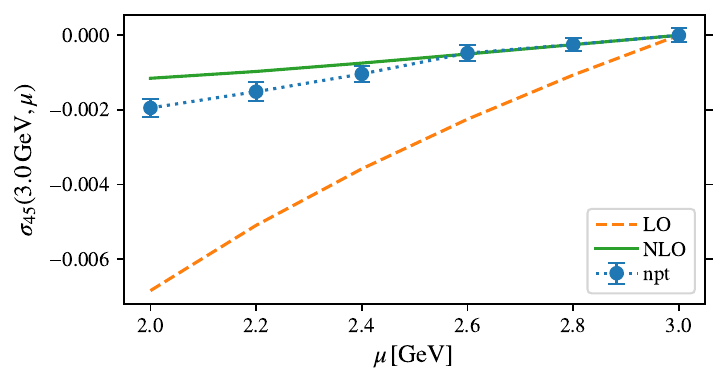}}\\
    \vspace{-0.3cm}
    \subfloat{\includegraphics[width=0.49\textwidth]{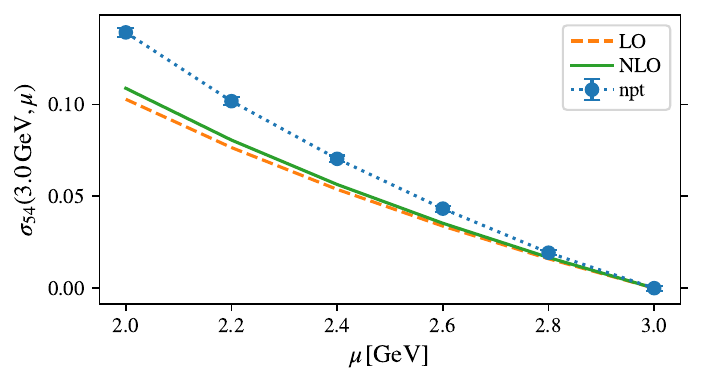}}
    \hfill
    \subfloat{\includegraphics[width=0.49\textwidth]{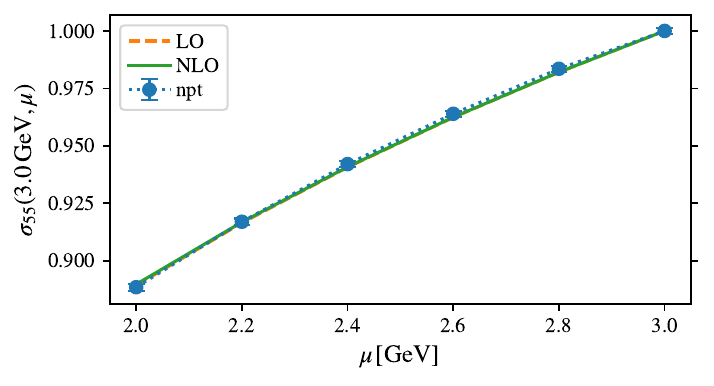}}
    \caption{\label{fig:npr-vs-pt-scaling} Comparison of the scale evolution
      matrix $\sigma(3\,\mathrm{GeV},\mu)$ (see~Eq.\eqref{eq:stepscaling}) in
      the RI-SMOM$^{(\gamma_\mu,\gamma_\mu)}$ scheme and NPR basis evaluated
      non-perturbatively (blue circles), perturbatively at leading order (orange
      dashed lines) and next-to-leading order (green solid lines).}
\end{figure*}

We choose a lower scale $\mu$ for non-perturbative renormalisation to reduce
cutoff effects, and a higher scale $\mu'$ for matching to
$\overline{\text{MS}}$. This distance is bridged using non-perturbative running
\begin{align}
    O_i^{\overline{\text{MS}}} (\mu') = R_{ij}^{\overline{\text{MS}} \leftarrow \text{RI}}(\mu')\sigma_{jk}(\mu',\mu)O_k^\text{RI}(\mu),
\end{align}
 $O_i$ being any of the quantities in Eq.~\eqref{eq:renormfactors}. The matching
factors $R_{ij}^{\overline{\text{MS}} \leftarrow \text{RI}}$ are computed in
next-to-leading order perturbation theory and presented in
Ref.~\cite{Boyle:2017skn} for both schemes. The non-perturbative scale evolution
matrix is given by
\begin{align}
    \sigma_{ij}(\mu', \mu) &= \lim_{a\to 0} \sigma_{ij}(\mu', \mu, a) \nonumber \\
    &= \lim_{a\to 0} Z_{ik}(\mu',a)Z_{kj}(\mu, a)^{-1}\,.
\end{align}
We perform this continuum limit including the chirally extrapolated
renormalisation constants from all lattice spacings as a fit linear in $a^2$. In
the few cases where the quality of fit does not lead to an acceptable $p$-value,
we rescale the uncertainty by $\sqrt{\chi^2/\mathrm{d.o.f.}}$. In
Figure~\ref{fig:npr-vs-pt-scaling} we compare our non-perturbative step-scaling
results to leading and next-to-leading order perturbation theory. For
completeness we also provide the step-scaling matrices, $\sigma_R$, for the
$R_i$
\begin{widetext}
\begin{align}
    \sigma_R(3\,\mathrm{GeV},2\,\mathrm{GeV}) &= \begin{bmatrix}
1.0 & 0.0 & 0.0 & 0.0 & 0.0 \\
0.0 & 1.2153(28) & -0.08396(60) & 0.0 & 0.0 \\
0.0 & -0.00426(52) & 0.90868(42) & 0.0 & 0.0 \\
0.0 & 0.0 & 0.0 & 1.3186(42) & 0.1018(15) \\
0.0 & 0.0 & 0.0 & 0.00976(68) & 0.99984(59) \\
\end{bmatrix} \label{eq:step2_3_rat},\\
    \sigma_R(3\,\mathrm{GeV}\xleftarrow{\Delta=0.5}2\,\mathrm{GeV}) &= \begin{bmatrix}
1.0 & 0.0 & 0.0 & 0.0 & 0.0 \\
0.0 & 1.2149(25) & -0.08315(50) & 0.0 & 0.0 \\
0.0 & -0.00374(43) & 0.90889(40) & 0.0 & 0.0 \\
0.0 & 0.0 & 0.0 & 1.3190(35) & 0.1023(14) \\
0.0 & 0.0 & 0.0 & 0.01083(44) & 0.99940(59) \\
\end{bmatrix}. \label{eq:step2_3_multi_rat}
\end{align}
\end{widetext}

\begin{table*}
    \caption{\label{tab:running_values} Chirally-allowed elements of the
      non-perturbative scaling matrix $\sigma(3\,\mathrm{GeV},2\,\mathrm{GeV})$
      using chirally extrapolated $Z$-factors.}  \centering
    \begin{tabular}{c|ccccc}
\hline
\hline
$a^{-1}$ [GeV] & 1.7848(50) & 1.7295(38) & 2.3833(86) & 2.3586(70) & 2.708(10) \\
\hline
$\sigma_{22}$ & 1.1866(15) & 1.18862(99) & 1.2055(27) & 1.1980(16) & 1.2047(16) \\
$\sigma_{23}$ & -0.020560(36) & -0.02201(14) & -0.01408(33) & -0.01294(22) & -0.01165(18) \\
$\sigma_{32}$ & -0.09661(26) & -0.09870(19) & -0.09318(55) & -0.09034(41) & -0.08994(17) \\
$\sigma_{33}$ & 0.94800(88) & 0.95149(38) & 0.92958(89) & 0.93208(35) & 0.92586(49) \\
\hline
$\sigma_{44}$ & 1.2441(30) & 1.2447(15) & 1.2844(46) & 1.2748(21) & 1.2910(23) \\
$\sigma_{45}$ & -0.00982(21) & -0.01235(22) & -0.00146(54) & -0.00159(23) & 0.00067(30) \\
$\sigma_{54}$ & 0.0439(12) & 0.04063(45) & 0.0711(17) & 0.06801(83) & 0.0774(10) \\
$\sigma_{55}$ & 1.017341(75) & 1.018409(16) & 1.00827(44) & 1.00994(39) & 1.00838(48) \\
\hline
\hline
\end{tabular}
\end{table*}

\section{Relations between basis conventions} \label{sec:Fierz}

We distinguish between operators $O_i$ in the ``SUSY" basis, defined in
Eq.~\eqref{eq:opbasis_orig_susy}, and operators $Q_i$ in the ``NPR'' or
``lattice'' basis, in Eq.~\eqref{eq:opbasis_orig_npr}. The SUSY basis contains
both colour-unmixed and colour-mixed operators, while the NPR basis comprises
only colour-unmixed operators, more convenient for lattice computations. For the
$K^0\bar K^0$ matrix elements of these operators we need only the parity-even
parts, $O_i^+$ or $Q_i^+$, shown in Eqs.~\eqref{eq:opbasis_susy}
and~\eqref{eq:opbasis_unmixed_even}. Since we work with the $Q_i^+$ on the
lattice, we quote here the matrix $T$ which relates the $Q_i^+$ to the $O_i^+$:
\begin{widetext}
\begin{equation}
    O^+ = T Q^+ = 
    \begin{pmatrix}
    1 & 0 & 0  & 0 & 0 \\
    0 & 0 & 0 & 1 & 0 \\
    0 & 0 & 0 & -\frac12 & \frac12 \\
    0 & 0 & 1 & 0 & 0 \\
    0 & -\frac12 & 0 & 0 & 0 \\
    \end{pmatrix}Q^+ \qquad\text{or}\qquad
    \begin{pmatrix}
    O^+_1 \\ O^+_2 \\ O^+_3 \\ O^+_4 \\ O^+_5
    \end{pmatrix}
    = 
    \begin{pmatrix}
    Q_1^+ \\
    Q_4^+ \\
    (Q_5^+ - Q_4^+)/2 \\
    Q_3^+ \\
    -Q_2^+/2
    \end{pmatrix}\,.
\end{equation}
\end{widetext}
We perform our non-perturbative calculations of the matrix elements and
renormalisation constants in the NPR basis and subsequently convert to the SUSY
basis. Matrix elements of the renormalised $Q_i^+$ in some scheme $X$ at scale
$\mu$ are given by,
\begin{equation}
    \matrixel{K}{Q_i^+}{\bar{K}}^X(\mu) = Z^X_{ij}(\mu) \matrixel{K}{Q_{j,\mathrm{bare}}^+}{\bar{K}}\,.
\end{equation}
When chiral symmetry is maintained, $Z_{ij}^X(\mu)$ is block-diagonal and the
matrix elements of the renormalised operators $O_i^+$ in the SUSY basis in
scheme $X$ at scale $\mu$ are related to matrix elements of the bare operators
in the NPR basis by,
\begin{widetext}
\begin{equation}
    \begin{pmatrix}
    \matrixel{K}{O_1^+}{\bar{K}} \\
    \matrixel{K}{O_2^+}{\bar{K}} \\
    \matrixel{K}{O_3^+}{\bar{K}} \\
    \matrixel{K}{O_4^+}{\bar{K}} \\
    \matrixel{K}{O_5^+}{\bar{K}}
    \end{pmatrix}
    = 
    \begin{pmatrix}
    Z^X_{11} & 0 & 0  & 0 & 0 \\
    0 & 0 & 0 & Z^X_{44} & Z^X_{45} \\
    0 & 0 & 0 & \frac{-Z^X_{44}+Z^X_{54}}{2} & \frac{-Z^X_{45}+Z^X_{55}}{2} \\
    0 & Z^X_{32} & Z^X_{33} & 0 & 0 \\
    0 & -\frac{Z^X_{22}}{2} & -\frac{Z^X_{23}}{2}& 0 & 0 \\
    \end{pmatrix}
    \begin{pmatrix}
    \matrixel{K}{Q_{1,\mathrm{bare}}^+}{\bar{K}} \\
    \matrixel{K}{Q_{2,\mathrm{bare}}^+}{\bar{K}} \\
    \matrixel{K}{Q_{3,\mathrm{bare}}^+}{\bar{K}} \\
    \matrixel{K}{Q_{4,\mathrm{bare}}^+}{\bar{K}} \\
    \matrixel{K}{Q_{5,\mathrm{bare}}^+}{\bar{K}}
    \end{pmatrix}\,.
\end{equation}
\end{widetext}
The NPR-basis operators are written in colour-unmixed form. They are related by
Fierz transformations to the same operators written in colour-mixed form by
$Q_i^{\mathrm{mix}} = F_{ij} Q_j^{\mathrm{unmixed}}$, where
\begin{equation}
F = \begin{pmatrix}
1 & 0 & 0  & 0 & 0 \\
0 & 0 & -2 & 0 & 0 \\
0 & -\frac{1}{2} & 0 & 0 & 0 \\
0 & 0 & 0 & -\frac{1}{2} & \frac{1}{2} \\
0 & 0 & 0 & \frac{3}{2} & \frac{1}{2} \\
\end{pmatrix}.
\end{equation}
The matrix $F$ is used when working
out the matrix $T$ which transforms from the NPR basis to the SUSY basis.

\section{Perturbative scaling}

We compare our non-perturbative scaling results with perturbation theory in
Figure~\ref{fig:npr-vs-pt-scaling} and hence include here our notation and
definitions for the perturbative computations to next-to-leading order. Full
details are provided in a Mathematica~\cite{Mathematica} notebook as
supplementary information.  This is well-covered
ground~\cite{Ciuchini:1997bw,Buras:2000if,Papinutto:2016xpq}, but by explicitly
including both $g^2$ and $g^2\log(g)$ terms at NLO we are able to avoid having
to take a limit to compute the scaling matrix at NLO.

The (matrix) operator renormalisation constants and scaling matrix are
related by
\begin{equation}
  \label{eq:sigmadef}
Z(\mu') = \sigma(\mu',\mu) Z(\mu).
\end{equation}
The anomalous dimension matrix, $\gamma$ is defined by
\begin{equation}
  \mu\frac{dZ}{d\mu}= - \gamma Z.
\end{equation}
Using $\mu\, dg/d\mu = \beta(g)$, we have
\begin{equation}
  \sigma(\mu',\mu) = \text{T}_g \exp\left( \int_{g(\mu)}^{g(\mu')}
  dg'\, \frac{-\gamma(g)}{\beta(g)} \right),
\end{equation}
where $\text{T}_g$ denotes $g$-ordering. We let
\begin{equation}
  \bar a = \frac{\alpha}{4\pi} = \frac{g^2}{16\pi^2}
\end{equation}
and expand
\begin{equation}
\begin{aligned}
  \beta(g) &= -\beta^0 g\bar a - \beta^1 g\bar a^2 + \cdots\\
  \gamma(g) &= \gamma^0 \bar a + \gamma^1 \bar a^2 + \cdots
\end{aligned}
\end{equation}
so that
\begin{equation}
  \int_{g(\mu)}^{g(\mu')}
  dg'\, \frac{-\gamma(g')}{\beta(g')} =
  \int_{\bar a(\mu)}^{\bar a(\mu')}\frac{d\bar a'}{\bar a'}\,
  \frac{(\gamma^0 + \gamma^1 \bar a' + \cdots)}
       {2(\beta^0 + \beta^1 \bar a' + \cdots)}.
\end{equation}
With our conventions, $\beta^0 = 11-2N_f/3$ for $N_f$ flavours. We work with the
operators $Q_i^+$ defined in Eq.~(\ref{eq:opbasis_unmixed_even}). They are
related to the positive parity parts of the basis $Q_\text{BMU} =
\{Q_1^\text{VLL}, Q_1^\text{LR}, Q_2^\text{LR}, Q_1^\text{SLL},
Q_2^\text{SLL}\}$ used in Ref.~\cite{Buras:2000if} by $Q^+ = R\cdot
Q_\text{BMU}^+$ with $R=\text{diag}(4,4,4,4,1)$. This changes the off-diagonal
elements of the anomalous dimension matrices in the bottom-right $2\times2$
block. In Ref.~\cite{Papinutto:2016xpq}, the anomalous dimension is defined with
the opposite sign and expanded as $\gamma^\text{PPP} =
-g^2(\gamma^{0,\text{PPP}} + \gamma^{1,\text{PPP}}g^2 + \cdots)$, which means
that $\gamma^{0,\text{PPP}} = \gamma^0/(16\pi^2)$.

\subsection{LO scaling}

If we diagonalise $\gamma^0$,
\begin{equation}
  \label{eq:Vdef}
  V^{-1} \gamma^0 V = \gamma^0_D,
\end{equation}
where $\gamma^0_D$ is the diagonal matrix of eigenvalues, then the
scaling matrix at leading order is
\begin{equation}
  \sigma^0(\mu',\mu) = V
  \left(\frac{a(\mu')}{a(\mu)}\right)^{\gamma^0_D/2\beta^0} V^{-1}.
\end{equation}
To match notation from Ref.~\cite{Buras:2000if}, we write
\begin{equation}
  \label{eq:adef}
  (V^{-1}\gamma^0V)_{ij} = 2\beta^0 a_i\,\delta_{ij}
  \qquad\text{where}\qquad
  a = \frac{\gamma^0_D}{2\beta^0}.
\end{equation}
Under $\text{SU}(3)_L\times \text{SU}(3)_R$ flavour, $Q_1^+$ is $(27,1)$ and
renormalises multiplicatively, $Q_{2,3}^+$ are $(8,8)$ and mix, and $Q_{4,5}^+$
are $(6,\bar6)$ and also mix. When we diagonalise $\gamma^0$ we permute the
eigenvalues and eigenvectors to preserve the block-diagonal structure of the
$5\times5$ anomalous dimension matrix.  Our choice leads to
\begin{equation}
  \gamma^0_D = \text{diag}\left(4,2,-16,\frac23\big(1+\sqrt{241}\big),
  \frac23\big(1-\sqrt{241}\big)\right).
\end{equation}

\subsection{NLO scaling}

Beyond leading order we write
\begin{equation}
  \label{eq:scalingmatrix}
  \sigma(\mu',\mu) \equiv K(\mu') \sigma^0(\mu',\mu) K^{-1}(\mu).
\end{equation}
The equation satisfied by $K$ is
\begin{equation}
  \pdv Kg - \frac1g \left[\frac{\gamma^0}{\beta^0},K\right] =
  -\left(\frac{\gamma(g)}{\beta(g)}+
  \frac{\gamma^0}{\beta^0 g}\right)K,
\end{equation}
where $K=K(\mu)$ and $g=g(\mu)$.
Expanding to NLO in the coupling and writing
\begin{equation}
  \label{eq:expandK}
  K = 1 + \frac{g^2}{16\pi^2}\,J + \frac{g^2}{16\pi^2}\,\log (g) \,L,
\end{equation}
we find that $J$ and $L$ satisfy
\begin{equation}
  \begin{aligned}
    L &= \left[\frac{\gamma^0}{2\beta^0},L\right]\\
    J+\frac12 L &= \left[\frac{\gamma^0}{2\beta^0},J\right] +
    \frac1{2\beta^0}\,\gamma^1 -
    \frac{\beta^1}{2(\beta^0)^2}\,\gamma^0.
    \label{eq:LJeqn}
  \end{aligned}
\end{equation}
Using the matrix $V$ from equation~(\ref{eq:Vdef}) which diagonalises
$\gamma^0$, we define
\begin{equation}
\begin{aligned}
  G &= V^{-1} \gamma^1 V\\
  S &= V^{-1} J V\\
  T &= V^{-1} L V
\end{aligned}
\end{equation}
and recalling the $a_i$ from equation~(\ref{eq:adef}), we find
\begin{equation}
  T_{ij} = (a_i-a_j) T_{ij}.
\end{equation}
This shows that $T_{ij}$ vanishes, \emph{except} when
$a_i-a_j=1$. We also have
\begin{equation}
  S_{ij} + \frac12 T_{ij} - (a_i-a_j) S_{ij} =
  \frac1{2\beta^0} G_{ij} - \frac{\beta^1}{\beta^0} a_i \delta_{ij}.
\end{equation}
If $a_i-a_j=1$ then $S_{ij}$ drops out of this equation and its value
is arbitrary (we choose to make it zero), but then $T_{ij}$ is nonzero
from the previous equation and its value is determined by this
equation. If $a_i-a_j \neq 1$, then $T_{ij}=0$ and the equation
determines $S_{ij}$. Once we know $S$ and $T$ we can find $J$ and $L$
and hence determine the $K$ matrix and the NLO expression for the
scaling matrix $\sigma$.

The leading order anomalous dimension matrix for the $4$-quark operators has
$a_2-a_3=1$ when the number of flavours is $N_f=3$. This means that
$T_{23}\neq0$ and there is a $g^2\log g$ term in $K_{23}$. Including
the $g^2\log g$ term allows us to avoid expanding the solution for
the scaling matrix around $N_f=3$. We checked that either method gives
the same result for $\sigma$. For our check we shifted $a_2\to
a_2+\delta$, computed the scaling matrix $\sigma$ and took the limit
$\delta\to0$. As an additional check we also did the limiting
procedure by shifting $\beta^0 \to 11-2(3+\delta)/3$ when constructing
$a=\gamma^0_D/2\beta^0$ in equation~(\ref{eq:adef}) and again found
the same result. We also learn and checked that we can add an
arbitrary shift to $S_{23}$ or $J_{23}$ without changing the result
for $\sigma$.

From equations~\eqref{eq:LJeqn} we know that the only nonzero element of $T$ in
any scheme is $T_{23} = G_{23}/\beta^0$ and hence in any scheme,
\begin{equation}
  L_{23} = \frac{(V^{-1}\gamma^1 V)_{23}}{\beta^0 (V^{-1})_{22} V_{33}}
   = -\frac{40}{27} = -1.48148.
\end{equation}
We checked this for $\overline{\text{MS}}$, RI-MOM,
RI-SMOM$^{(\gamma_\mu,\gamma_\mu)}$ and RI-SMOM$^{(\fslash q,\fslash q)}$.

For $\gamma^1$ in $\overline{\text{MS}}$ we used Ref.~\cite{Buras:2000if} and
for RI-MOM we used Ref.~\cite{Papinutto:2016xpq} (that paper does not give the
$11$ element; for this we used the value in Ref.~\cite{Aoki:2010pe}). For the
RI-SMOM schemes we used $\gamma^{\overline{\text{MS}},1}$ together with the
conversion factors $\Delta r$ from RI-SMOM to $\overline{\text{MS}}$ in
Ref.~\cite{Boyle:2017skn} and applied Eq.~\eqref{eq:schemeconversion} below.

To evaluate the perturbative scaling numerically, we used the 5-loop
expression for the running strong coupling to evolve its value from
the $Z$-mass to the charm mass, with quark-flavour thresholds at $\bar
m_b(\bar m_b)$ and $\bar m_c(\bar m_c)$, the $\overline{\text{MS}}$
bottom and charm masses evaluated at their own
scales~\cite{Baikov:2016tgj,Herzog:2017ohr,Luthe:2017ttc,Chetyrkin:2005ia,Schroder:2005hy,Liu:2015fxa,Chetyrkin:2000yt,Schmidt:2012az,Herren:2017osy,Baikov:2014qja,Luthe:2016xec,Baikov:2017ujl}.
We then evaluated $\alpha_s(\mu)$ for $2\,\text{GeV} \leq \mu \leq
3\,\text{GeV}$ for three flavours, corresponding to our $2+1$ flavour
simulations. Our inputs were~\cite{ParticleDataGroup:2022pth}
\begin{equation}
\begin{aligned}
\alpha_s(M_Z) &= 0.1180, \\
M_Z &= 91.1876\,\text{GeV}, \\
\bar m_b(\bar m_b) &= 4.18\,\text{GeV}, \\
\bar m_c(\bar m_c) & = 1.28\,\text{GeV}, 
\end{aligned}
\end{equation}
and we determine
\begin{equation}
\alpha_s(2\,\text{GeV}) = 0.293347, \quad
\alpha_s(3\,\text{GeV}) = 0.243580.
\end{equation}

Changing from operators $Q^+(\mu)$ renormalised in some scheme at scale
$\mu$ to RGI operators $\hat Q^+$, is done by~\cite{Papinutto:2016xpq}:
\begin{equation}
  \label{eq:rgi-ops}
  \hat Q^+ = [\alpha(\mu)]^{-\gamma^0/2\beta^0} K^{-1}(\mu)Q^+(\mu).
\end{equation}
The $11$ element of this relation converts the kaon bag parameter $B_K(\mu)$ to the RGI $\hat B_K$ (see for example the discussion in the 2019 FLAG review~\cite{FlavourLatticeAveragingGroup:2019iem}). To compute RGI $\hat B_i$'s for the BSM operators,  we would in addition have to take into account the quark-mass combination $m_s(\mu)+m_d(\mu)$ appearing in the $B_i(\mu)$'s definition.

\subsection{Logarithmic term in NLO scaling expression}

Here we will let $g'=g(\mu')$, $\alpha'=\alpha(\mu')$ and $g=g(\mu)$,
$\alpha=\alpha(\mu)$. From equations~(\ref{eq:scalingmatrix})
and~(\ref{eq:expandK}), the $\log(g)$ terms in $\sigma(\mu',\mu)$ at NLO are
  \begin{multline}
  \frac1{4\pi}\left(
  \alpha' \log g'\,L \sigma^0(\mu',\mu) -
  \alpha \log g\, \sigma^0(\mu',\mu) L \right)\\
  = \frac1{4\pi}V\left(
  \alpha' \log g'\,T \Big(\frac{\alpha'}\alpha\Big)^a -
  \alpha \log g\,\Big(\frac{\alpha'}\alpha\Big)^a T\right)V^{-1}.
  \end{multline}
From above, the only nonzero element of $T$ is $T_{23}$, so the term
in parentheses becomes
\begin{multline}
  T_{23} \left(
  \alpha' \log g' \Big(\frac{\alpha'}\alpha\Big)^{a_3} -
  \alpha \log g \Big(\frac{\alpha'}\alpha\Big)^{a_2}
  \right) \\
  =
  T_{23}\Big(\frac{\alpha'}\alpha\Big)^{1/9}\alpha
  \log\Big(\frac{g'}g\Big)
\end{multline}
where we have used $a_2=1/9$ and $a_3=-8/9$. Switching back to $L$ and
expressing everything in terms of $\alpha'$ and $\alpha$ we find
\begin{equation}
  \text{NLO logs in $\sigma(\mu',\mu)$} =
  -\frac5{(27\pi)} \Big(\frac{\alpha'}\alpha\Big)^{1/9}\alpha
  \log\Big(\frac{\alpha'}\alpha\Big).
\end{equation}
We checked explicitly for $\overline{\text{MS}}$, RI-MOM,
RI-SMOM$^{(\gamma_\mu,\gamma_\mu)}$ and RI-SMOM$^{(\fslash q,\fslash q)}$ that the same
term arises when we calculate the NLO scaling by the $\delta$-shift
and limit procedure.

\subsection{Scheme conversion}

Following Ref.~\cite{Boyle:2017skn} we define the matrix $R$ to convert from
renormalisation scheme $A$ to scheme $B$ by \def\rba{R^{B\leftarrow A}}
\def\drba{\Delta r^{B\leftarrow A}}
\begin{equation}
  Z^B = \rba Z^A = (1 - \bar a \drba) Z^A\,,
\end{equation}
where $\bar a=g^2/16\pi^2$. From $\beta(g)\, dZ^B/dg = - \gamma^B(g) Z^B$ we find
\begin{equation}
  \gamma^A = (\rba)^{-1} \gamma^B \rba +
  \beta(\rba)^{-1} \frac{d\rba}{dg}.
\end{equation}
Expanding to $O(\bar a^2)$ and noting that $\gamma^0$ is universal, we can
relate the NLO anomalous dimensions in schemes $A$ and $B$ using the
$1$-loop conversion factor $\drba$,
\begin{equation}
  \label{eq:schemeconversion}
  \gamma^{A,1} = \gamma^{B,1} - [\gamma^0,\drba]+2\beta^0\drba.
\end{equation}
In particular we can determine the NLO anomalous dimensions where $A$ is
RI-SMOM$^{(\gamma_\mu,\gamma_\mu)}$ or RI-SMOM$^{(\fslash q,\fslash q)}$
respectively, from the $\overline{\text{MS}}$ NLO anomalous dimension and
$\Delta r^{\overline{\text{MS}}\leftarrow A}$
\begin{equation}
  \gamma^{A,1} = \gamma^{\overline{\text{MS}},1} -
    [\gamma^0,\Delta r^{\overline{\text{MS}}\leftarrow A}] +
   2\beta^0\Delta r^{\overline{\text{MS}}\leftarrow A}.
\end{equation}
By combining the scheme conversion equation~(\ref{eq:schemeconversion}) with
equations~\eqref{eq:LJeqn} for $J$ and $L$, we can show that
\begin{equation}
  \label{eq:DeltaJ}
  J^A - J^{\overline{\text{MS}}} =
  \Delta r^{\overline{\text{MS}}\leftarrow A} \qquad
  \text{except for the $23$ element}
\end{equation}
and that $L^A = L^{\overline{\text{MS}}}$. We checked
equation~(\ref{eq:DeltaJ}) when $A$ is RI-MOM, RI-SMOM$^{(\gamma_\mu,\gamma_\mu)}$
or RI-SMOM$^{(\fslash q,\fslash q)}$.

\section{Further detailed numerical fit results}
\label{sec:chiCLresults}
  
Our central value for the $\mathcal{B}_i$ and $R_i$ in $\overline{\mathrm{MS}}$
at $\mu=3\,\mathrm{GeV}$ is taken as the mean of the conversion from
RI-SMOM$^{(\gamma_\mu,\gamma_\mu)}$ and RI-SMOM$^{(\slashed q,\slashed q)}$ to
$\overline{\mathrm{MS}}$. For completeness we also seperately list these
conversions with their full uncertainty budget in Table~\ref{tab:MSconv}.
\begin{table*}
    \centering
    \caption{Central values and combined systematic errors for ratio and bag
      parameters at $\mu=3\,\mathrm{GeV}$ in $\overline{\text{MS}}$ after
      converting from the two RI-SMOM schemes --- $(\gamma_\mu, \gamma_\mu)$ and
      $(\slashed{q}, \slashed{q})$, in the SUSY basis. We list the errors
      arising from statistics, chiral extrapolation, residual chiral symmetry
      breaking, and discretisation and combine it into total
      uncertainties. \label{tab:MSconv}} \begin{tabular}{c|c|cccc|ccccc}
\hline
\hline
scheme & & $R_2$ & $R_3$ & $R_4$ & $R_5$ & $\mathcal{B}_1$ & $\mathcal{B}_2$ & $\mathcal{B}_3$ & $\mathcal{B}_4$ & $\mathcal{B}_5$\\
\hline
\multirow{6}{*}{$\overline{\text{MS}}\leftarrow\text{RI-SMOM}^{(\gamma_\mu, \gamma_\mu)}$} & central & $-18.73$ & $5.781$ & $41.45$ & $10.80$ & $0.5185$ & $0.4759$ & $0.728$ & $0.8862$ & $0.6977$ \\
\cline{2-11}
 & stat & 0.60\% & 0.69\% & 0.72\% & 0.43\% & 0.28\% & 0.24\% & 0.72\% & 0.21\% & 1.02\% \\
 & chiral & 0.21\% & 0.42\% & 0.61\% & 0.46\% & 0.20\% & 0.17\% & 0.29\% & 0.17\% & 0.25\% \\
 & rcsb & 0.10\% & 0.15\% & 0.09\% & 0.03\% & 0.04\% & 0.03\% & 0.06\% & 0.01\% & 0.00\% \\
 & discr & 0.16\% & 0.53\% & 0.49\% & 1.23\% & 0.01\% & 0.44\% & 1.61\% & 0.16\% & 0.38\% \\
\cline{2-11}
 & total & 0.66\% & 0.98\% & 1.07\% & 1.38\% & 0.35\% & 0.53\% & 1.79\% & 0.32\% & 1.12\% \\
\hline
\multirow{6}{*}{$\overline{\text{MS}}\leftarrow\text{RI-SMOM}^{(\slashed{q}, \slashed{q})}$} & central & $-19.07$ & $6.059$ & $42.43$ & $10.49$ & $0.5295$ & $0.4829$ & $0.764$ & $0.9070$ & $0.6788$ \\
\cline{2-11}
 & stat & 0.68\% & 0.92\% & 0.81\% & 0.83\% & 0.29\% & 0.43\% & 1.24\% & 0.36\% & 2.21\% \\
 & chiral & 0.48\% & 0.78\% & 1.25\% & 1.26\% & 0.24\% & 0.27\% & 0.44\% & 0.29\% & 0.51\% \\
 & rcsb & 0.29\% & 0.21\% & 0.23\% & 0.13\% & 0.08\% & 0.19\% & 0.29\% & 0.03\% & 0.01\% \\
 & discr & 0.34\% & 0.65\% & 0.20\% & 2.30\% & 0.10\% & 0.64\% & 1.92\% & 0.19\% & 0.10\% \\
\cline{2-11}
 & total & 0.95\% & 1.39\% & 1.52\% & 2.75\% & 0.40\% & 0.83\% & 2.34\% & 0.50\% & 2.27\% \\
\hline
\hline
\end{tabular}
\end{table*}

\FloatBarrier
\bibliography{paper.bib}

\begin{thebibliography}{108}%
\makeatletter
\providecommand \@ifxundefined [1]{%
 \@ifx{#1\undefined}
}%
\providecommand \@ifnum [1]{%
 \ifnum #1\expandafter \@firstoftwo
 \else \expandafter \@secondoftwo
 \fi
}%
\providecommand \@ifx [1]{%
 \ifx #1\expandafter \@firstoftwo
 \else \expandafter \@secondoftwo
 \fi
}%
\providecommand \natexlab [1]{#1}%
\providecommand \enquote  [1]{``#1''}%
\providecommand \bibnamefont  [1]{#1}%
\providecommand \bibfnamefont [1]{#1}%
\providecommand \citenamefont [1]{#1}%
\providecommand \href@noop [0]{\@secondoftwo}%
\providecommand \href [0]{\begingroup \@sanitize@url \@href}%
\providecommand \@href[1]{\@@startlink{#1}\@@href}%
\providecommand \@@href[1]{\endgroup#1\@@endlink}%
\providecommand \@sanitize@url [0]{\catcode `\\12\catcode `\$12\catcode
  `\&12\catcode `\#12\catcode `\^12\catcode `\_12\catcode `\%12\relax}%
\providecommand \@@startlink[1]{}%
\providecommand \@@endlink[0]{}%
\providecommand \url  [0]{\begingroup\@sanitize@url \@url }%
\providecommand \@url [1]{\endgroup\@href {#1}{\urlprefix }}%
\providecommand \urlprefix  [0]{URL }%
\providecommand \Eprint [0]{\href }%
\providecommand \doibase [0]{http://dx.doi.org/}%
\providecommand \selectlanguage [0]{\@gobble}%
\providecommand \bibinfo  [0]{\@secondoftwo}%
\providecommand \bibfield  [0]{\@secondoftwo}%
\providecommand \translation [1]{[#1]}%
\providecommand \BibitemOpen [0]{}%
\providecommand \bibitemStop [0]{}%
\providecommand \bibitemNoStop [0]{.\EOS\space}%
\providecommand \EOS [0]{\spacefactor3000\relax}%
\providecommand \BibitemShut  [1]{\csname bibitem#1\endcsname}%
\let\auto@bib@innerbib\@empty
\bibitem [{\citenamefont {Christenson}\ \emph {et~al.}(1964)\citenamefont
  {Christenson}, \citenamefont {Cronin}, \citenamefont {Fitch},\ and\
  \citenamefont {Turlay}}]{PhysRevLett.13.138}%
  \BibitemOpen
  \bibfield  {author} {\bibinfo {author} {\bibfnamefont {J.~H.}\ \bibnamefont
  {Christenson}}, \bibinfo {author} {\bibfnamefont {J.~W.}\ \bibnamefont
  {Cronin}}, \bibinfo {author} {\bibfnamefont {V.~L.}\ \bibnamefont {Fitch}}, \
  and\ \bibinfo {author} {\bibfnamefont {R.}~\bibnamefont {Turlay}},\
  }\bibfield  {title} {\enquote {\bibinfo {title} {Evidence for the
  $2\ensuremath{\pi}$ decay of the $k_{2}^{0}$ meson},}\ }\href {\doibase
  10.1103/PhysRevLett.13.138} {\bibfield  {journal} {\bibinfo  {journal} {Phys.
  Rev. Lett.}\ }\textbf {\bibinfo {volume} {13}},\ \bibinfo {pages} {138--140}
  (\bibinfo {year} {1964})}\BibitemShut {NoStop}%
\bibitem [{\citenamefont {Buras}(2020)}]{Buras:2020xsm}%
  \BibitemOpen
  \bibfield  {author} {\bibinfo {author} {\bibfnamefont {Andrzej}\ \bibnamefont
  {Buras}},\ }\href {\doibase 10.1017/9781139524100} {\emph {\bibinfo {title}
  {{Gauge Theory of Weak Decays}}}}\ (\bibinfo  {publisher} {Cambridge
  University Press},\ \bibinfo {year} {2020})\BibitemShut {NoStop}%
\bibitem [{\citenamefont {Bigi}\ and\ \citenamefont
  {Sanda}(2009)}]{Bigi:2000yz}%
  \BibitemOpen
  \bibfield  {author} {\bibinfo {author} {\bibfnamefont {Ikaros~I.}\
  \bibnamefont {Bigi}}\ and\ \bibinfo {author} {\bibfnamefont {A.~I.}\
  \bibnamefont {Sanda}},\ }\href {\doibase 10.1017/CBO9780511581014} {\emph
  {\bibinfo {title} {{CP violation}}}},\ Vol.~\bibinfo {volume} {9}\ (\bibinfo
  {publisher} {Cambridge University Press},\ \bibinfo {year}
  {2009})\BibitemShut {NoStop}%
\bibitem [{\citenamefont {Branco}\ \emph {et~al.}(1999)\citenamefont {Branco},
  \citenamefont {Lavoura},\ and\ \citenamefont {Silva}}]{Branco:1999fs}%
  \BibitemOpen
  \bibfield  {author} {\bibinfo {author} {\bibfnamefont {Gustavo~C.}\
  \bibnamefont {Branco}}, \bibinfo {author} {\bibfnamefont {Luis}\ \bibnamefont
  {Lavoura}}, \ and\ \bibinfo {author} {\bibfnamefont {Joao~P.}\ \bibnamefont
  {Silva}},\ }\href@noop {} {\emph {\bibinfo {title} {{CP Violation}}}},\ Vol.\
  \bibinfo {volume} {103}\ (\bibinfo  {publisher} {Oxford University Press},\
  \bibinfo {year} {1999})\BibitemShut {NoStop}%
\bibitem [{\citenamefont {Inami}\ and\ \citenamefont
  {Lim}(1981)}]{Inami:1980fz}%
  \BibitemOpen
  \bibfield  {author} {\bibinfo {author} {\bibfnamefont {T.}~\bibnamefont
  {Inami}}\ and\ \bibinfo {author} {\bibfnamefont {C.~S.}\ \bibnamefont
  {Lim}},\ }\bibfield  {title} {\enquote {\bibinfo {title} {{Effects of
  Superheavy Quarks and Leptons in Low-Energy Weak Processes k(L)
  ---\ensuremath{>} mu anti-mu, K+ ---\ensuremath{>} pi+ Neutrino anti-neutrino
  and K0 \ensuremath{<}---\ensuremath{>} anti-K0}},}\ }\href {\doibase
  10.1143/PTP.65.297} {\bibfield  {journal} {\bibinfo  {journal} {Prog. Theor.
  Phys.}\ }\textbf {\bibinfo {volume} {65}},\ \bibinfo {pages} {297} (\bibinfo
  {year} {1981})},\ \bibinfo {note} {[Erratum: Prog.Theor.Phys. 65, 1772
  (1981)]}\BibitemShut {NoStop}%
\bibitem [{\citenamefont {Aoki}\ \emph
  {et~al.}(2011{\natexlab{a}})\citenamefont {Aoki}, \citenamefont {Arthur},
  \citenamefont {Blum}, \citenamefont {Boyle}, \citenamefont {Brommel} \emph
  {et~al.}}]{Aoki:2010pe}%
  \BibitemOpen
  \bibfield  {author} {\bibinfo {author} {\bibfnamefont {Y.}~\bibnamefont
  {Aoki}}, \bibinfo {author} {\bibfnamefont {R.}~\bibnamefont {Arthur}},
  \bibinfo {author} {\bibfnamefont {T.}~\bibnamefont {Blum}}, \bibinfo {author}
  {\bibfnamefont {P.A.}\ \bibnamefont {Boyle}}, \bibinfo {author}
  {\bibfnamefont {D.}~\bibnamefont {Brommel}},  \emph {et~al.},\ }\bibfield
  {title} {\enquote {\bibinfo {title} {{Continuum Limit of $B_K$ from 2+1
  Flavor Domain Wall QCD}},}\ }\href {\doibase 10.1103/PhysRevD.84.014503}
  {\bibfield  {journal} {\bibinfo  {journal} {Phys.Rev.}\ }\textbf {\bibinfo
  {volume} {D84}},\ \bibinfo {pages} {014503} (\bibinfo {year}
  {2011}{\natexlab{a}})},\ \Eprint {http://arxiv.org/abs/1012.4178}
  {arXiv:1012.4178 [hep-lat]} \BibitemShut {NoStop}%
\bibitem [{\citenamefont {Blum}\ \emph {et~al.}(2016)\citenamefont {Blum} \emph
  {et~al.}}]{RBC:2014ntl}%
  \BibitemOpen
  \bibfield  {author} {\bibinfo {author} {\bibfnamefont {T.}~\bibnamefont
  {Blum}} \emph {et~al.} (\bibinfo {collaboration} {RBC, UKQCD}),\ }\bibfield
  {title} {\enquote {\bibinfo {title} {{Domain wall QCD with physical quark
  masses}},}\ }\href {\doibase 10.1103/PhysRevD.93.074505} {\bibfield
  {journal} {\bibinfo  {journal} {Phys. Rev. D}\ }\textbf {\bibinfo {volume}
  {93}},\ \bibinfo {pages} {074505} (\bibinfo {year} {2016})},\ \Eprint
  {http://arxiv.org/abs/1411.7017} {arXiv:1411.7017 [hep-lat]} \BibitemShut
  {NoStop}%
\bibitem [{\citenamefont {Bae}\ \emph {et~al.}(2013{\natexlab{a}})\citenamefont
  {Bae} \emph {et~al.}}]{Bae:2013tca}%
  \BibitemOpen
  \bibfield  {author} {\bibinfo {author} {\bibfnamefont {Taegil}\ \bibnamefont
  {Bae}} \emph {et~al.} (\bibinfo {collaboration} {SWME}),\ }\bibfield  {title}
  {\enquote {\bibinfo {title} {{Neutral kaon mixing from new physics: matrix
  elements in $N_f=2+1$ lattice QCD}},}\ }\href {\doibase
  10.1103/PhysRevD.88.071503} {\bibfield  {journal} {\bibinfo  {journal} {Phys.
  Rev.}\ }\textbf {\bibinfo {volume} {D88}},\ \bibinfo {pages} {071503}
  (\bibinfo {year} {2013}{\natexlab{a}})},\ \Eprint
  {http://arxiv.org/abs/1309.2040} {arXiv:1309.2040 [hep-lat]} \BibitemShut
  {NoStop}%
\bibitem [{\citenamefont {Choi}\ \emph {et~al.}(2016)\citenamefont {Choi} \emph
  {et~al.}}]{SWME:2015oos}%
  \BibitemOpen
  \bibfield  {author} {\bibinfo {author} {\bibfnamefont {Benjamin~J.}\
  \bibnamefont {Choi}} \emph {et~al.} (\bibinfo {collaboration} {SWME}),\
  }\bibfield  {title} {\enquote {\bibinfo {title} {{Kaon BSM B-parameters using
  improved staggered fermions from $N_f=2+1$ unquenched QCD}},}\ }\href
  {\doibase 10.1103/PhysRevD.93.014511} {\bibfield  {journal} {\bibinfo
  {journal} {Phys. Rev. D}\ }\textbf {\bibinfo {volume} {93}},\ \bibinfo
  {pages} {014511} (\bibinfo {year} {2016})},\ \Eprint
  {http://arxiv.org/abs/1509.00592} {arXiv:1509.00592 [hep-lat]} \BibitemShut
  {NoStop}%
\bibitem [{\citenamefont {Bae}\ \emph {et~al.}(2014)\citenamefont {Bae} \emph
  {et~al.}}]{Bae:2014sja}%
  \BibitemOpen
  \bibfield  {author} {\bibinfo {author} {\bibfnamefont {Taegil}\ \bibnamefont
  {Bae}} \emph {et~al.} (\bibinfo {collaboration} {SWME}),\ }\bibfield  {title}
  {\enquote {\bibinfo {title} {{Improved determination of BK with staggered
  quarks}},}\ }\href {\doibase 10.1103/PhysRevD.89.074504} {\bibfield
  {journal} {\bibinfo  {journal} {Phys. Rev.}\ }\textbf {\bibinfo {volume}
  {D89}},\ \bibinfo {pages} {074504} (\bibinfo {year} {2014})},\ \Eprint
  {http://arxiv.org/abs/1402.0048} {arXiv:1402.0048 [hep-lat]} \BibitemShut
  {NoStop}%
\bibitem [{\citenamefont {Carrasco}\ \emph {et~al.}(2015)\citenamefont
  {Carrasco}, \citenamefont {Dimopoulos}, \citenamefont {Frezzotti},
  \citenamefont {Lubicz}, \citenamefont {Rossi}, \citenamefont {Simula},\ and\
  \citenamefont {Tarantino}}]{Carrasco:2015pra}%
  \BibitemOpen
  \bibfield  {author} {\bibinfo {author} {\bibfnamefont {N.}~\bibnamefont
  {Carrasco}}, \bibinfo {author} {\bibfnamefont {P.}~\bibnamefont
  {Dimopoulos}}, \bibinfo {author} {\bibfnamefont {R.}~\bibnamefont
  {Frezzotti}}, \bibinfo {author} {\bibfnamefont {V.}~\bibnamefont {Lubicz}},
  \bibinfo {author} {\bibfnamefont {G.~C}\ \bibnamefont {Rossi}}, \bibinfo
  {author} {\bibfnamefont {S.}~\bibnamefont {Simula}}, \ and\ \bibinfo {author}
  {\bibfnamefont {C.}~\bibnamefont {Tarantino}} (\bibinfo {collaboration}
  {ETM}),\ }\bibfield  {title} {\enquote {\bibinfo {title}
  {{\ensuremath{\Delta}S=2 and \ensuremath{\Delta}C=2 bag parameters in the
  standard model and beyond from N$_f$=2+1+1 twisted-mass lattice QCD}},}\
  }\href {\doibase 10.1103/PhysRevD.92.034516} {\bibfield  {journal} {\bibinfo
  {journal} {Phys. Rev. D}\ }\textbf {\bibinfo {volume} {92}},\ \bibinfo
  {pages} {034516} (\bibinfo {year} {2015})},\ \Eprint
  {http://arxiv.org/abs/1505.06639} {arXiv:1505.06639 [hep-lat]} \BibitemShut
  {NoStop}%
\bibitem [{\citenamefont {Bertone}\ \emph
  {et~al.}(2013{\natexlab{a}})\citenamefont {Bertone} \emph
  {et~al.}}]{Bertone:2012cu}%
  \BibitemOpen
  \bibfield  {author} {\bibinfo {author} {\bibfnamefont {V.}~\bibnamefont
  {Bertone}} \emph {et~al.} (\bibinfo {collaboration} {ETM}),\ }\bibfield
  {title} {\enquote {\bibinfo {title} {{Kaon Mixing Beyond the SM from
  N$_{f}$=2 tmQCD and model independent constraints from the UTA}},}\ }\href
  {\doibase 10.1007/JHEP07(2013)143, 10.1007/JHEP03(2013)089} {\bibfield
  {journal} {\bibinfo  {journal} {JHEP}\ }\textbf {\bibinfo {volume} {03}},\
  \bibinfo {pages} {089} (\bibinfo {year} {2013}{\natexlab{a}})},\ \bibinfo
  {note} {[Erratum: JHEP07,143(2013)]},\ \Eprint
  {http://arxiv.org/abs/1207.1287} {arXiv:1207.1287 [hep-lat]} \BibitemShut
  {NoStop}%
\bibitem [{\citenamefont {Aoki}\ \emph {et~al.}(2022)\citenamefont {Aoki} \emph
  {et~al.}}]{FlavourLatticeAveragingGroupFLAG:2021npn}%
  \BibitemOpen
  \bibfield  {author} {\bibinfo {author} {\bibfnamefont {Y.}~\bibnamefont
  {Aoki}} \emph {et~al.} (\bibinfo {collaboration} {Flavour Lattice Averaging
  Group (FLAG)}),\ }\bibfield  {title} {\enquote {\bibinfo {title} {{FLAG
  Review 2021}},}\ }\href {\doibase 10.1140/epjc/s10052-022-10536-1} {\bibfield
   {journal} {\bibinfo  {journal} {Eur. Phys. J. C}\ }\textbf {\bibinfo
  {volume} {82}},\ \bibinfo {pages} {869} (\bibinfo {year} {2022})},\ \Eprint
  {http://arxiv.org/abs/2111.09849} {arXiv:2111.09849 [hep-lat]} \BibitemShut
  {NoStop}%
\bibitem [{\citenamefont {Buchalla}\ \emph {et~al.}(1996)\citenamefont
  {Buchalla}, \citenamefont {Buras},\ and\ \citenamefont
  {Lautenbacher}}]{Buchalla:1995vs}%
  \BibitemOpen
  \bibfield  {author} {\bibinfo {author} {\bibfnamefont {Gerhard}\ \bibnamefont
  {Buchalla}}, \bibinfo {author} {\bibfnamefont {Andrzej~J.}\ \bibnamefont
  {Buras}}, \ and\ \bibinfo {author} {\bibfnamefont {Markus~E.}\ \bibnamefont
  {Lautenbacher}},\ }\bibfield  {title} {\enquote {\bibinfo {title} {Weak
  decays beyond leading logarithms},}\ }\href {\doibase
  10.1103/RevModPhys.68.1125} {\bibfield  {journal} {\bibinfo  {journal} {Rev.
  Mod. Phys.}\ }\textbf {\bibinfo {volume} {68}},\ \bibinfo {pages}
  {1125--1144} (\bibinfo {year} {1996})},\ \Eprint
  {http://arxiv.org/abs/hep-ph/9512380} {arXiv:hep-ph/9512380} \BibitemShut
  {NoStop}%
\bibitem [{\citenamefont {Christ}(2011)}]{Christ:2012np}%
  \BibitemOpen
  \bibfield  {author} {\bibinfo {author} {\bibfnamefont {Norman~H.}\
  \bibnamefont {Christ}},\ }\bibfield  {title} {\enquote {\bibinfo {title}
  {{Computing the long-distance contribution to the kaon mixing parameter
  {$\epsilon_K$}}},}\ }\bibfield  {booktitle} {\emph {\bibinfo {booktitle}
  {{Proceedings, 29th International Symposium on Lattice field theory (Lattice
  2011): Squaw Valley, Lake Tahoe, USA, July 10-16, 2011}}},\ }\href@noop {}
  {\bibfield  {journal} {\bibinfo  {journal} {PoS}\ }\textbf {\bibinfo {volume}
  {LATTICE2011}},\ \bibinfo {pages} {277} (\bibinfo {year} {2011})},\ \Eprint
  {http://arxiv.org/abs/1201.2065} {arXiv:1201.2065 [hep-lat]} \BibitemShut
  {NoStop}%
\bibitem [{\citenamefont {Christ}\ and\ \citenamefont
  {Bai}(2016)}]{Christ:2015phf}%
  \BibitemOpen
  \bibfield  {author} {\bibinfo {author} {\bibfnamefont {Norman~H.}\
  \bibnamefont {Christ}}\ and\ \bibinfo {author} {\bibfnamefont {Ziyuan}\
  \bibnamefont {Bai}},\ }\bibfield  {title} {\enquote {\bibinfo {title}
  {{Computing the long-distance contributions to $\varepsilon_K$}},}\
  }\bibfield  {booktitle} {\emph {\bibinfo {booktitle} {{Proceedings, 33rd
  International Symposium on Lattice Field Theory (Lattice 2015): Kobe, Japan,
  July 14-18, 2015}}},\ }\href@noop {} {\bibfield  {journal} {\bibinfo
  {journal} {PoS}\ }\textbf {\bibinfo {volume} {LATTICE2015}},\ \bibinfo
  {pages} {342} (\bibinfo {year} {2016})}\BibitemShut {NoStop}%
\bibitem [{\citenamefont {Bai}(2017)}]{Bai:2016gzv}%
  \BibitemOpen
  \bibfield  {author} {\bibinfo {author} {\bibfnamefont {Ziyuan}\ \bibnamefont
  {Bai}},\ }\bibfield  {title} {\enquote {\bibinfo {title} {{Long distance part
  of $\epsilon_K$ from lattice QCD}},}\ }\bibfield  {booktitle} {\emph
  {\bibinfo {booktitle} {{Proceedings, 34th International Symposium on Lattice
  Field Theory (Lattice 2016): Southampton, UK, July 24-30, 2016}}},\
  }\href@noop {} {\bibfield  {journal} {\bibinfo  {journal} {PoS}\ }\textbf
  {\bibinfo {volume} {LATTICE2016}},\ \bibinfo {pages} {309} (\bibinfo {year}
  {2017})},\ \Eprint {http://arxiv.org/abs/1611.06601} {arXiv:1611.06601
  [hep-lat]} \BibitemShut {NoStop}%
\bibitem [{\citenamefont {Wang}(2022)}]{Wang:2022lfq}%
  \BibitemOpen
  \bibfield  {author} {\bibinfo {author} {\bibfnamefont {Bigeng}\ \bibnamefont
  {Wang}},\ }\bibfield  {title} {\enquote {\bibinfo {title} {{Calculating
  $\Delta m_K$ with lattice QCD}},}\ }\href {\doibase 10.22323/1.396.0141}
  {\bibfield  {journal} {\bibinfo  {journal} {PoS}\ }\textbf {\bibinfo {volume}
  {LATTICE2021}},\ \bibinfo {pages} {141} (\bibinfo {year} {2022})},\ \Eprint
  {http://arxiv.org/abs/2301.01387} {arXiv:2301.01387 [hep-lat]} \BibitemShut
  {NoStop}%
\bibitem [{\citenamefont {Bai}\ \emph {et~al.}(2024)\citenamefont {Bai},
  \citenamefont {Christ}, \citenamefont {Karpie}, \citenamefont {Sachrajda},
  \citenamefont {Soni},\ and\ \citenamefont {Wang}}]{Bai:2023lkr}%
  \BibitemOpen
  \bibfield  {author} {\bibinfo {author} {\bibfnamefont {Ziyuan}\ \bibnamefont
  {Bai}}, \bibinfo {author} {\bibfnamefont {Norman~H.}\ \bibnamefont {Christ}},
  \bibinfo {author} {\bibfnamefont {Joseph~M.}\ \bibnamefont {Karpie}},
  \bibinfo {author} {\bibfnamefont {Christopher~T.}\ \bibnamefont {Sachrajda}},
  \bibinfo {author} {\bibfnamefont {Amarjit}\ \bibnamefont {Soni}}, \ and\
  \bibinfo {author} {\bibfnamefont {Bigeng}\ \bibnamefont {Wang}},\ }\bibfield
  {title} {\enquote {\bibinfo {title} {{Long-distance contribution to
  \ensuremath{\varepsilon}K from lattice QCD}},}\ }\href {\doibase
  10.1103/PhysRevD.109.054501} {\bibfield  {journal} {\bibinfo  {journal}
  {Phys. Rev. D}\ }\textbf {\bibinfo {volume} {109}},\ \bibinfo {pages}
  {054501} (\bibinfo {year} {2024})},\ \Eprint
  {http://arxiv.org/abs/2309.01193} {arXiv:2309.01193 [hep-lat]} \BibitemShut
  {NoStop}%
\bibitem [{\citenamefont {Anzivino}\ \emph {et~al.}(2023)\citenamefont
  {Anzivino} \emph {et~al.}}]{Anzivino:2023bhp}%
  \BibitemOpen
  \bibfield  {author} {\bibinfo {author} {\bibfnamefont {G.}~\bibnamefont
  {Anzivino}} \emph {et~al.},\ }\bibfield  {title} {\enquote {\bibinfo {title}
  {{Workshop summary -- Kaons@CERN 2023}},}\ }in\ \href@noop {} {\emph
  {\bibinfo {booktitle} {{Kaons@CERN 2023}}}}\ (\bibinfo {year} {2023})\
  \Eprint {http://arxiv.org/abs/2311.02923} {arXiv:2311.02923 [hep-ph]}
  \BibitemShut {NoStop}%
\bibitem [{\citenamefont {Buras}(2023)}]{Buras:2023qaf}%
  \BibitemOpen
  \bibfield  {author} {\bibinfo {author} {\bibfnamefont {Andrzej~J.}\
  \bibnamefont {Buras}},\ }\bibfield  {title} {\enquote {\bibinfo {title}
  {{Kaon Theory: 50 Years Later}},}\ \ }(\bibinfo {year} {2023})\ \Eprint
  {http://arxiv.org/abs/2307.15737} {arXiv:2307.15737 [hep-ph]} \BibitemShut
  {NoStop}%
\bibitem [{\citenamefont {Goudzovski}\ \emph {et~al.}(2022)\citenamefont
  {Goudzovski} \emph {et~al.}}]{Goudzovski:2022scl}%
  \BibitemOpen
  \bibfield  {author} {\bibinfo {author} {\bibfnamefont {Evgueni}\ \bibnamefont
  {Goudzovski}} \emph {et~al.},\ }\bibfield  {title} {\enquote {\bibinfo
  {title} {{Weak Decays of Strange and Light Quarks}},}\ }\href@noop {} {\
  (\bibinfo {year} {2022})},\ \Eprint {http://arxiv.org/abs/2209.07156}
  {arXiv:2209.07156 [hep-ex]} \BibitemShut {NoStop}%
\bibitem [{\citenamefont {Bertone}\ \emph
  {et~al.}(2013{\natexlab{b}})\citenamefont {Bertone} \emph
  {et~al.}}]{ETM:2012vvy}%
  \BibitemOpen
  \bibfield  {author} {\bibinfo {author} {\bibfnamefont {V.}~\bibnamefont
  {Bertone}} \emph {et~al.} (\bibinfo {collaboration} {ETM}),\ }\bibfield
  {title} {\enquote {\bibinfo {title} {{Kaon Mixing Beyond the SM from
  N$_{f}$=2 tmQCD and model independent constraints from the UTA}},}\ }\href
  {\doibase 10.1007/JHEP03(2013)089} {\bibfield  {journal} {\bibinfo  {journal}
  {JHEP}\ }\textbf {\bibinfo {volume} {03}},\ \bibinfo {pages} {089} (\bibinfo
  {year} {2013}{\natexlab{b}})},\ \bibinfo {note} {[Erratum: JHEP 07, 143
  (2013)]},\ \Eprint {http://arxiv.org/abs/1207.1287} {arXiv:1207.1287
  [hep-lat]} \BibitemShut {NoStop}%
\bibitem [{\citenamefont {Lellouch}(2011)}]{Lellouch:2011qw}%
  \BibitemOpen
  \bibfield  {author} {\bibinfo {author} {\bibfnamefont {Laurent}\ \bibnamefont
  {Lellouch}},\ }\bibfield  {title} {\enquote {\bibinfo {title} {{Flavor
  physics and lattice quantum chromodynamics}},}\ }in\ \href@noop {} {\emph
  {\bibinfo {booktitle} {{Les Houches Summer School: Session 93: Modern
  perspectives in lattice QCD: Quantum field theory and high performance
  computing}}}}\ (\bibinfo {year} {2011})\ pp.\ \bibinfo {pages} {629--698},\
  \Eprint {http://arxiv.org/abs/1104.5484} {arXiv:1104.5484 [hep-lat]}
  \BibitemShut {NoStop}%
\bibitem [{\citenamefont {Allton}\ \emph {et~al.}(1999)\citenamefont {Allton},
  \citenamefont {Conti}, \citenamefont {Donini}, \citenamefont {Gimenez},
  \citenamefont {Giusti}, \citenamefont {Martinelli}, \citenamefont {Talevi},\
  and\ \citenamefont {Vladikas}}]{Allton:1998sm}%
  \BibitemOpen
  \bibfield  {author} {\bibinfo {author} {\bibfnamefont {C.~R.}\ \bibnamefont
  {Allton}}, \bibinfo {author} {\bibfnamefont {L.}~\bibnamefont {Conti}},
  \bibinfo {author} {\bibfnamefont {A.}~\bibnamefont {Donini}}, \bibinfo
  {author} {\bibfnamefont {V.}~\bibnamefont {Gimenez}}, \bibinfo {author}
  {\bibfnamefont {Leonardo}\ \bibnamefont {Giusti}}, \bibinfo {author}
  {\bibfnamefont {G.}~\bibnamefont {Martinelli}}, \bibinfo {author}
  {\bibfnamefont {M.}~\bibnamefont {Talevi}}, \ and\ \bibinfo {author}
  {\bibfnamefont {A.}~\bibnamefont {Vladikas}},\ }\bibfield  {title} {\enquote
  {\bibinfo {title} {{B parameters for Delta S = 2 supersymmetric
  operators}},}\ }\href {\doibase 10.1016/S0370-2693(99)00283-X} {\bibfield
  {journal} {\bibinfo  {journal} {Phys. Lett.}\ }\textbf {\bibinfo {volume}
  {B453}},\ \bibinfo {pages} {30--39} (\bibinfo {year} {1999})},\ \Eprint
  {http://arxiv.org/abs/hep-lat/9806016} {arXiv:hep-lat/9806016 [hep-lat]}
  \BibitemShut {NoStop}%
\bibitem [{\citenamefont {Donini}\ \emph {et~al.}(1999)\citenamefont {Donini},
  \citenamefont {Gimenez}, \citenamefont {Giusti},\ and\ \citenamefont
  {Martinelli}}]{Donini:1999nn}%
  \BibitemOpen
  \bibfield  {author} {\bibinfo {author} {\bibfnamefont {A.}~\bibnamefont
  {Donini}}, \bibinfo {author} {\bibfnamefont {V.}~\bibnamefont {Gimenez}},
  \bibinfo {author} {\bibfnamefont {Leonardo}\ \bibnamefont {Giusti}}, \ and\
  \bibinfo {author} {\bibfnamefont {G.}~\bibnamefont {Martinelli}},\ }\bibfield
   {title} {\enquote {\bibinfo {title} {{Renormalization group invariant matrix
  elements of Delta S = 2 and Delta I = 3/2 four fermion operators without
  quark masses}},}\ }\href {\doibase 10.1016/S0370-2693(99)01300-3} {\bibfield
  {journal} {\bibinfo  {journal} {Phys. Lett.}\ }\textbf {\bibinfo {volume}
  {B470}},\ \bibinfo {pages} {233--242} (\bibinfo {year} {1999})},\ \Eprint
  {http://arxiv.org/abs/hep-lat/9910017} {arXiv:hep-lat/9910017 [hep-lat]}
  \BibitemShut {NoStop}%
\bibitem [{\citenamefont {Babich}\ \emph {et~al.}(2006)\citenamefont {Babich},
  \citenamefont {Garron}, \citenamefont {Hoelbling}, \citenamefont {Howard},
  \citenamefont {Lellouch},\ and\ \citenamefont {Rebbi}}]{Babich:2006bh}%
  \BibitemOpen
  \bibfield  {author} {\bibinfo {author} {\bibfnamefont {Ronald}\ \bibnamefont
  {Babich}}, \bibinfo {author} {\bibfnamefont {Nicolas}\ \bibnamefont
  {Garron}}, \bibinfo {author} {\bibfnamefont {Christian}\ \bibnamefont
  {Hoelbling}}, \bibinfo {author} {\bibfnamefont {Joseph}\ \bibnamefont
  {Howard}}, \bibinfo {author} {\bibfnamefont {Laurent}\ \bibnamefont
  {Lellouch}}, \ and\ \bibinfo {author} {\bibfnamefont {Claudio}\ \bibnamefont
  {Rebbi}},\ }\bibfield  {title} {\enquote {\bibinfo {title} {{K0 - anti-0
  mixing beyond the standard model and CP-violating electroweak penguins in
  quenched QCD with exact chiral symmetry}},}\ }\href {\doibase
  10.1103/PhysRevD.74.073009} {\bibfield  {journal} {\bibinfo  {journal} {Phys.
  Rev.}\ }\textbf {\bibinfo {volume} {D74}},\ \bibinfo {pages} {073009}
  (\bibinfo {year} {2006})},\ \Eprint {http://arxiv.org/abs/hep-lat/0605016}
  {arXiv:hep-lat/0605016 [hep-lat]} \BibitemShut {NoStop}%
\bibitem [{\citenamefont {Boyle}\ \emph {et~al.}(2012)\citenamefont {Boyle},
  \citenamefont {Garron},\ and\ \citenamefont {Hudspith}}]{Boyle:2012qb}%
  \BibitemOpen
  \bibfield  {author} {\bibinfo {author} {\bibfnamefont {P.~A.}\ \bibnamefont
  {Boyle}}, \bibinfo {author} {\bibfnamefont {N.}~\bibnamefont {Garron}}, \
  and\ \bibinfo {author} {\bibfnamefont {R.~J.}\ \bibnamefont {Hudspith}}
  (\bibinfo {collaboration} {RBC, UKQCD}),\ }\bibfield  {title} {\enquote
  {\bibinfo {title} {{Neutral kaon mixing beyond the standard model with $N_f =
  2+1$ chiral fermions}},}\ }\href {\doibase 10.1103/PhysRevD.86.054028}
  {\bibfield  {journal} {\bibinfo  {journal} {Phys. Rev.}\ }\textbf {\bibinfo
  {volume} {D86}},\ \bibinfo {pages} {054028} (\bibinfo {year} {2012})},\
  \Eprint {http://arxiv.org/abs/1206.5737} {arXiv:1206.5737 [hep-lat]}
  \BibitemShut {NoStop}%
\bibitem [{\citenamefont {Garron}\ \emph {et~al.}(2016)\citenamefont {Garron},
  \citenamefont {Hudspith},\ and\ \citenamefont {Lytle}}]{Garron:2016mva}%
  \BibitemOpen
  \bibfield  {author} {\bibinfo {author} {\bibfnamefont {Nicolas}\ \bibnamefont
  {Garron}}, \bibinfo {author} {\bibfnamefont {Renwick~J.}\ \bibnamefont
  {Hudspith}}, \ and\ \bibinfo {author} {\bibfnamefont {Andrew~T.}\
  \bibnamefont {Lytle}} (\bibinfo {collaboration} {RBC/UKQCD}),\ }\bibfield
  {title} {\enquote {\bibinfo {title} {{Neutral Kaon Mixing Beyond the Standard
  Model with $N_f=2+1$ Chiral Fermions Part 1: Bare Matrix Elements and
  Physical Results}},}\ }\href {\doibase 10.1007/JHEP11(2016)001} {\bibfield
  {journal} {\bibinfo  {journal} {JHEP}\ }\textbf {\bibinfo {volume} {11}},\
  \bibinfo {pages} {001} (\bibinfo {year} {2016})},\ \Eprint
  {http://arxiv.org/abs/1609.03334} {arXiv:1609.03334 [hep-lat]} \BibitemShut
  {NoStop}%
\bibitem [{\citenamefont {Leem}\ \emph {et~al.}(2014)\citenamefont {Leem} \emph
  {et~al.}}]{Jang:2014aea}%
  \BibitemOpen
  \bibfield  {author} {\bibinfo {author} {\bibfnamefont {Jaehoon}\ \bibnamefont
  {Leem}} \emph {et~al.} (\bibinfo {collaboration} {SWME}),\ }\bibfield
  {title} {\enquote {\bibinfo {title} {{Calculation of BSM Kaon B-parameters
  using Staggered Quarks}},}\ }\bibfield  {booktitle} {\emph {\bibinfo
  {booktitle} {{Proceedings, 32nd International Symposium on Lattice Field
  Theory (Lattice 2014): Brookhaven, NY, USA, June 23-28, 2014}}},\ }\href@noop
  {} {\bibfield  {journal} {\bibinfo  {journal} {PoS}\ }\textbf {\bibinfo
  {volume} {LATTICE2014}},\ \bibinfo {pages} {370} (\bibinfo {year} {2014})},\
  \Eprint {http://arxiv.org/abs/1411.1501} {arXiv:1411.1501 [hep-lat]}
  \BibitemShut {NoStop}%
\bibitem [{\citenamefont {Tsang}\ and\ \citenamefont
  {Della~Morte}(2023)}]{Tsang:2023nay}%
  \BibitemOpen
  \bibfield  {author} {\bibinfo {author} {\bibfnamefont {J.~Tobias}\
  \bibnamefont {Tsang}}\ and\ \bibinfo {author} {\bibfnamefont {Michele}\
  \bibnamefont {Della~Morte}},\ }\bibfield  {title} {\enquote {\bibinfo {title}
  {{B-physics from lattice gauge theory}},}\ }\href {\doibase
  10.1140/epjs/s11734-023-01011-3} {\  (\bibinfo {year} {2023}),\
  10.1140/epjs/s11734-023-01011-3},\ \Eprint {http://arxiv.org/abs/2310.02705}
  {arXiv:2310.02705 [hep-lat]} \BibitemShut {NoStop}%
\bibitem [{\citenamefont {Boyle}\ \emph
  {et~al.}(2017{\natexlab{a}})\citenamefont {Boyle}, \citenamefont
  {Del~Debbio}, \citenamefont {Juttner}, \citenamefont {Khamseh}, \citenamefont
  {Sanfilippo},\ and\ \citenamefont {Tsang}}]{Boyle:2017jwu}%
  \BibitemOpen
  \bibfield  {author} {\bibinfo {author} {\bibfnamefont {Peter~A.}\
  \bibnamefont {Boyle}}, \bibinfo {author} {\bibfnamefont {Luigi}\ \bibnamefont
  {Del~Debbio}}, \bibinfo {author} {\bibfnamefont {Andreas}\ \bibnamefont
  {Juttner}}, \bibinfo {author} {\bibfnamefont {Ava}\ \bibnamefont {Khamseh}},
  \bibinfo {author} {\bibfnamefont {Francesco}\ \bibnamefont {Sanfilippo}}, \
  and\ \bibinfo {author} {\bibfnamefont {Justus~Tobias}\ \bibnamefont
  {Tsang}},\ }\bibfield  {title} {\enquote {\bibinfo {title} {{The decay
  constants $\mathbf{f_D}$ and $\mathbf{f_{D_s}}$ in the continuum limit of
  $\mathbf{N_f=2+1}$ domain wall lattice QCD}},}\ }\href {\doibase
  10.1007/JHEP12(2017)008} {\bibfield  {journal} {\bibinfo  {journal} {JHEP}\
  }\textbf {\bibinfo {volume} {12}},\ \bibinfo {pages} {008} (\bibinfo {year}
  {2017}{\natexlab{a}})},\ \Eprint {http://arxiv.org/abs/1701.02644}
  {arXiv:1701.02644 [hep-lat]} \BibitemShut {NoStop}%
\bibitem [{\citenamefont {Boyle}\ \emph
  {et~al.}(2017{\natexlab{b}})\citenamefont {Boyle}, \citenamefont {Garron},
  \citenamefont {Hudspith}, \citenamefont {Lehner},\ and\ \citenamefont
  {Lytle}}]{Boyle:2017skn}%
  \BibitemOpen
  \bibfield  {author} {\bibinfo {author} {\bibfnamefont {Peter~A.}\
  \bibnamefont {Boyle}}, \bibinfo {author} {\bibfnamefont {Nicolas}\
  \bibnamefont {Garron}}, \bibinfo {author} {\bibfnamefont {Renwick~J.}\
  \bibnamefont {Hudspith}}, \bibinfo {author} {\bibfnamefont {Christoph}\
  \bibnamefont {Lehner}}, \ and\ \bibinfo {author} {\bibfnamefont {Andrew~T.}\
  \bibnamefont {Lytle}} (\bibinfo {collaboration} {RBC, UKQCD}),\ }\bibfield
  {title} {\enquote {\bibinfo {title} {{Neutral kaon mixing beyond the Standard
  Model with n$_{f}$ = 2 + 1 chiral fermions. Part 2: non perturbative
  renormalisation of the $\Delta F=2$ four-quark operators}},}\ }\href
  {\doibase 10.1007/JHEP10(2017)054} {\bibfield  {journal} {\bibinfo  {journal}
  {JHEP}\ }\textbf {\bibinfo {volume} {10}},\ \bibinfo {pages} {054} (\bibinfo
  {year} {2017}{\natexlab{b}})},\ \Eprint {http://arxiv.org/abs/1708.03552}
  {arXiv:1708.03552 [hep-lat]} \BibitemShut {NoStop}%
\bibitem [{\citenamefont {Boyle}\ \emph
  {et~al.}(2018{\natexlab{a}})\citenamefont {Boyle}, \citenamefont {Garron},
  \citenamefont {Kettle}, \citenamefont {Khamseh},\ and\ \citenamefont
  {Tsang}}]{Boyle:2017ssm}%
  \BibitemOpen
  \bibfield  {author} {\bibinfo {author} {\bibfnamefont {Peter}\ \bibnamefont
  {Boyle}}, \bibinfo {author} {\bibfnamefont {Nicolas}\ \bibnamefont {Garron}},
  \bibinfo {author} {\bibfnamefont {Julia}\ \bibnamefont {Kettle}}, \bibinfo
  {author} {\bibfnamefont {Ava}\ \bibnamefont {Khamseh}}, \ and\ \bibinfo
  {author} {\bibfnamefont {Justus~Tobias}\ \bibnamefont {Tsang}},\ }\bibfield
  {title} {\enquote {\bibinfo {title} {{BSM Kaon Mixing at the Physical
  Point}},}\ }\href {\doibase 10.1051/epjconf/201817513010} {\bibfield
  {journal} {\bibinfo  {journal} {EPJ Web Conf.}\ }\textbf {\bibinfo {volume}
  {175}},\ \bibinfo {pages} {13010} (\bibinfo {year} {2018}{\natexlab{a}})},\
  \Eprint {http://arxiv.org/abs/1710.09176} {arXiv:1710.09176 [hep-lat]}
  \BibitemShut {NoStop}%
\bibitem [{\citenamefont {Boyle}\ \emph {et~al.}(2019)\citenamefont {Boyle},
  \citenamefont {Garron}, \citenamefont {Hudspith}, \citenamefont {Juttner},
  \citenamefont {Kettle}, \citenamefont {Khamseh},\ and\ \citenamefont
  {Tsang}}]{Boyle:2018eor}%
  \BibitemOpen
  \bibfield  {author} {\bibinfo {author} {\bibfnamefont {Peter}\ \bibnamefont
  {Boyle}}, \bibinfo {author} {\bibfnamefont {Nicolas}\ \bibnamefont {Garron}},
  \bibinfo {author} {\bibfnamefont {Renwick~James}\ \bibnamefont {Hudspith}},
  \bibinfo {author} {\bibfnamefont {Andreas}\ \bibnamefont {Juttner}}, \bibinfo
  {author} {\bibfnamefont {Julia}\ \bibnamefont {Kettle}}, \bibinfo {author}
  {\bibfnamefont {Ava}\ \bibnamefont {Khamseh}}, \ and\ \bibinfo {author}
  {\bibfnamefont {Justus~Tobias}\ \bibnamefont {Tsang}},\ }\bibfield  {title}
  {\enquote {\bibinfo {title} {{Beyond the Standard Model Kaon Mixing with
  Physical Masses}},}\ }\href {\doibase 10.22323/1.334.0285} {\bibfield
  {journal} {\bibinfo  {journal} {PoS}\ }\textbf {\bibinfo {volume}
  {LATTICE2018}},\ \bibinfo {pages} {285} (\bibinfo {year} {2019})},\ \Eprint
  {http://arxiv.org/abs/1812.04981} {arXiv:1812.04981 [hep-lat]} \BibitemShut
  {NoStop}%
\bibitem [{\citenamefont {Nicholson}\ \emph {et~al.}(2018)\citenamefont
  {Nicholson} \emph {et~al.}}]{Nicholson:2018mwc}%
  \BibitemOpen
  \bibfield  {author} {\bibinfo {author} {\bibfnamefont {A.}~\bibnamefont
  {Nicholson}} \emph {et~al.},\ }\bibfield  {title} {\enquote {\bibinfo {title}
  {{Heavy physics contributions to neutrinoless double beta decay from QCD}},}\
  }\href {\doibase 10.1103/PhysRevLett.121.172501} {\bibfield  {journal}
  {\bibinfo  {journal} {Phys. Rev. Lett.}\ }\textbf {\bibinfo {volume} {121}},\
  \bibinfo {pages} {172501} (\bibinfo {year} {2018})},\ \Eprint
  {http://arxiv.org/abs/1805.02634} {arXiv:1805.02634 [nucl-th]} \BibitemShut
  {NoStop}%
\bibitem [{\citenamefont {Detmold}\ \emph {et~al.}(2023)\citenamefont
  {Detmold}, \citenamefont {Jay}, \citenamefont {Murphy}, \citenamefont
  {Oare},\ and\ \citenamefont {Shanahan}}]{Detmold:2022jwu}%
  \BibitemOpen
  \bibfield  {author} {\bibinfo {author} {\bibfnamefont {William}\ \bibnamefont
  {Detmold}}, \bibinfo {author} {\bibfnamefont {William~I.}\ \bibnamefont
  {Jay}}, \bibinfo {author} {\bibfnamefont {David~J.}\ \bibnamefont {Murphy}},
  \bibinfo {author} {\bibfnamefont {Patrick~R.}\ \bibnamefont {Oare}}, \ and\
  \bibinfo {author} {\bibfnamefont {Phiala~E.}\ \bibnamefont {Shanahan}},\
  }\bibfield  {title} {\enquote {\bibinfo {title} {{Neutrinoless double beta
  decay from lattice QCD: The short-distance
  \ensuremath{\pi}-\textrightarrow{}\ensuremath{\pi}+e-e- amplitude}},}\ }\href
  {\doibase 10.1103/PhysRevD.107.094501} {\bibfield  {journal} {\bibinfo
  {journal} {Phys. Rev. D}\ }\textbf {\bibinfo {volume} {107}},\ \bibinfo
  {pages} {094501} (\bibinfo {year} {2023})},\ \Eprint
  {http://arxiv.org/abs/2208.05322} {arXiv:2208.05322 [hep-lat]} \BibitemShut
  {NoStop}%
\bibitem [{\citenamefont {Gabbiani}\ \emph {et~al.}(1996)\citenamefont
  {Gabbiani}, \citenamefont {Gabrielli}, \citenamefont {Masiero},\ and\
  \citenamefont {Silvestrini}}]{Gabbiani:1996hi}%
  \BibitemOpen
  \bibfield  {author} {\bibinfo {author} {\bibfnamefont {F.}~\bibnamefont
  {Gabbiani}}, \bibinfo {author} {\bibfnamefont {E.}~\bibnamefont {Gabrielli}},
  \bibinfo {author} {\bibfnamefont {A.}~\bibnamefont {Masiero}}, \ and\
  \bibinfo {author} {\bibfnamefont {L.}~\bibnamefont {Silvestrini}},\
  }\bibfield  {title} {\enquote {\bibinfo {title} {{A Complete analysis of FCNC
  and CP constraints in general SUSY extensions of the standard model}},}\
  }\href {\doibase 10.1016/0550-3213(96)00390-2} {\bibfield  {journal}
  {\bibinfo  {journal} {Nucl. Phys.}\ }\textbf {\bibinfo {volume} {B477}},\
  \bibinfo {pages} {321--352} (\bibinfo {year} {1996})},\ \Eprint
  {http://arxiv.org/abs/hep-ph/9604387} {arXiv:hep-ph/9604387 [hep-ph]}
  \BibitemShut {NoStop}%
\bibitem [{\citenamefont {Bae}\ \emph {et~al.}(2013{\natexlab{b}})\citenamefont
  {Bae} \emph {et~al.}}]{SWME:2013laf}%
  \BibitemOpen
  \bibfield  {author} {\bibinfo {author} {\bibfnamefont {Taegil}\ \bibnamefont
  {Bae}} \emph {et~al.} (\bibinfo {collaboration} {SWME}),\ }\bibfield  {title}
  {\enquote {\bibinfo {title} {{Neutral kaon mixing from new physics: matrix
  elements in $N_f=2+1$ lattice QCD}},}\ }\href {\doibase
  10.1103/PhysRevD.88.071503} {\bibfield  {journal} {\bibinfo  {journal} {Phys.
  Rev. D}\ }\textbf {\bibinfo {volume} {88}},\ \bibinfo {pages} {071503}
  (\bibinfo {year} {2013}{\natexlab{b}})},\ \Eprint
  {http://arxiv.org/abs/1309.2040} {arXiv:1309.2040 [hep-lat]} \BibitemShut
  {NoStop}%
\bibitem [{\citenamefont {Becirevic}\ and\ \citenamefont
  {Villadoro}(2004{\natexlab{a}})}]{Becirevic:2004qd}%
  \BibitemOpen
  \bibfield  {author} {\bibinfo {author} {\bibfnamefont {Damir}\ \bibnamefont
  {Becirevic}}\ and\ \bibinfo {author} {\bibfnamefont {Giovanni}\ \bibnamefont
  {Villadoro}},\ }\bibfield  {title} {\enquote {\bibinfo {title} {{Remarks on
  the hadronic matrix elements relevant to the SUSY K0 - anti-K0 mixing
  amplitude}},}\ }\href {\doibase 10.1103/PhysRevD.70.094036} {\bibfield
  {journal} {\bibinfo  {journal} {Phys. Rev. D}\ }\textbf {\bibinfo {volume}
  {70}},\ \bibinfo {pages} {094036} (\bibinfo {year} {2004}{\natexlab{a}})},\
  \Eprint {http://arxiv.org/abs/hep-lat/0408029} {arXiv:hep-lat/0408029}
  \BibitemShut {NoStop}%
\bibitem [{\citenamefont {Allton}\ \emph {et~al.}(2008)\citenamefont {Allton}
  \emph {et~al.}}]{Allton:2008pn}%
  \BibitemOpen
  \bibfield  {author} {\bibinfo {author} {\bibfnamefont {C.}~\bibnamefont
  {Allton}} \emph {et~al.} (\bibinfo {collaboration} {RBC-UKQCD
  Collaboration}),\ }\bibfield  {title} {\enquote {\bibinfo {title} {{Physical
  Results from 2+1 Flavor Domain Wall QCD and SU(2) Chiral Perturbation
  Theory}},}\ }\href {\doibase 10.1103/PhysRevD.78.114509} {\bibfield
  {journal} {\bibinfo  {journal} {Phys.Rev.}\ }\textbf {\bibinfo {volume}
  {D78}},\ \bibinfo {pages} {114509} (\bibinfo {year} {2008})},\ \Eprint
  {http://arxiv.org/abs/0804.0473} {arXiv:0804.0473 [hep-lat]} \BibitemShut
  {NoStop}%
\bibitem [{\citenamefont {Aoki}\ \emph
  {et~al.}(2011{\natexlab{b}})\citenamefont {Aoki} \emph
  {et~al.}}]{Aoki:2010dy}%
  \BibitemOpen
  \bibfield  {author} {\bibinfo {author} {\bibfnamefont {Y.}~\bibnamefont
  {Aoki}} \emph {et~al.} (\bibinfo {collaboration} {RBC Collaboration, UKQCD
  Collaboration}),\ }\bibfield  {title} {\enquote {\bibinfo {title} {{Continuum
  Limit Physics from 2+1 Flavor Domain Wall QCD}},}\ }\href {\doibase
  10.1103/PhysRevD.83.074508} {\bibfield  {journal} {\bibinfo  {journal}
  {Phys.Rev.}\ }\textbf {\bibinfo {volume} {D83}},\ \bibinfo {pages} {074508}
  (\bibinfo {year} {2011}{\natexlab{b}})},\ \Eprint
  {http://arxiv.org/abs/1011.0892} {arXiv:1011.0892 [hep-lat]} \BibitemShut
  {NoStop}%
\bibitem [{\citenamefont {Boyle}\ \emph
  {et~al.}(2018{\natexlab{b}})\citenamefont {Boyle}, \citenamefont
  {Del~Debbio}, \citenamefont {Garron}, \citenamefont {Juttner}, \citenamefont
  {Soni}, \citenamefont {Tsang},\ and\ \citenamefont {Witzel}}]{Boyle:2018knm}%
  \BibitemOpen
  \bibfield  {author} {\bibinfo {author} {\bibfnamefont {Peter~A.}\
  \bibnamefont {Boyle}}, \bibinfo {author} {\bibfnamefont {Luigi}\ \bibnamefont
  {Del~Debbio}}, \bibinfo {author} {\bibfnamefont {Nicolas}\ \bibnamefont
  {Garron}}, \bibinfo {author} {\bibfnamefont {Andreas}\ \bibnamefont
  {Juttner}}, \bibinfo {author} {\bibfnamefont {Amarjit}\ \bibnamefont {Soni}},
  \bibinfo {author} {\bibfnamefont {Justus~Tobias}\ \bibnamefont {Tsang}}, \
  and\ \bibinfo {author} {\bibfnamefont {Oliver}\ \bibnamefont {Witzel}}
  (\bibinfo {collaboration} {RBC/UKQCD}),\ }\bibfield  {title} {\enquote
  {\bibinfo {title} {{SU(3)-breaking ratios for $D_{(s)}$ and $B_{(s)}$
  mesons}},}\ }\href@noop {} {\  (\bibinfo {year} {2018}{\natexlab{b}})},\
  \Eprint {http://arxiv.org/abs/1812.08791} {arXiv:1812.08791 [hep-lat]}
  \BibitemShut {NoStop}%
\bibitem [{\citenamefont {Iwasaki}\ and\ \citenamefont
  {Yoshie}(1984)}]{Iwasaki:1984cj}%
  \BibitemOpen
  \bibfield  {author} {\bibinfo {author} {\bibfnamefont {Y.}~\bibnamefont
  {Iwasaki}}\ and\ \bibinfo {author} {\bibfnamefont {T.}~\bibnamefont
  {Yoshie}},\ }\bibfield  {title} {\enquote {\bibinfo {title} {{Renormalization
  Group Improved Action for SU(3) Lattice Gauge Theory and the String
  Tension}},}\ }\href {\doibase 10.1016/0370-2693(84)91500-4} {\bibfield
  {journal} {\bibinfo  {journal} {Phys. Lett.}\ }\textbf {\bibinfo {volume}
  {143B}},\ \bibinfo {pages} {449--452} (\bibinfo {year} {1984})}\BibitemShut
  {NoStop}%
\bibitem [{\citenamefont {Iwasaki}(1985)}]{Iwasaki:1985we}%
  \BibitemOpen
  \bibfield  {author} {\bibinfo {author} {\bibfnamefont {Y.}~\bibnamefont
  {Iwasaki}},\ }\bibfield  {title} {\enquote {\bibinfo {title}
  {{Renormalization Group Analysis of Lattice Theories and Improved Lattice
  Action: Two-Dimensional Nonlinear O(N) Sigma Model}},}\ }\href {\doibase
  10.1016/0550-3213(85)90606-6} {\bibfield  {journal} {\bibinfo  {journal}
  {Nucl. Phys.}\ }\textbf {\bibinfo {volume} {B258}},\ \bibinfo {pages}
  {141--156} (\bibinfo {year} {1985})}\BibitemShut {NoStop}%
\bibitem [{\citenamefont {Brower}\ \emph {et~al.}(2005)\citenamefont {Brower},
  \citenamefont {Neff},\ and\ \citenamefont {Orginos}}]{Brower:2004xi}%
  \BibitemOpen
  \bibfield  {author} {\bibinfo {author} {\bibfnamefont {Richard~C.}\
  \bibnamefont {Brower}}, \bibinfo {author} {\bibfnamefont {Hartmut}\
  \bibnamefont {Neff}}, \ and\ \bibinfo {author} {\bibfnamefont {Kostas}\
  \bibnamefont {Orginos}},\ }\bibfield  {title} {\enquote {\bibinfo {title}
  {{Mobius fermions: Improved domain wall chiral fermions}},}\ }\href {\doibase
  10.1016/j.nuclphysbps.2004.11.180} {\bibfield  {journal} {\bibinfo  {journal}
  {Nucl.Phys.Proc.Suppl.}\ }\textbf {\bibinfo {volume} {140}},\ \bibinfo
  {pages} {686--688} (\bibinfo {year} {2005})},\ \Eprint
  {http://arxiv.org/abs/hep-lat/0409118} {arXiv:hep-lat/0409118 [hep-lat]}
  \BibitemShut {NoStop}%
\bibitem [{\citenamefont {Brower}\ \emph {et~al.}(2006)\citenamefont {Brower},
  \citenamefont {Neff},\ and\ \citenamefont {Orginos}}]{Brower:2005qw}%
  \BibitemOpen
  \bibfield  {author} {\bibinfo {author} {\bibfnamefont {R.C.}\ \bibnamefont
  {Brower}}, \bibinfo {author} {\bibfnamefont {H.}~\bibnamefont {Neff}}, \ and\
  \bibinfo {author} {\bibfnamefont {K.}~\bibnamefont {Orginos}},\ }\bibfield
  {title} {\enquote {\bibinfo {title} {{Mobius fermions}},}\ }\href {\doibase
  10.1016/j.nuclphysbps.2006.01.047} {\bibfield  {journal} {\bibinfo  {journal}
  {Nucl.Phys.Proc.Suppl.}\ }\textbf {\bibinfo {volume} {153}},\ \bibinfo
  {pages} {191--198} (\bibinfo {year} {2006})},\ \Eprint
  {http://arxiv.org/abs/hep-lat/0511031} {arXiv:hep-lat/0511031 [hep-lat]}
  \BibitemShut {NoStop}%
\bibitem [{\citenamefont {Brower}\ \emph {et~al.}(2017)\citenamefont {Brower},
  \citenamefont {Neff},\ and\ \citenamefont {Orginos}}]{Brower:2012vk}%
  \BibitemOpen
  \bibfield  {author} {\bibinfo {author} {\bibfnamefont {Richard~C.}\
  \bibnamefont {Brower}}, \bibinfo {author} {\bibfnamefont {Harmut}\
  \bibnamefont {Neff}}, \ and\ \bibinfo {author} {\bibfnamefont {Kostas}\
  \bibnamefont {Orginos}},\ }\bibfield  {title} {\enquote {\bibinfo {title}
  {{The M\"obius domain wall fermion algorithm}},}\ }\href {\doibase
  10.1016/j.cpc.2017.01.024} {\bibfield  {journal} {\bibinfo  {journal}
  {Comput. Phys. Commun.}\ }\textbf {\bibinfo {volume} {220}},\ \bibinfo
  {pages} {1--19} (\bibinfo {year} {2017})},\ \Eprint
  {http://arxiv.org/abs/1206.5214} {arXiv:1206.5214 [hep-lat]} \BibitemShut
  {NoStop}%
\bibitem [{\citenamefont {Shamir}(1993)}]{Shamir:1993zy}%
  \BibitemOpen
  \bibfield  {author} {\bibinfo {author} {\bibfnamefont {Yigal}\ \bibnamefont
  {Shamir}},\ }\bibfield  {title} {\enquote {\bibinfo {title} {{Chiral fermions
  from lattice boundaries}},}\ }\href {\doibase 10.1016/0550-3213(93)90162-I}
  {\bibfield  {journal} {\bibinfo  {journal} {Nucl.Phys.}\ }\textbf {\bibinfo
  {volume} {B406}},\ \bibinfo {pages} {90--106} (\bibinfo {year} {1993})},\
  \Eprint {http://arxiv.org/abs/hep-lat/9303005} {arXiv:hep-lat/9303005
  [hep-lat]} \BibitemShut {NoStop}%
\bibitem [{\citenamefont {Boyle}\ \emph {et~al.}(2016)\citenamefont {Boyle},
  \citenamefont {Cossu}, \citenamefont {Yamaguchi},\ and\ \citenamefont
  {Portelli}}]{Boyle:2016lbp}%
  \BibitemOpen
  \bibfield  {author} {\bibinfo {author} {\bibfnamefont {Peter~A.}\
  \bibnamefont {Boyle}}, \bibinfo {author} {\bibfnamefont {Guido}\ \bibnamefont
  {Cossu}}, \bibinfo {author} {\bibfnamefont {Azusa}\ \bibnamefont
  {Yamaguchi}}, \ and\ \bibinfo {author} {\bibfnamefont {Antonin}\ \bibnamefont
  {Portelli}},\ }\bibfield  {title} {\enquote {\bibinfo {title} {{Grid: A next
  generation data parallel C++ QCD library}},}\ }\href {\doibase
  10.22323/1.251.0023} {\bibfield  {journal} {\bibinfo  {journal} {PoS}\
  }\textbf {\bibinfo {volume} {LATTICE2015}},\ \bibinfo {pages} {023} (\bibinfo
  {year} {2016})}\BibitemShut {NoStop}%
\bibitem [{\citenamefont {Yamaguchi}\ \emph {et~al.}(2022)\citenamefont
  {Yamaguchi}, \citenamefont {Boyle}, \citenamefont {Cossu}, \citenamefont
  {Filaci}, \citenamefont {Lehner},\ and\ \citenamefont
  {Portelli}}]{Yamaguchi:2022feu}%
  \BibitemOpen
  \bibfield  {author} {\bibinfo {author} {\bibfnamefont {Azusa}\ \bibnamefont
  {Yamaguchi}}, \bibinfo {author} {\bibfnamefont {Peter}\ \bibnamefont
  {Boyle}}, \bibinfo {author} {\bibfnamefont {Guido}\ \bibnamefont {Cossu}},
  \bibinfo {author} {\bibfnamefont {Gianluca}\ \bibnamefont {Filaci}}, \bibinfo
  {author} {\bibfnamefont {Christoph}\ \bibnamefont {Lehner}}, \ and\ \bibinfo
  {author} {\bibfnamefont {Antonin}\ \bibnamefont {Portelli}},\ }\bibfield
  {title} {\enquote {\bibinfo {title} {{Grid: OneCode and FourAPIs}},}\ }\href
  {\doibase 10.22323/1.396.0035} {\bibfield  {journal} {\bibinfo  {journal}
  {PoS}\ }\textbf {\bibinfo {volume} {LATTICE2021}},\ \bibinfo {pages} {035}
  (\bibinfo {year} {2022})},\ \Eprint {http://arxiv.org/abs/2203.06777}
  {arXiv:2203.06777 [hep-lat]} \BibitemShut {NoStop}%
\bibitem [{\citenamefont {Portelli}\ \emph {et~al.}(2023)\citenamefont
  {Portelli}, \citenamefont {Lachini}, \citenamefont {Erben}, \citenamefont
  {Marshall}, \citenamefont {Joswig}, \citenamefont {Hodgson}, \citenamefont
  {{\'O h\'Og\'ain}}, \citenamefont {G\"ulpers}, \citenamefont {Boyle},
  \citenamefont {Asmussen}, \citenamefont {Hill}, \citenamefont {Barone},
  \citenamefont {Richings}, \citenamefont {Abbott}, \citenamefont {B\"urger},\
  and\ \citenamefont {Lee}}]{antonin_portelli_2023_8023716}%
  \BibitemOpen
  \bibfield  {author} {\bibinfo {author} {\bibfnamefont {Antonin}\ \bibnamefont
  {Portelli}}, \bibinfo {author} {\bibfnamefont {Nelson}\ \bibnamefont
  {Lachini}}, \bibinfo {author} {\bibfnamefont {Felix}\ \bibnamefont {Erben}},
  \bibinfo {author} {\bibfnamefont {Michael}\ \bibnamefont {Marshall}},
  \bibinfo {author} {\bibfnamefont {Fabian}\ \bibnamefont {Joswig}}, \bibinfo
  {author} {\bibfnamefont {Raoul}\ \bibnamefont {Hodgson}}, \bibinfo {author}
  {\bibfnamefont {Fionn}\ \bibnamefont {{\'O h\'Og\'ain}}}, \bibinfo {author}
  {\bibfnamefont {Vera}\ \bibnamefont {G\"ulpers}}, \bibinfo {author}
  {\bibfnamefont {Peter}\ \bibnamefont {Boyle}}, \bibinfo {author}
  {\bibfnamefont {Nils}\ \bibnamefont {Asmussen}}, \bibinfo {author}
  {\bibfnamefont {Ryan}\ \bibnamefont {Hill}}, \bibinfo {author} {\bibfnamefont
  {Alessandro}\ \bibnamefont {Barone}}, \bibinfo {author} {\bibfnamefont
  {James}\ \bibnamefont {Richings}}, \bibinfo {author} {\bibfnamefont {Ryan}\
  \bibnamefont {Abbott}}, \bibinfo {author} {\bibfnamefont {Simon}\
  \bibnamefont {B\"urger}}, \ and\ \bibinfo {author} {\bibfnamefont {Joseph}\
  \bibnamefont {Lee}},\ }\href {\doibase 10.5281/zenodo.8023716} {\enquote
  {\bibinfo {title} {aportelli/hadrons: Hadrons v1.4},}\ } (\bibinfo {year}
  {2023})\BibitemShut {NoStop}%
\bibitem [{\citenamefont {Foster}\ and\ \citenamefont
  {Michael}(1999)}]{Foster:1998vw}%
  \BibitemOpen
  \bibfield  {author} {\bibinfo {author} {\bibfnamefont {M.}~\bibnamefont
  {Foster}}\ and\ \bibinfo {author} {\bibfnamefont {Christopher}\ \bibnamefont
  {Michael}} (\bibinfo {collaboration} {UKQCD}),\ }\bibfield  {title} {\enquote
  {\bibinfo {title} {{Quark mass dependence of hadron masses from lattice
  QCD}},}\ }\href {\doibase 10.1103/PhysRevD.59.074503} {\bibfield  {journal}
  {\bibinfo  {journal} {Phys. Rev. D}\ }\textbf {\bibinfo {volume} {59}},\
  \bibinfo {pages} {074503} (\bibinfo {year} {1999})},\ \Eprint
  {http://arxiv.org/abs/hep-lat/9810021} {arXiv:hep-lat/9810021} \BibitemShut
  {NoStop}%
\bibitem [{\citenamefont {McNeile}\ and\ \citenamefont
  {Michael}(2006)}]{McNeile:2006bz}%
  \BibitemOpen
  \bibfield  {author} {\bibinfo {author} {\bibfnamefont {C.}~\bibnamefont
  {McNeile}}\ and\ \bibinfo {author} {\bibfnamefont {Christopher}\ \bibnamefont
  {Michael}} (\bibinfo {collaboration} {UKQCD}),\ }\bibfield  {title} {\enquote
  {\bibinfo {title} {{Decay width of light quark hybrid meson from the
  lattice}},}\ }\href {\doibase 10.1103/PhysRevD.73.074506} {\bibfield
  {journal} {\bibinfo  {journal} {Phys. Rev. D}\ }\textbf {\bibinfo {volume}
  {73}},\ \bibinfo {pages} {074506} (\bibinfo {year} {2006})},\ \Eprint
  {http://arxiv.org/abs/hep-lat/0603007} {arXiv:hep-lat/0603007} \BibitemShut
  {NoStop}%
\bibitem [{\citenamefont {Boyle}\ \emph {et~al.}(2008)\citenamefont {Boyle},
  \citenamefont {J{\"u}ttner}, \citenamefont {Kelly},\ and\ \citenamefont
  {Kenway}}]{Boyle:2008rh}%
  \BibitemOpen
  \bibfield  {author} {\bibinfo {author} {\bibfnamefont {P.~A.}\ \bibnamefont
  {Boyle}}, \bibinfo {author} {\bibfnamefont {A.}~\bibnamefont {J{\"u}ttner}},
  \bibinfo {author} {\bibfnamefont {C.}~\bibnamefont {Kelly}}, \ and\ \bibinfo
  {author} {\bibfnamefont {R.~D.}\ \bibnamefont {Kenway}},\ }\bibfield  {title}
  {\enquote {\bibinfo {title} {{Use of stochastic sources for the lattice
  determination of light quark physics}},}\ }\href {\doibase
  10.1088/1126-6708/2008/08/086} {\bibfield  {journal} {\bibinfo  {journal}
  {JHEP}\ }\textbf {\bibinfo {volume} {08}},\ \bibinfo {pages} {086} (\bibinfo
  {year} {2008})},\ \Eprint {http://arxiv.org/abs/0804.1501} {arXiv:0804.1501
  [hep-lat]} \BibitemShut {NoStop}%
\bibitem [{\citenamefont {Martinelli}\ \emph {et~al.}(1995)\citenamefont
  {Martinelli}, \citenamefont {Pittori}, \citenamefont {Sachrajda},
  \citenamefont {Testa},\ and\ \citenamefont {Vladikas}}]{Martinelli:1994ty}%
  \BibitemOpen
  \bibfield  {author} {\bibinfo {author} {\bibfnamefont {G.}~\bibnamefont
  {Martinelli}}, \bibinfo {author} {\bibfnamefont {C.}~\bibnamefont {Pittori}},
  \bibinfo {author} {\bibfnamefont {Christopher~T.}\ \bibnamefont {Sachrajda}},
  \bibinfo {author} {\bibfnamefont {M.}~\bibnamefont {Testa}}, \ and\ \bibinfo
  {author} {\bibfnamefont {A.}~\bibnamefont {Vladikas}},\ }\bibfield  {title}
  {\enquote {\bibinfo {title} {{A General method for nonperturbative
  renormalization of lattice operators}},}\ }\href {\doibase
  10.1016/0550-3213(95)00126-D} {\bibfield  {journal} {\bibinfo  {journal}
  {Nucl.Phys.}\ }\textbf {\bibinfo {volume} {B445}},\ \bibinfo {pages}
  {81--108} (\bibinfo {year} {1995})},\ \Eprint
  {http://arxiv.org/abs/hep-lat/9411010} {arXiv:hep-lat/9411010 [hep-lat]}
  \BibitemShut {NoStop}%
\bibitem [{\citenamefont {Sturm}\ \emph {et~al.}(2009)\citenamefont {Sturm},
  \citenamefont {Aoki}, \citenamefont {Christ}, \citenamefont {Izubuchi},
  \citenamefont {Sachrajda} \emph {et~al.}}]{Sturm:2009kb}%
  \BibitemOpen
  \bibfield  {author} {\bibinfo {author} {\bibfnamefont {C.}~\bibnamefont
  {Sturm}}, \bibinfo {author} {\bibfnamefont {Y.}~\bibnamefont {Aoki}},
  \bibinfo {author} {\bibfnamefont {N.H.}\ \bibnamefont {Christ}}, \bibinfo
  {author} {\bibfnamefont {T.}~\bibnamefont {Izubuchi}}, \bibinfo {author}
  {\bibfnamefont {C.T.C.}\ \bibnamefont {Sachrajda}},  \emph {et~al.},\
  }\bibfield  {title} {\enquote {\bibinfo {title} {{Renormalization of quark
  bilinear operators in a momentum-subtraction scheme with a nonexceptional
  subtraction point}},}\ }\href {\doibase 10.1103/PhysRevD.80.014501}
  {\bibfield  {journal} {\bibinfo  {journal} {Phys.Rev.}\ }\textbf {\bibinfo
  {volume} {D80}},\ \bibinfo {pages} {014501} (\bibinfo {year} {2009})},\
  \Eprint {http://arxiv.org/abs/0901.2599} {arXiv:0901.2599 [hep-ph]}
  \BibitemShut {NoStop}%
\bibitem [{\citenamefont {Arthur}\ and\ \citenamefont
  {Boyle}(2011)}]{Arthur:2010ht}%
  \BibitemOpen
  \bibfield  {author} {\bibinfo {author} {\bibfnamefont {R.}~\bibnamefont
  {Arthur}}\ and\ \bibinfo {author} {\bibfnamefont {P.A.}\ \bibnamefont
  {Boyle}} (\bibinfo {collaboration} {RBC Collaboration, UKQCD
  Collaboration}),\ }\bibfield  {title} {\enquote {\bibinfo {title} {{Step
  Scaling with off-shell renormalisation}},}\ }\href {\doibase
  10.1103/PhysRevD.83.114511} {\bibfield  {journal} {\bibinfo  {journal}
  {Phys.Rev.}\ }\textbf {\bibinfo {volume} {D83}},\ \bibinfo {pages} {114511}
  (\bibinfo {year} {2011})},\ \Eprint {http://arxiv.org/abs/1006.0422}
  {arXiv:1006.0422 [hep-lat]} \BibitemShut {NoStop}%
\bibitem [{\citenamefont {Arthur}\ \emph {et~al.}(2012)\citenamefont {Arthur},
  \citenamefont {Boyle}, \citenamefont {Garron}, \citenamefont {Kelly},\ and\
  \citenamefont {Lytle}}]{Arthur:2011cn}%
  \BibitemOpen
  \bibfield  {author} {\bibinfo {author} {\bibfnamefont {R.}~\bibnamefont
  {Arthur}}, \bibinfo {author} {\bibfnamefont {P.A.}\ \bibnamefont {Boyle}},
  \bibinfo {author} {\bibfnamefont {N.}~\bibnamefont {Garron}}, \bibinfo
  {author} {\bibfnamefont {C.}~\bibnamefont {Kelly}}, \ and\ \bibinfo {author}
  {\bibfnamefont {A.T.}\ \bibnamefont {Lytle}} (\bibinfo {collaboration} {RBC
  Collaboration, UKQCD Collaboration}),\ }\bibfield  {title} {\enquote
  {\bibinfo {title} {{Opening the Rome-Southampton window for operator mixing
  matrices}},}\ }\href {\doibase 10.1103/PhysRevD.85.014501} {\bibfield
  {journal} {\bibinfo  {journal} {Phys.Rev.}\ }\textbf {\bibinfo {volume}
  {D85}},\ \bibinfo {pages} {014501} (\bibinfo {year} {2012})},\ \Eprint
  {http://arxiv.org/abs/1109.1223} {arXiv:1109.1223 [hep-lat]} \BibitemShut
  {NoStop}%
\bibitem [{\citenamefont {Boyle}\ \emph {et~al.}(2011)\citenamefont {Boyle},
  \citenamefont {Garron},\ and\ \citenamefont {Lytle}}]{Boyle:2011cc}%
  \BibitemOpen
  \bibfield  {author} {\bibinfo {author} {\bibfnamefont {P.A.}\ \bibnamefont
  {Boyle}}, \bibinfo {author} {\bibfnamefont {N.}~\bibnamefont {Garron}}, \
  and\ \bibinfo {author} {\bibfnamefont {A.T.}\ \bibnamefont {Lytle}} (\bibinfo
  {collaboration} {RBC-UKQCD Collaboration}),\ }\bibfield  {title} {\enquote
  {\bibinfo {title} {{Non-perturbative running and renormalization of kaon
  four-quark operators with nf=2+1 domain-wall fermions}},}\ }\href@noop {}
  {\bibfield  {journal} {\bibinfo  {journal} {PoS}\ }\textbf {\bibinfo {volume}
  {LATTICE2011}},\ \bibinfo {pages} {227} (\bibinfo {year} {2011})},\ \Eprint
  {http://arxiv.org/abs/1112.0537} {arXiv:1112.0537 [hep-lat]} \BibitemShut
  {NoStop}%
\bibitem [{\citenamefont {Workman}\ \emph {et~al.}(2022)\citenamefont {Workman}
  \emph {et~al.}}]{ParticleDataGroup:2022pth}%
  \BibitemOpen
  \bibfield  {author} {\bibinfo {author} {\bibfnamefont {R.~L.}\ \bibnamefont
  {Workman}} \emph {et~al.} (\bibinfo {collaboration} {Particle Data Group}),\
  }\bibfield  {title} {\enquote {\bibinfo {title} {{Review of Particle
  Physics}},}\ }\href {\doibase 10.1093/ptep/ptac097} {\bibfield  {journal}
  {\bibinfo  {journal} {PTEP}\ }\textbf {\bibinfo {volume} {2022}},\ \bibinfo
  {pages} {083C01} (\bibinfo {year} {2022})}\BibitemShut {NoStop}%
\bibitem [{\citenamefont {Becirevic}\ and\ \citenamefont
  {Villadoro}(2004{\natexlab{b}})}]{Becirevic:2003wk}%
  \BibitemOpen
  \bibfield  {author} {\bibinfo {author} {\bibfnamefont {Damir}\ \bibnamefont
  {Becirevic}}\ and\ \bibinfo {author} {\bibfnamefont {Giovanni}\ \bibnamefont
  {Villadoro}},\ }\bibfield  {title} {\enquote {\bibinfo {title} {{Impact of
  the finite volume effects on the chiral behavior of f(K) and B(K)}},}\ }\href
  {\doibase 10.1103/PhysRevD.69.054010} {\bibfield  {journal} {\bibinfo
  {journal} {Phys. Rev. D}\ }\textbf {\bibinfo {volume} {69}},\ \bibinfo
  {pages} {054010} (\bibinfo {year} {2004}{\natexlab{b}})},\ \Eprint
  {http://arxiv.org/abs/hep-lat/0311028} {arXiv:hep-lat/0311028} \BibitemShut
  {NoStop}%
\bibitem [{\citenamefont {Alexandrou}\ \emph {et~al.}(2021)\citenamefont
  {Alexandrou} \emph {et~al.}}]{ExtendedTwistedMass:2021gbo}%
  \BibitemOpen
  \bibfield  {author} {\bibinfo {author} {\bibfnamefont {C.}~\bibnamefont
  {Alexandrou}} \emph {et~al.} (\bibinfo {collaboration} {Extended Twisted
  Mass}),\ }\bibfield  {title} {\enquote {\bibinfo {title} {{Quark masses using
  twisted-mass fermion gauge ensembles}},}\ }\href {\doibase
  10.1103/PhysRevD.104.074515} {\bibfield  {journal} {\bibinfo  {journal}
  {Phys. Rev. D}\ }\textbf {\bibinfo {volume} {104}},\ \bibinfo {pages}
  {074515} (\bibinfo {year} {2021})},\ \Eprint
  {http://arxiv.org/abs/2104.13408} {arXiv:2104.13408 [hep-lat]} \BibitemShut
  {NoStop}%
\bibitem [{\citenamefont {Bazavov}\ \emph {et~al.}(2018)\citenamefont {Bazavov}
  \emph {et~al.}}]{FermilabLattice:2018est}%
  \BibitemOpen
  \bibfield  {author} {\bibinfo {author} {\bibfnamefont {A.}~\bibnamefont
  {Bazavov}} \emph {et~al.} (\bibinfo {collaboration} {Fermilab Lattice, MILC,
  TUMQCD}),\ }\bibfield  {title} {\enquote {\bibinfo {title} {{Up-, down-,
  strange-, charm-, and bottom-quark masses from four-flavor lattice QCD}},}\
  }\href {\doibase 10.1103/PhysRevD.98.054517} {\bibfield  {journal} {\bibinfo
  {journal} {Phys. Rev. D}\ }\textbf {\bibinfo {volume} {98}},\ \bibinfo
  {pages} {054517} (\bibinfo {year} {2018})},\ \Eprint
  {http://arxiv.org/abs/1802.04248} {arXiv:1802.04248 [hep-lat]} \BibitemShut
  {NoStop}%
\bibitem [{\citenamefont {Carrasco}\ \emph {et~al.}(2014)\citenamefont
  {Carrasco} \emph {et~al.}}]{EuropeanTwistedMass:2014osg}%
  \BibitemOpen
  \bibfield  {author} {\bibinfo {author} {\bibfnamefont {N.}~\bibnamefont
  {Carrasco}} \emph {et~al.} (\bibinfo {collaboration} {European Twisted
  Mass}),\ }\bibfield  {title} {\enquote {\bibinfo {title} {{Up, down, strange
  and charm quark masses with N$_f$ = 2+1+1 twisted mass lattice QCD}},}\
  }\href {\doibase 10.1016/j.nuclphysb.2014.07.025} {\bibfield  {journal}
  {\bibinfo  {journal} {Nucl. Phys. B}\ }\textbf {\bibinfo {volume} {887}},\
  \bibinfo {pages} {19--68} (\bibinfo {year} {2014})},\ \Eprint
  {http://arxiv.org/abs/1403.4504} {arXiv:1403.4504 [hep-lat]} \BibitemShut
  {NoStop}%
\bibitem [{\citenamefont {Giusti}\ \emph {et~al.}(2017)\citenamefont {Giusti},
  \citenamefont {Lubicz}, \citenamefont {Tarantino}, \citenamefont
  {Martinelli}, \citenamefont {Sanfilippo}, \citenamefont {Simula},\ and\
  \citenamefont {Tantalo}}]{Giusti:2017dmp}%
  \BibitemOpen
  \bibfield  {author} {\bibinfo {author} {\bibfnamefont {D.}~\bibnamefont
  {Giusti}}, \bibinfo {author} {\bibfnamefont {V.}~\bibnamefont {Lubicz}},
  \bibinfo {author} {\bibfnamefont {C.}~\bibnamefont {Tarantino}}, \bibinfo
  {author} {\bibfnamefont {G.}~\bibnamefont {Martinelli}}, \bibinfo {author}
  {\bibfnamefont {F.}~\bibnamefont {Sanfilippo}}, \bibinfo {author}
  {\bibfnamefont {S.}~\bibnamefont {Simula}}, \ and\ \bibinfo {author}
  {\bibfnamefont {N.}~\bibnamefont {Tantalo}},\ }\bibfield  {title} {\enquote
  {\bibinfo {title} {{Leading isospin-breaking corrections to pion, kaon and
  charmed-meson masses with Twisted-Mass fermions}},}\ }\href {\doibase
  10.1103/PhysRevD.95.114504} {\bibfield  {journal} {\bibinfo  {journal} {Phys.
  Rev. D}\ }\textbf {\bibinfo {volume} {95}},\ \bibinfo {pages} {114504}
  (\bibinfo {year} {2017})},\ \Eprint {http://arxiv.org/abs/1704.06561}
  {arXiv:1704.06561 [hep-lat]} \BibitemShut {NoStop}%
\bibitem [{\citenamefont {Lytle}\ \emph {et~al.}(2018)\citenamefont {Lytle},
  \citenamefont {Davies}, \citenamefont {Hatton}, \citenamefont {Lepage},\ and\
  \citenamefont {Sturm}}]{Lytle:2018evc}%
  \BibitemOpen
  \bibfield  {author} {\bibinfo {author} {\bibfnamefont {A.~T.}\ \bibnamefont
  {Lytle}}, \bibinfo {author} {\bibfnamefont {C.~T.~H.}\ \bibnamefont
  {Davies}}, \bibinfo {author} {\bibfnamefont {D.}~\bibnamefont {Hatton}},
  \bibinfo {author} {\bibfnamefont {G.~P.}\ \bibnamefont {Lepage}}, \ and\
  \bibinfo {author} {\bibfnamefont {C.}~\bibnamefont {Sturm}} (\bibinfo
  {collaboration} {HPQCD}),\ }\bibfield  {title} {\enquote {\bibinfo {title}
  {{Determination of quark masses from $\mathbf{n_f=4}$ lattice QCD and the
  RI-SMOM intermediate scheme}},}\ }\href {\doibase 10.1103/PhysRevD.98.014513}
  {\bibfield  {journal} {\bibinfo  {journal} {Phys. Rev. D}\ }\textbf {\bibinfo
  {volume} {98}},\ \bibinfo {pages} {014513} (\bibinfo {year} {2018})},\
  \Eprint {http://arxiv.org/abs/1805.06225} {arXiv:1805.06225 [hep-lat]}
  \BibitemShut {NoStop}%
\bibitem [{\citenamefont {Chakraborty}\ \emph {et~al.}(2015)\citenamefont
  {Chakraborty}, \citenamefont {Davies}, \citenamefont {Galloway},
  \citenamefont {Knecht}, \citenamefont {Koponen}, \citenamefont {Donald},
  \citenamefont {Dowdall}, \citenamefont {Lepage},\ and\ \citenamefont
  {McNeile}}]{Chakraborty:2014aca}%
  \BibitemOpen
  \bibfield  {author} {\bibinfo {author} {\bibfnamefont {Bipasha}\ \bibnamefont
  {Chakraborty}}, \bibinfo {author} {\bibfnamefont {C.~T.~H.}\ \bibnamefont
  {Davies}}, \bibinfo {author} {\bibfnamefont {B.}~\bibnamefont {Galloway}},
  \bibinfo {author} {\bibfnamefont {P.}~\bibnamefont {Knecht}}, \bibinfo
  {author} {\bibfnamefont {J.}~\bibnamefont {Koponen}}, \bibinfo {author}
  {\bibfnamefont {G.~C.}\ \bibnamefont {Donald}}, \bibinfo {author}
  {\bibfnamefont {R.~J.}\ \bibnamefont {Dowdall}}, \bibinfo {author}
  {\bibfnamefont {G.~P.}\ \bibnamefont {Lepage}}, \ and\ \bibinfo {author}
  {\bibfnamefont {C.}~\bibnamefont {McNeile}},\ }\bibfield  {title} {\enquote
  {\bibinfo {title} {{High-precision quark masses and QCD coupling from $n_f=4$
  lattice QCD}},}\ }\href {\doibase 10.1103/PhysRevD.91.054508} {\bibfield
  {journal} {\bibinfo  {journal} {Phys. Rev. D}\ }\textbf {\bibinfo {volume}
  {91}},\ \bibinfo {pages} {054508} (\bibinfo {year} {2015})},\ \Eprint
  {http://arxiv.org/abs/1408.4169} {arXiv:1408.4169 [hep-lat]} \BibitemShut
  {NoStop}%
\bibitem [{\citenamefont {Bruno}\ \emph {et~al.}(2020)\citenamefont {Bruno},
  \citenamefont {Campos}, \citenamefont {Fritzsch}, \citenamefont {Koponen},
  \citenamefont {Pena}, \citenamefont {Preti}, \citenamefont {Ramos},\ and\
  \citenamefont {Vladikas}}]{Bruno:2019vup}%
  \BibitemOpen
  \bibfield  {author} {\bibinfo {author} {\bibfnamefont {Mattia}\ \bibnamefont
  {Bruno}}, \bibinfo {author} {\bibfnamefont {Isabel}\ \bibnamefont {Campos}},
  \bibinfo {author} {\bibfnamefont {Patrick}\ \bibnamefont {Fritzsch}},
  \bibinfo {author} {\bibfnamefont {Jonna}\ \bibnamefont {Koponen}}, \bibinfo
  {author} {\bibfnamefont {Carlos}\ \bibnamefont {Pena}}, \bibinfo {author}
  {\bibfnamefont {David}\ \bibnamefont {Preti}}, \bibinfo {author}
  {\bibfnamefont {Alberto}\ \bibnamefont {Ramos}}, \ and\ \bibinfo {author}
  {\bibfnamefont {Anastassios}\ \bibnamefont {Vladikas}} (\bibinfo
  {collaboration} {ALPHA}),\ }\bibfield  {title} {\enquote {\bibinfo {title}
  {{Light quark masses in ${N_\mathrm{f}=2+1}$ lattice QCD with Wilson
  fermions}},}\ }\href {\doibase 10.1140/epjc/s10052-020-7698-z} {\bibfield
  {journal} {\bibinfo  {journal} {Eur. Phys. J. C}\ }\textbf {\bibinfo {volume}
  {80}},\ \bibinfo {pages} {169} (\bibinfo {year} {2020})},\ \Eprint
  {http://arxiv.org/abs/1911.08025} {arXiv:1911.08025 [hep-lat]} \BibitemShut
  {NoStop}%
\bibitem [{\citenamefont {Durr}\ \emph
  {et~al.}(2011{\natexlab{a}})\citenamefont {Durr}, \citenamefont {Fodor},
  \citenamefont {Hoelbling}, \citenamefont {Katz}, \citenamefont {Krieg},
  \citenamefont {Kurth}, \citenamefont {Lellouch}, \citenamefont {Lippert},
  \citenamefont {Szabo},\ and\ \citenamefont {Vulvert}}]{Durr:2010aw}%
  \BibitemOpen
  \bibfield  {author} {\bibinfo {author} {\bibfnamefont {S.}~\bibnamefont
  {Durr}}, \bibinfo {author} {\bibfnamefont {Z.}~\bibnamefont {Fodor}},
  \bibinfo {author} {\bibfnamefont {C.}~\bibnamefont {Hoelbling}}, \bibinfo
  {author} {\bibfnamefont {S.~D.}\ \bibnamefont {Katz}}, \bibinfo {author}
  {\bibfnamefont {S.}~\bibnamefont {Krieg}}, \bibinfo {author} {\bibfnamefont
  {T.}~\bibnamefont {Kurth}}, \bibinfo {author} {\bibfnamefont
  {L.}~\bibnamefont {Lellouch}}, \bibinfo {author} {\bibfnamefont
  {T.}~\bibnamefont {Lippert}}, \bibinfo {author} {\bibfnamefont {K.~K.}\
  \bibnamefont {Szabo}}, \ and\ \bibinfo {author} {\bibfnamefont
  {G.}~\bibnamefont {Vulvert}},\ }\bibfield  {title} {\enquote {\bibinfo
  {title} {{Lattice QCD at the physical point: Simulation and analysis
  details}},}\ }\href {\doibase 10.1007/JHEP08(2011)148} {\bibfield  {journal}
  {\bibinfo  {journal} {JHEP}\ }\textbf {\bibinfo {volume} {08}},\ \bibinfo
  {pages} {148} (\bibinfo {year} {2011}{\natexlab{a}})},\ \Eprint
  {http://arxiv.org/abs/1011.2711} {arXiv:1011.2711 [hep-lat]} \BibitemShut
  {NoStop}%
\bibitem [{\citenamefont {Durr}\ \emph
  {et~al.}(2011{\natexlab{b}})\citenamefont {Durr}, \citenamefont {Fodor},
  \citenamefont {Hoelbling}, \citenamefont {Katz}, \citenamefont {Krieg},
  \citenamefont {Kurth}, \citenamefont {Lellouch}, \citenamefont {Lippert},
  \citenamefont {Szabo},\ and\ \citenamefont {Vulvert}}]{Durr:2010vn}%
  \BibitemOpen
  \bibfield  {author} {\bibinfo {author} {\bibfnamefont {S.}~\bibnamefont
  {Durr}}, \bibinfo {author} {\bibfnamefont {Z.}~\bibnamefont {Fodor}},
  \bibinfo {author} {\bibfnamefont {C.}~\bibnamefont {Hoelbling}}, \bibinfo
  {author} {\bibfnamefont {S.~D.}\ \bibnamefont {Katz}}, \bibinfo {author}
  {\bibfnamefont {S.}~\bibnamefont {Krieg}}, \bibinfo {author} {\bibfnamefont
  {T.}~\bibnamefont {Kurth}}, \bibinfo {author} {\bibfnamefont
  {L.}~\bibnamefont {Lellouch}}, \bibinfo {author} {\bibfnamefont
  {T.}~\bibnamefont {Lippert}}, \bibinfo {author} {\bibfnamefont {K.~K.}\
  \bibnamefont {Szabo}}, \ and\ \bibinfo {author} {\bibfnamefont
  {G.}~\bibnamefont {Vulvert}},\ }\bibfield  {title} {\enquote {\bibinfo
  {title} {{Lattice QCD at the physical point: light quark masses}},}\ }\href
  {\doibase 10.1016/j.physletb.2011.05.053} {\bibfield  {journal} {\bibinfo
  {journal} {Phys. Lett. B}\ }\textbf {\bibinfo {volume} {701}},\ \bibinfo
  {pages} {265--268} (\bibinfo {year} {2011}{\natexlab{b}})},\ \Eprint
  {http://arxiv.org/abs/1011.2403} {arXiv:1011.2403 [hep-lat]} \BibitemShut
  {NoStop}%
\bibitem [{\citenamefont {Bazavov}\ \emph {et~al.}(2010)\citenamefont {Bazavov}
  \emph {et~al.}}]{Bazavov:2010yq}%
  \BibitemOpen
  \bibfield  {author} {\bibinfo {author} {\bibfnamefont {A.}~\bibnamefont
  {Bazavov}} \emph {et~al.},\ }\bibfield  {title} {\enquote {\bibinfo {title}
  {{Staggered chiral perturbation theory in the two-flavor case and SU(2)
  analysis of the MILC data}},}\ }\href@noop {} {\bibfield  {journal} {\bibinfo
   {journal} {PoS}\ }\textbf {\bibinfo {volume} {LATTICE2010}},\ \bibinfo
  {pages} {083} (\bibinfo {year} {2010})},\ \Eprint
  {http://arxiv.org/abs/1011.1792} {arXiv:1011.1792 [hep-lat]} \BibitemShut
  {NoStop}%
\bibitem [{\citenamefont {Bruno}\ \emph {et~al.}(2019)\citenamefont {Bruno},
  \citenamefont {Campos}, \citenamefont {Koponen}, \citenamefont {Pena},
  \citenamefont {Preti}, \citenamefont {Ramos},\ and\ \citenamefont
  {Vladikas}}]{Bruno:2019xed}%
  \BibitemOpen
  \bibfield  {author} {\bibinfo {author} {\bibfnamefont {M.}~\bibnamefont
  {Bruno}}, \bibinfo {author} {\bibfnamefont {I.}~\bibnamefont {Campos}},
  \bibinfo {author} {\bibfnamefont {J.}~\bibnamefont {Koponen}}, \bibinfo
  {author} {\bibfnamefont {Carlos}\ \bibnamefont {Pena}}, \bibinfo {author}
  {\bibfnamefont {David}\ \bibnamefont {Preti}}, \bibinfo {author}
  {\bibfnamefont {Alberto}\ \bibnamefont {Ramos}}, \ and\ \bibinfo {author}
  {\bibfnamefont {Anastassios}\ \bibnamefont {Vladikas}} (\bibinfo
  {collaboration} {ALPHA}),\ }\bibfield  {title} {\enquote {\bibinfo {title}
  {{Light and strange quark masses from $N_f=2+1$ simulations with Wilson
  fermions}},}\ }\href {\doibase 10.22323/1.334.0220} {\bibfield  {journal}
  {\bibinfo  {journal} {PoS}\ }\textbf {\bibinfo {volume} {LATTICE2018}},\
  \bibinfo {pages} {220} (\bibinfo {year} {2019})},\ \Eprint
  {http://arxiv.org/abs/1903.04094} {arXiv:1903.04094 [hep-lat]} \BibitemShut
  {NoStop}%
\bibitem [{\citenamefont {Bazavov}\ \emph {et~al.}(2009)\citenamefont {Bazavov}
  \emph {et~al.}}]{MILC:2009ltw}%
  \BibitemOpen
  \bibfield  {author} {\bibinfo {author} {\bibfnamefont {A.}~\bibnamefont
  {Bazavov}} \emph {et~al.} (\bibinfo {collaboration} {MILC}),\ }\bibfield
  {title} {\enquote {\bibinfo {title} {{MILC results for light
  pseudoscalars}},}\ }\href {\doibase 10.22323/1.086.0007} {\bibfield
  {journal} {\bibinfo  {journal} {PoS}\ }\textbf {\bibinfo {volume} {CD09}},\
  \bibinfo {pages} {007} (\bibinfo {year} {2009})},\ \Eprint
  {http://arxiv.org/abs/0910.2966} {arXiv:0910.2966 [hep-ph]} \BibitemShut
  {NoStop}%
\bibitem [{\citenamefont {McNeile}\ \emph {et~al.}(2010)\citenamefont
  {McNeile}, \citenamefont {Davies}, \citenamefont {Follana}, \citenamefont
  {Hornbostel},\ and\ \citenamefont {Lepage}}]{McNeile:2010ji}%
  \BibitemOpen
  \bibfield  {author} {\bibinfo {author} {\bibfnamefont {C.}~\bibnamefont
  {McNeile}}, \bibinfo {author} {\bibfnamefont {C.~T.~H.}\ \bibnamefont
  {Davies}}, \bibinfo {author} {\bibfnamefont {E.}~\bibnamefont {Follana}},
  \bibinfo {author} {\bibfnamefont {K.}~\bibnamefont {Hornbostel}}, \ and\
  \bibinfo {author} {\bibfnamefont {G.~P.}\ \bibnamefont {Lepage}},\ }\bibfield
   {title} {\enquote {\bibinfo {title} {{High-Precision c and b Masses, and QCD
  Coupling from Current-Current Correlators in Lattice and Continuum QCD}},}\
  }\href {\doibase 10.1103/PhysRevD.82.034512} {\bibfield  {journal} {\bibinfo
  {journal} {Phys. Rev. D}\ }\textbf {\bibinfo {volume} {82}},\ \bibinfo
  {pages} {034512} (\bibinfo {year} {2010})},\ \Eprint
  {http://arxiv.org/abs/1004.4285} {arXiv:1004.4285 [hep-lat]} \BibitemShut
  {NoStop}%
\bibitem [{\citenamefont {Durr}\ \emph
  {et~al.}(2011{\natexlab{c}})\citenamefont {Durr} \emph
  {et~al.}}]{BMW:2011zrh}%
  \BibitemOpen
  \bibfield  {author} {\bibinfo {author} {\bibfnamefont {S.}~\bibnamefont
  {Durr}} \emph {et~al.} (\bibinfo {collaboration} {BMW}),\ }\bibfield  {title}
  {\enquote {\bibinfo {title} {{Precision computation of the kaon bag
  parameter}},}\ }\href {\doibase 10.1016/j.physletb.2011.10.043} {\bibfield
  {journal} {\bibinfo  {journal} {Phys. Lett. B}\ }\textbf {\bibinfo {volume}
  {705}},\ \bibinfo {pages} {477--481} (\bibinfo {year}
  {2011}{\natexlab{c}})},\ \Eprint {http://arxiv.org/abs/1106.3230}
  {arXiv:1106.3230 [hep-lat]} \BibitemShut {NoStop}%
\bibitem [{\citenamefont {Laiho}\ and\ \citenamefont {Van~de
  Water}(2011)}]{Laiho:2011np}%
  \BibitemOpen
  \bibfield  {author} {\bibinfo {author} {\bibfnamefont {Jack}\ \bibnamefont
  {Laiho}}\ and\ \bibinfo {author} {\bibfnamefont {Ruth~S.}\ \bibnamefont
  {Van~de Water}},\ }\bibfield  {title} {\enquote {\bibinfo {title}
  {{Pseudoscalar decay constants, light-quark masses, and $B_K$ from
  mixed-action lattice QCD}},}\ }\href {\doibase 10.22323/1.139.0293}
  {\bibfield  {journal} {\bibinfo  {journal} {PoS}\ }\textbf {\bibinfo {volume}
  {LATTICE2011}},\ \bibinfo {pages} {293} (\bibinfo {year} {2011})},\ \Eprint
  {http://arxiv.org/abs/1112.4861} {arXiv:1112.4861 [hep-lat]} \BibitemShut
  {NoStop}%
\bibitem [{\citenamefont {Tsang}(2023)}]{Tsang:2022smt}%
  \BibitemOpen
  \bibfield  {author} {\bibinfo {author} {\bibfnamefont {Justus~Tobias}\
  \bibnamefont {Tsang}},\ }\bibfield  {title} {\enquote {\bibinfo {title}
  {{Neutral meson mixing in the $B^0_{(s)}$ sector from Lattice QCD}},}\ }\href
  {\doibase 10.22323/1.411.0107} {\bibfield  {journal} {\bibinfo  {journal}
  {PoS}\ }\textbf {\bibinfo {volume} {CKM2021}},\ \bibinfo {pages} {107}
  (\bibinfo {year} {2023})},\ \Eprint {http://arxiv.org/abs/2204.01259}
  {arXiv:2204.01259 [hep-lat]} \BibitemShut {NoStop}%
\bibitem [{\citenamefont {Boyle}\ \emph {et~al.}(2022)\citenamefont {Boyle},
  \citenamefont {Erben}, \citenamefont {J\"uttner}, \citenamefont {Kaneko},
  \citenamefont {Marshall}, \citenamefont {Portelli}, \citenamefont {Witzel},
  \citenamefont {Del~Debbio},\ and\ \citenamefont {Tsang}}]{Boyle:2021kqn}%
  \BibitemOpen
  \bibfield  {author} {\bibinfo {author} {\bibfnamefont {Peter}\ \bibnamefont
  {Boyle}}, \bibinfo {author} {\bibfnamefont {Felix}\ \bibnamefont {Erben}},
  \bibinfo {author} {\bibfnamefont {Andreas}\ \bibnamefont {J\"uttner}},
  \bibinfo {author} {\bibfnamefont {Takashi}\ \bibnamefont {Kaneko}}, \bibinfo
  {author} {\bibfnamefont {Michael}\ \bibnamefont {Marshall}}, \bibinfo
  {author} {\bibfnamefont {Antonin}\ \bibnamefont {Portelli}}, \bibinfo
  {author} {\bibfnamefont {Oliver}\ \bibnamefont {Witzel}}, \bibinfo {author}
  {\bibfnamefont {Luigi}\ \bibnamefont {Del~Debbio}}, \ and\ \bibinfo {author}
  {\bibfnamefont {Justus~Tobias}\ \bibnamefont {Tsang}},\ }\bibfield  {title}
  {\enquote {\bibinfo {title} {{BSM $B - \bar{B}$ mixing on JLQCD and RBC/UKQCD
  $N_f=2+1$ DWF ensembles}},}\ }\href {\doibase 10.22323/1.396.0224} {\bibfield
   {journal} {\bibinfo  {journal} {PoS}\ }\textbf {\bibinfo {volume}
  {LATTICE2021}},\ \bibinfo {pages} {224} (\bibinfo {year} {2022})},\ \Eprint
  {http://arxiv.org/abs/2111.11287} {arXiv:2111.11287 [hep-lat]} \BibitemShut
  {NoStop}%
\bibitem [{\citenamefont {Aoki}\ \emph
  {et~al.}(2011{\natexlab{c}})\citenamefont {Aoki} \emph
  {et~al.}}]{RBC:2010qam}%
  \BibitemOpen
  \bibfield  {author} {\bibinfo {author} {\bibfnamefont {Y.}~\bibnamefont
  {Aoki}} \emph {et~al.} (\bibinfo {collaboration} {RBC, UKQCD}),\ }\bibfield
  {title} {\enquote {\bibinfo {title} {{Continuum Limit Physics from 2+1 Flavor
  Domain Wall QCD}},}\ }\href {\doibase 10.1103/PhysRevD.83.074508} {\bibfield
  {journal} {\bibinfo  {journal} {Phys. Rev. D}\ }\textbf {\bibinfo {volume}
  {83}},\ \bibinfo {pages} {074508} (\bibinfo {year} {2011}{\natexlab{c}})},\
  \Eprint {http://arxiv.org/abs/1011.0892} {arXiv:1011.0892 [hep-lat]}
  \BibitemShut {NoStop}%
\bibitem [{\citenamefont {Urbach}\ \emph {et~al.}(2006)\citenamefont {Urbach},
  \citenamefont {Jansen}, \citenamefont {Shindler},\ and\ \citenamefont
  {Wenger}}]{Urbach:2005ji}%
  \BibitemOpen
  \bibfield  {author} {\bibinfo {author} {\bibfnamefont {C.}~\bibnamefont
  {Urbach}}, \bibinfo {author} {\bibfnamefont {K.}~\bibnamefont {Jansen}},
  \bibinfo {author} {\bibfnamefont {A.}~\bibnamefont {Shindler}}, \ and\
  \bibinfo {author} {\bibfnamefont {U.}~\bibnamefont {Wenger}},\ }\bibfield
  {title} {\enquote {\bibinfo {title} {{HMC algorithm with multiple time scale
  integration and mass preconditioning}},}\ }\href {\doibase
  10.1016/j.cpc.2005.08.006} {\bibfield  {journal} {\bibinfo  {journal}
  {Comput. Phys. Commun.}\ }\textbf {\bibinfo {volume} {174}},\ \bibinfo
  {pages} {87--98} (\bibinfo {year} {2006})},\ \Eprint
  {http://arxiv.org/abs/hep-lat/0506011} {arXiv:hep-lat/0506011} \BibitemShut
  {NoStop}%
\bibitem [{\citenamefont {Chen}\ and\ \citenamefont
  {Chiu}(2014)}]{Chen:2014hyy}%
  \BibitemOpen
  \bibfield  {author} {\bibinfo {author} {\bibfnamefont {Yu-Chih}\ \bibnamefont
  {Chen}}\ and\ \bibinfo {author} {\bibfnamefont {Ting-Wai}\ \bibnamefont
  {Chiu}} (\bibinfo {collaboration} {TWQCD}),\ }\bibfield  {title} {\enquote
  {\bibinfo {title} {{Exact Pseudofermion Action for Monte Carlo Simulation of
  Domain-Wall Fermion}},}\ }\href {\doibase 10.1016/j.physletb.2014.09.016}
  {\bibfield  {journal} {\bibinfo  {journal} {Phys. Lett. B}\ }\textbf
  {\bibinfo {volume} {738}},\ \bibinfo {pages} {55--60} (\bibinfo {year}
  {2014})},\ \Eprint {http://arxiv.org/abs/1403.1683} {arXiv:1403.1683
  [hep-lat]} \BibitemShut {NoStop}%
\bibitem [{\citenamefont {Jung}\ \emph {et~al.}(2018)\citenamefont {Jung},
  \citenamefont {Kelly}, \citenamefont {Mawhinney},\ and\ \citenamefont
  {Murphy}}]{Jung:2017xef}%
  \BibitemOpen
  \bibfield  {author} {\bibinfo {author} {\bibfnamefont {C.}~\bibnamefont
  {Jung}}, \bibinfo {author} {\bibfnamefont {C.}~\bibnamefont {Kelly}},
  \bibinfo {author} {\bibfnamefont {R.~D.}\ \bibnamefont {Mawhinney}}, \ and\
  \bibinfo {author} {\bibfnamefont {D.~J.}\ \bibnamefont {Murphy}},\ }\bibfield
   {title} {\enquote {\bibinfo {title} {{Domain Wall Fermion QCD with the Exact
  One Flavor Algorithm}},}\ }\href {\doibase 10.1103/PhysRevD.97.054503}
  {\bibfield  {journal} {\bibinfo  {journal} {Phys. Rev. D}\ }\textbf {\bibinfo
  {volume} {97}},\ \bibinfo {pages} {054503} (\bibinfo {year} {2018})},\
  \Eprint {http://arxiv.org/abs/1706.05843} {arXiv:1706.05843 [hep-lat]}
  \BibitemShut {NoStop}%
\bibitem [{\citenamefont {Clark}\ \emph {et~al.}(2011)\citenamefont {Clark},
  \citenamefont {Joo}, \citenamefont {Kennedy},\ and\ \citenamefont
  {Silva}}]{Clark:2011ir}%
  \BibitemOpen
  \bibfield  {author} {\bibinfo {author} {\bibfnamefont {M.~A.}\ \bibnamefont
  {Clark}}, \bibinfo {author} {\bibfnamefont {Balint}\ \bibnamefont {Joo}},
  \bibinfo {author} {\bibfnamefont {A.~D.}\ \bibnamefont {Kennedy}}, \ and\
  \bibinfo {author} {\bibfnamefont {P.~J.}\ \bibnamefont {Silva}},\ }\bibfield
  {title} {\enquote {\bibinfo {title} {{Improving dynamical lattice QCD
  simulations through integrator tuning using Poisson brackets and a
  force-gradient integrator}},}\ }\href {\doibase 10.1103/PhysRevD.84.071502}
  {\bibfield  {journal} {\bibinfo  {journal} {Phys. Rev. D}\ }\textbf {\bibinfo
  {volume} {84}},\ \bibinfo {pages} {071502} (\bibinfo {year} {2011})},\
  \Eprint {http://arxiv.org/abs/1108.1828} {arXiv:1108.1828 [hep-lat]}
  \BibitemShut {NoStop}%
\bibitem [{\citenamefont {Yin}\ and\ \citenamefont
  {Mawhinney}(2011)}]{Yin:2011np}%
  \BibitemOpen
  \bibfield  {author} {\bibinfo {author} {\bibfnamefont {Hantao}\ \bibnamefont
  {Yin}}\ and\ \bibinfo {author} {\bibfnamefont {Robert~D.}\ \bibnamefont
  {Mawhinney}},\ }\bibfield  {title} {\enquote {\bibinfo {title} {{Improving
  DWF Simulations: the Force Gradient Integrator and the M\"obius Accelerated
  DWF Solver}},}\ }\href {\doibase 10.22323/1.139.0051} {\bibfield  {journal}
  {\bibinfo  {journal} {PoS}\ }\textbf {\bibinfo {volume} {LATTICE2011}},\
  \bibinfo {pages} {051} (\bibinfo {year} {2011})},\ \Eprint
  {http://arxiv.org/abs/1111.5059} {arXiv:1111.5059 [hep-lat]} \BibitemShut
  {NoStop}%
\bibitem [{\citenamefont {Fierz}(1937)}]{Fierz:1937wjm}%
  \BibitemOpen
  \bibfield  {author} {\bibinfo {author} {\bibfnamefont {Markus}\ \bibnamefont
  {Fierz}},\ }\bibfield  {title} {\enquote {\bibinfo {title} {{Zur Fermischen
  Theorie des $\beta$-Zerfalls}},}\ }\href {\doibase 10.1007/bf01330070}
  {\bibfield  {journal} {\bibinfo  {journal} {Z. Phys.}\ }\textbf {\bibinfo
  {volume} {104}},\ \bibinfo {pages} {553--565} (\bibinfo {year}
  {1937})}\BibitemShut {NoStop}%
\bibitem [{\citenamefont {Garron}(2018)}]{Garron:2018tst}%
  \BibitemOpen
  \bibfield  {author} {\bibinfo {author} {\bibfnamefont {Nicolas}\ \bibnamefont
  {Garron}},\ }\bibfield  {title} {\enquote {\bibinfo {title} {{Fierz
  transformations and renormalization schemes for fourquark operators}},}\
  }\href {\doibase 10.1051/epjconf/201817510005} {\bibfield  {journal}
  {\bibinfo  {journal} {EPJ Web Conf.}\ }\textbf {\bibinfo {volume} {175}},\
  \bibinfo {pages} {10005} (\bibinfo {year} {2018})}\BibitemShut {NoStop}%
\bibitem [{\citenamefont {Aoki}\ \emph {et~al.}(2008)\citenamefont {Aoki},
  \citenamefont {Boyle}, \citenamefont {Christ}, \citenamefont {Dawson},
  \citenamefont {Donnellan} \emph {et~al.}}]{Aoki:2007xm}%
  \BibitemOpen
  \bibfield  {author} {\bibinfo {author} {\bibfnamefont {Y.}~\bibnamefont
  {Aoki}}, \bibinfo {author} {\bibfnamefont {P.A.}\ \bibnamefont {Boyle}},
  \bibinfo {author} {\bibfnamefont {N.H.}\ \bibnamefont {Christ}}, \bibinfo
  {author} {\bibfnamefont {C.}~\bibnamefont {Dawson}}, \bibinfo {author}
  {\bibfnamefont {M.A.}\ \bibnamefont {Donnellan}},  \emph {et~al.},\
  }\bibfield  {title} {\enquote {\bibinfo {title} {{Non-perturbative
  renormalization of quark bilinear operators and B(K) using domain wall
  fermions}},}\ }\href {\doibase 10.1103/PhysRevD.78.054510} {\bibfield
  {journal} {\bibinfo  {journal} {Phys.Rev.}\ }\textbf {\bibinfo {volume}
  {D78}},\ \bibinfo {pages} {054510} (\bibinfo {year} {2008})},\ \Eprint
  {http://arxiv.org/abs/0712.1061} {arXiv:0712.1061 [hep-lat]} \BibitemShut
  {NoStop}%
\bibitem [{\citenamefont {Boyle}\ \emph
  {et~al.}(2017{\natexlab{c}})\citenamefont {Boyle}, \citenamefont
  {Del~Debbio},\ and\ \citenamefont {Khamseh}}]{Boyle:2016wis}%
  \BibitemOpen
  \bibfield  {author} {\bibinfo {author} {\bibfnamefont {Peter}\ \bibnamefont
  {Boyle}}, \bibinfo {author} {\bibfnamefont {Luigi}\ \bibnamefont
  {Del~Debbio}}, \ and\ \bibinfo {author} {\bibfnamefont {Ava}\ \bibnamefont
  {Khamseh}},\ }\bibfield  {title} {\enquote {\bibinfo {title} {{Massive
  momentum-subtraction scheme}},}\ }\href {\doibase 10.1103/PhysRevD.95.054505}
  {\bibfield  {journal} {\bibinfo  {journal} {Phys. Rev. D}\ }\textbf {\bibinfo
  {volume} {95}},\ \bibinfo {pages} {054505} (\bibinfo {year}
  {2017}{\natexlab{c}})},\ \Eprint {http://arxiv.org/abs/1611.06908}
  {arXiv:1611.06908 [hep-lat]} \BibitemShut {NoStop}%
\bibitem [{\citenamefont {Del~Debbio}\ \emph {et~al.}(2023)\citenamefont
  {Del~Debbio}, \citenamefont {Erben}, \citenamefont {Flynn}, \citenamefont
  {Mukherjee},\ and\ \citenamefont {Tsang}}]{DelDebbio:2023naa}%
  \BibitemOpen
  \bibfield  {author} {\bibinfo {author} {\bibfnamefont {Luigi}\ \bibnamefont
  {Del~Debbio}}, \bibinfo {author} {\bibfnamefont {Felix}\ \bibnamefont
  {Erben}}, \bibinfo {author} {\bibfnamefont {Jonathan}\ \bibnamefont {Flynn}},
  \bibinfo {author} {\bibfnamefont {Rajnandini}\ \bibnamefont {Mukherjee}}, \
  and\ \bibinfo {author} {\bibfnamefont {J.~Tobias}\ \bibnamefont {Tsang}},\
  }\bibfield  {title} {\enquote {\bibinfo {title} {{Charm quark mass using a
  massive nonperturbative renormalisation scheme}},}\ }\href@noop {} {\
  (\bibinfo {year} {2023})},\ \Eprint {http://arxiv.org/abs/2312.16537}
  {arXiv:2312.16537 [hep-lat]} \BibitemShut {NoStop}%
\bibitem [{\citenamefont {Garron}\ \emph {et~al.}(2022)\citenamefont {Garron},
  \citenamefont {Cahill}, \citenamefont {Gorbahn}, \citenamefont {Gracey},\
  and\ \citenamefont {Rakow}}]{Garron:2022fnn}%
  \BibitemOpen
  \bibfield  {author} {\bibinfo {author} {\bibfnamefont {Nicolas}\ \bibnamefont
  {Garron}}, \bibinfo {author} {\bibfnamefont {Caroline}\ \bibnamefont
  {Cahill}}, \bibinfo {author} {\bibfnamefont {Martin}\ \bibnamefont
  {Gorbahn}}, \bibinfo {author} {\bibfnamefont {J.~A.}\ \bibnamefont {Gracey}},
  \ and\ \bibinfo {author} {\bibfnamefont {P.~E.~L.}\ \bibnamefont {Rakow}},\
  }\bibfield  {title} {\enquote {\bibinfo {title} {{Exploring interpolating
  momentum schemes}},}\ }\href {\doibase 10.22323/1.396.0438} {\bibfield
  {journal} {\bibinfo  {journal} {PoS}\ }\textbf {\bibinfo {volume}
  {LATTICE2021}},\ \bibinfo {pages} {438} (\bibinfo {year} {2022})},\ \Eprint
  {http://arxiv.org/abs/2202.04394} {arXiv:2202.04394 [hep-lat]} \BibitemShut
  {NoStop}%
\bibitem [{Mat(2024)}]{Mathematica}%
  \BibitemOpen
  \href {https://www.wolfram.com/mathematica} {\enquote {\bibinfo {title}
  {Mathematica, {V}ersion 14.0},}\ } (\bibinfo {year} {2024}),\ \bibinfo {note}
  {{Wolfram Research} Inc., Champaign, IL}\BibitemShut {NoStop}%
\bibitem [{\citenamefont {Ciuchini}\ \emph {et~al.}(1998)\citenamefont
  {Ciuchini}, \citenamefont {Franco}, \citenamefont {Lubicz}, \citenamefont
  {Martinelli}, \citenamefont {Scimemi},\ and\ \citenamefont
  {Silvestrini}}]{Ciuchini:1997bw}%
  \BibitemOpen
  \bibfield  {author} {\bibinfo {author} {\bibfnamefont {Marco}\ \bibnamefont
  {Ciuchini}}, \bibinfo {author} {\bibfnamefont {E.}~\bibnamefont {Franco}},
  \bibinfo {author} {\bibfnamefont {V.}~\bibnamefont {Lubicz}}, \bibinfo
  {author} {\bibfnamefont {G.}~\bibnamefont {Martinelli}}, \bibinfo {author}
  {\bibfnamefont {I.}~\bibnamefont {Scimemi}}, \ and\ \bibinfo {author}
  {\bibfnamefont {L.}~\bibnamefont {Silvestrini}},\ }\bibfield  {title}
  {\enquote {\bibinfo {title} {Next-to-leading order {QCD} corrections to
  {$\Delta F = 2$} effective {Hamiltonians}},}\ }\href {\doibase
  10.1016/S0550-3213(98)00161-8} {\bibfield  {journal} {\bibinfo  {journal}
  {Nucl. Phys. B}\ }\textbf {\bibinfo {volume} {523}},\ \bibinfo {pages}
  {501--525} (\bibinfo {year} {1998})},\ \Eprint
  {http://arxiv.org/abs/hep-ph/9711402} {arXiv:hep-ph/9711402} \BibitemShut
  {NoStop}%
\bibitem [{\citenamefont {Buras}\ \emph {et~al.}(2000)\citenamefont {Buras},
  \citenamefont {Misiak},\ and\ \citenamefont {Urban}}]{Buras:2000if}%
  \BibitemOpen
  \bibfield  {author} {\bibinfo {author} {\bibfnamefont {Andrzej~J.}\
  \bibnamefont {Buras}}, \bibinfo {author} {\bibfnamefont {Mikolaj}\
  \bibnamefont {Misiak}}, \ and\ \bibinfo {author} {\bibfnamefont {Jorg}\
  \bibnamefont {Urban}},\ }\bibfield  {title} {\enquote {\bibinfo {title} {{Two
  loop QCD anomalous dimensions of flavor changing four quark operators within
  and beyond the standard model}},}\ }\href {\doibase
  10.1016/S0550-3213(00)00437-5} {\bibfield  {journal} {\bibinfo  {journal}
  {Nucl. Phys. B}\ }\textbf {\bibinfo {volume} {586}},\ \bibinfo {pages}
  {397--426} (\bibinfo {year} {2000})},\ \Eprint
  {http://arxiv.org/abs/hep-ph/0005183} {arXiv:hep-ph/0005183} \BibitemShut
  {NoStop}%
\bibitem [{\citenamefont {Papinutto}\ \emph {et~al.}(2017)\citenamefont
  {Papinutto}, \citenamefont {Pena},\ and\ \citenamefont
  {Preti}}]{Papinutto:2016xpq}%
  \BibitemOpen
  \bibfield  {author} {\bibinfo {author} {\bibfnamefont {Mauro}\ \bibnamefont
  {Papinutto}}, \bibinfo {author} {\bibfnamefont {Carlos}\ \bibnamefont
  {Pena}}, \ and\ \bibinfo {author} {\bibfnamefont {David}\ \bibnamefont
  {Preti}},\ }\bibfield  {title} {\enquote {\bibinfo {title} {On the
  perturbative renormalization of four-quark operators for new physics},}\
  }\href {\doibase 10.1140/epjc/s10052-017-4930-6} {\bibfield  {journal}
  {\bibinfo  {journal} {Eur. Phys. J. C}\ }\textbf {\bibinfo {volume} {77}},\
  \bibinfo {pages} {376} (\bibinfo {year} {2017})},\ \bibinfo {note} {[Erratum:
  Eur.Phys.J.C 78, 21 (2018)]},\ \Eprint {http://arxiv.org/abs/1612.06461}
  {arXiv:1612.06461 [hep-lat]} \BibitemShut {NoStop}%
\bibitem [{\citenamefont {Baikov}\ \emph
  {et~al.}(2017{\natexlab{a}})\citenamefont {Baikov}, \citenamefont
  {Chetyrkin},\ and\ \citenamefont {K\"uhn}}]{Baikov:2016tgj}%
  \BibitemOpen
  \bibfield  {author} {\bibinfo {author} {\bibfnamefont {P.~A.}\ \bibnamefont
  {Baikov}}, \bibinfo {author} {\bibfnamefont {K.~G.}\ \bibnamefont
  {Chetyrkin}}, \ and\ \bibinfo {author} {\bibfnamefont {J.~H.}\ \bibnamefont
  {K\"uhn}},\ }\bibfield  {title} {\enquote {\bibinfo {title} {{Five-Loop
  Running of the QCD coupling constant}},}\ }\href {\doibase
  10.1103/PhysRevLett.118.082002} {\bibfield  {journal} {\bibinfo  {journal}
  {Phys. Rev. Lett.}\ }\textbf {\bibinfo {volume} {118}},\ \bibinfo {pages}
  {082002} (\bibinfo {year} {2017}{\natexlab{a}})},\ \Eprint
  {http://arxiv.org/abs/1606.08659} {arXiv:1606.08659 [hep-ph]} \BibitemShut
  {NoStop}%
\bibitem [{\citenamefont {Herzog}\ \emph {et~al.}(2017)\citenamefont {Herzog},
  \citenamefont {Ruijl}, \citenamefont {Ueda}, \citenamefont {Vermaseren},\
  and\ \citenamefont {Vogt}}]{Herzog:2017ohr}%
  \BibitemOpen
  \bibfield  {author} {\bibinfo {author} {\bibfnamefont {F.}~\bibnamefont
  {Herzog}}, \bibinfo {author} {\bibfnamefont {B.}~\bibnamefont {Ruijl}},
  \bibinfo {author} {\bibfnamefont {T.}~\bibnamefont {Ueda}}, \bibinfo {author}
  {\bibfnamefont {J.~A.~M.}\ \bibnamefont {Vermaseren}}, \ and\ \bibinfo
  {author} {\bibfnamefont {A.}~\bibnamefont {Vogt}},\ }\bibfield  {title}
  {\enquote {\bibinfo {title} {{The five-loop beta function of Yang-Mills
  theory with fermions}},}\ }\href {\doibase 10.1007/JHEP02(2017)090}
  {\bibfield  {journal} {\bibinfo  {journal} {JHEP}\ }\textbf {\bibinfo
  {volume} {02}},\ \bibinfo {pages} {090} (\bibinfo {year} {2017})},\ \Eprint
  {http://arxiv.org/abs/1701.01404} {arXiv:1701.01404 [hep-ph]} \BibitemShut
  {NoStop}%
\bibitem [{\citenamefont {Luthe}\ \emph
  {et~al.}(2017{\natexlab{a}})\citenamefont {Luthe}, \citenamefont {Maier},
  \citenamefont {Marquard},\ and\ \citenamefont {Schroder}}]{Luthe:2017ttc}%
  \BibitemOpen
  \bibfield  {author} {\bibinfo {author} {\bibfnamefont {Thomas}\ \bibnamefont
  {Luthe}}, \bibinfo {author} {\bibfnamefont {Andreas}\ \bibnamefont {Maier}},
  \bibinfo {author} {\bibfnamefont {Peter}\ \bibnamefont {Marquard}}, \ and\
  \bibinfo {author} {\bibfnamefont {York}\ \bibnamefont {Schroder}},\
  }\bibfield  {title} {\enquote {\bibinfo {title} {{Complete renormalization of
  QCD at five loops}},}\ }\href {\doibase 10.1007/JHEP03(2017)020} {\bibfield
  {journal} {\bibinfo  {journal} {JHEP}\ }\textbf {\bibinfo {volume} {03}},\
  \bibinfo {pages} {020} (\bibinfo {year} {2017}{\natexlab{a}})},\ \Eprint
  {http://arxiv.org/abs/1701.07068} {arXiv:1701.07068 [hep-ph]} \BibitemShut
  {NoStop}%
\bibitem [{\citenamefont {Chetyrkin}\ \emph {et~al.}(2006)\citenamefont
  {Chetyrkin}, \citenamefont {Kuhn},\ and\ \citenamefont
  {Sturm}}]{Chetyrkin:2005ia}%
  \BibitemOpen
  \bibfield  {author} {\bibinfo {author} {\bibfnamefont {K.~G.}\ \bibnamefont
  {Chetyrkin}}, \bibinfo {author} {\bibfnamefont {Johann~H.}\ \bibnamefont
  {Kuhn}}, \ and\ \bibinfo {author} {\bibfnamefont {Christian}\ \bibnamefont
  {Sturm}},\ }\bibfield  {title} {\enquote {\bibinfo {title} {{QCD decoupling
  at four loops}},}\ }\href {\doibase 10.1016/j.nuclphysb.2006.03.020}
  {\bibfield  {journal} {\bibinfo  {journal} {Nucl. Phys. B}\ }\textbf
  {\bibinfo {volume} {744}},\ \bibinfo {pages} {121--135} (\bibinfo {year}
  {2006})},\ \Eprint {http://arxiv.org/abs/hep-ph/0512060}
  {arXiv:hep-ph/0512060} \BibitemShut {NoStop}%
\bibitem [{\citenamefont {Schroder}\ and\ \citenamefont
  {Steinhauser}(2006)}]{Schroder:2005hy}%
  \BibitemOpen
  \bibfield  {author} {\bibinfo {author} {\bibfnamefont {Y.}~\bibnamefont
  {Schroder}}\ and\ \bibinfo {author} {\bibfnamefont {M.}~\bibnamefont
  {Steinhauser}},\ }\bibfield  {title} {\enquote {\bibinfo {title} {{Four-loop
  decoupling relations for the strong coupling}},}\ }\href {\doibase
  10.1088/1126-6708/2006/01/051} {\bibfield  {journal} {\bibinfo  {journal}
  {JHEP}\ }\textbf {\bibinfo {volume} {01}},\ \bibinfo {pages} {051} (\bibinfo
  {year} {2006})},\ \Eprint {http://arxiv.org/abs/hep-ph/0512058}
  {arXiv:hep-ph/0512058} \BibitemShut {NoStop}%
\bibitem [{\citenamefont {Liu}\ and\ \citenamefont
  {Steinhauser}(2015)}]{Liu:2015fxa}%
  \BibitemOpen
  \bibfield  {author} {\bibinfo {author} {\bibfnamefont {Tao}\ \bibnamefont
  {Liu}}\ and\ \bibinfo {author} {\bibfnamefont {Matthias}\ \bibnamefont
  {Steinhauser}},\ }\bibfield  {title} {\enquote {\bibinfo {title} {{Decoupling
  of heavy quarks at four loops and effective Higgs-fermion coupling}},}\
  }\href {\doibase 10.1016/j.physletb.2015.05.023} {\bibfield  {journal}
  {\bibinfo  {journal} {Phys. Lett. B}\ }\textbf {\bibinfo {volume} {746}},\
  \bibinfo {pages} {330--334} (\bibinfo {year} {2015})},\ \Eprint
  {http://arxiv.org/abs/1502.04719} {arXiv:1502.04719 [hep-ph]} \BibitemShut
  {NoStop}%
\bibitem [{\citenamefont {Chetyrkin}\ \emph {et~al.}(2000)\citenamefont
  {Chetyrkin}, \citenamefont {Kuhn},\ and\ \citenamefont
  {Steinhauser}}]{Chetyrkin:2000yt}%
  \BibitemOpen
  \bibfield  {author} {\bibinfo {author} {\bibfnamefont {K.~G.}\ \bibnamefont
  {Chetyrkin}}, \bibinfo {author} {\bibfnamefont {Johann~H.}\ \bibnamefont
  {Kuhn}}, \ and\ \bibinfo {author} {\bibfnamefont {M.}~\bibnamefont
  {Steinhauser}},\ }\bibfield  {title} {\enquote {\bibinfo {title} {{RunDec: A
  Mathematica package for running and decoupling of the strong coupling and
  quark masses}},}\ }\href {\doibase 10.1016/S0010-4655(00)00155-7} {\bibfield
  {journal} {\bibinfo  {journal} {Comput. Phys. Commun.}\ }\textbf {\bibinfo
  {volume} {133}},\ \bibinfo {pages} {43--65} (\bibinfo {year} {2000})},\
  \Eprint {http://arxiv.org/abs/hep-ph/0004189} {arXiv:hep-ph/0004189}
  \BibitemShut {NoStop}%
\bibitem [{\citenamefont {Schmidt}\ and\ \citenamefont
  {Steinhauser}(2012)}]{Schmidt:2012az}%
  \BibitemOpen
  \bibfield  {author} {\bibinfo {author} {\bibfnamefont {Barbara}\ \bibnamefont
  {Schmidt}}\ and\ \bibinfo {author} {\bibfnamefont {Matthias}\ \bibnamefont
  {Steinhauser}},\ }\bibfield  {title} {\enquote {\bibinfo {title} {{CRunDec: a
  C++ package for running and decoupling of the strong coupling and quark
  masses}},}\ }\href {\doibase 10.1016/j.cpc.2012.03.023} {\bibfield  {journal}
  {\bibinfo  {journal} {Comput. Phys. Commun.}\ }\textbf {\bibinfo {volume}
  {183}},\ \bibinfo {pages} {1845--1848} (\bibinfo {year} {2012})},\ \Eprint
  {http://arxiv.org/abs/1201.6149} {arXiv:1201.6149 [hep-ph]} \BibitemShut
  {NoStop}%
\bibitem [{\citenamefont {Herren}\ and\ \citenamefont
  {Steinhauser}(2018)}]{Herren:2017osy}%
  \BibitemOpen
  \bibfield  {author} {\bibinfo {author} {\bibfnamefont {Florian}\ \bibnamefont
  {Herren}}\ and\ \bibinfo {author} {\bibfnamefont {Matthias}\ \bibnamefont
  {Steinhauser}},\ }\bibfield  {title} {\enquote {\bibinfo {title} {{Version 3
  of RunDec and CRunDec}},}\ }\href {\doibase 10.1016/j.cpc.2017.11.014}
  {\bibfield  {journal} {\bibinfo  {journal} {Comput. Phys. Commun.}\ }\textbf
  {\bibinfo {volume} {224}},\ \bibinfo {pages} {333--345} (\bibinfo {year}
  {2018})},\ \Eprint {http://arxiv.org/abs/1703.03751} {arXiv:1703.03751
  [hep-ph]} \BibitemShut {NoStop}%
\bibitem [{\citenamefont {Baikov}\ \emph {et~al.}(2014)\citenamefont {Baikov},
  \citenamefont {Chetyrkin},\ and\ \citenamefont {K\"uhn}}]{Baikov:2014qja}%
  \BibitemOpen
  \bibfield  {author} {\bibinfo {author} {\bibfnamefont {P.~A.}\ \bibnamefont
  {Baikov}}, \bibinfo {author} {\bibfnamefont {K.~G.}\ \bibnamefont
  {Chetyrkin}}, \ and\ \bibinfo {author} {\bibfnamefont {J.~H.}\ \bibnamefont
  {K\"uhn}},\ }\bibfield  {title} {\enquote {\bibinfo {title} {{Quark Mass and
  Field Anomalous Dimensions to ${\cal O}(\alpha_s^5)$}},}\ }\href {\doibase
  10.1007/JHEP10(2014)076} {\bibfield  {journal} {\bibinfo  {journal} {JHEP}\
  }\textbf {\bibinfo {volume} {10}},\ \bibinfo {pages} {076} (\bibinfo {year}
  {2014})},\ \Eprint {http://arxiv.org/abs/1402.6611} {arXiv:1402.6611
  [hep-ph]} \BibitemShut {NoStop}%
\bibitem [{\citenamefont {Luthe}\ \emph
  {et~al.}(2017{\natexlab{b}})\citenamefont {Luthe}, \citenamefont {Maier},
  \citenamefont {Marquard},\ and\ \citenamefont {Schr\"oder}}]{Luthe:2016xec}%
  \BibitemOpen
  \bibfield  {author} {\bibinfo {author} {\bibfnamefont {Thomas}\ \bibnamefont
  {Luthe}}, \bibinfo {author} {\bibfnamefont {Andreas}\ \bibnamefont {Maier}},
  \bibinfo {author} {\bibfnamefont {Peter}\ \bibnamefont {Marquard}}, \ and\
  \bibinfo {author} {\bibfnamefont {York}\ \bibnamefont {Schr\"oder}},\
  }\bibfield  {title} {\enquote {\bibinfo {title} {{Five-loop quark mass and
  field anomalous dimensions for a general gauge group}},}\ }\href {\doibase
  10.1007/JHEP01(2017)081} {\bibfield  {journal} {\bibinfo  {journal} {JHEP}\
  }\textbf {\bibinfo {volume} {01}},\ \bibinfo {pages} {081} (\bibinfo {year}
  {2017}{\natexlab{b}})},\ \Eprint {http://arxiv.org/abs/1612.05512}
  {arXiv:1612.05512 [hep-ph]} \BibitemShut {NoStop}%
\bibitem [{\citenamefont {Baikov}\ \emph
  {et~al.}(2017{\natexlab{b}})\citenamefont {Baikov}, \citenamefont
  {Chetyrkin},\ and\ \citenamefont {K\"uhn}}]{Baikov:2017ujl}%
  \BibitemOpen
  \bibfield  {author} {\bibinfo {author} {\bibfnamefont {P.~A.}\ \bibnamefont
  {Baikov}}, \bibinfo {author} {\bibfnamefont {K.~G.}\ \bibnamefont
  {Chetyrkin}}, \ and\ \bibinfo {author} {\bibfnamefont {J.~H.}\ \bibnamefont
  {K\"uhn}},\ }\bibfield  {title} {\enquote {\bibinfo {title} {{Five-loop
  fermion anomalous dimension for a general gauge group from four-loop massless
  propagators}},}\ }\href {\doibase 10.1007/JHEP04(2017)119} {\bibfield
  {journal} {\bibinfo  {journal} {JHEP}\ }\textbf {\bibinfo {volume} {04}},\
  \bibinfo {pages} {119} (\bibinfo {year} {2017}{\natexlab{b}})},\ \Eprint
  {http://arxiv.org/abs/1702.01458} {arXiv:1702.01458 [hep-ph]} \BibitemShut
  {NoStop}%
\bibitem [{\citenamefont {Aoki}\ \emph {et~al.}(2020)\citenamefont {Aoki} \emph
  {et~al.}}]{FlavourLatticeAveragingGroup:2019iem}%
  \BibitemOpen
  \bibfield  {author} {\bibinfo {author} {\bibfnamefont {S.}~\bibnamefont
  {Aoki}} \emph {et~al.} (\bibinfo {collaboration} {Flavour Lattice Averaging
  Group}),\ }\bibfield  {title} {\enquote {\bibinfo {title} {{FLAG} review
  2019: Flavour lattice averaging group ({FLAG})},}\ }\href {\doibase
  10.1140/epjc/s10052-019-7354-7} {\bibfield  {journal} {\bibinfo  {journal}
  {Eur. Phys. J. C}\ }\textbf {\bibinfo {volume} {80}},\ \bibinfo {pages} {113}
  (\bibinfo {year} {2020})},\ \Eprint {http://arxiv.org/abs/1902.08191}
  {arXiv:1902.08191 [hep-lat]} \BibitemShut {NoStop}%
\end{thebibliography}%

\end{document}